\documentclass[twocolumn]{aastex61}

\newcommand{\Ha}{H$\alpha$}
\newcommand{\Hb}{H$\beta$}

\newcommand{\NII}{[N\,{\sc ii}]}
\newcommand{\OIII}{[O\,{\sc iii}]}
\newcommand{\OII}{[O\,{\sc iii}]}
\newcommand{\SII}{[S\,{\sc ii}]}

\usepackage{verbatim}
\usepackage{graphicx}
\usepackage{grffile}
\usepackage{url}

\begin{document}
\title{The FMOS-COSMOS survey of star-forming galaxies at $z\sim1.6$ VI: Redshift and emission-line catalog and basic properties of star-forming galaxies}

\author[0000-0001-9044-1747]{Daichi Kashino}
\affiliation{Department of Physics, ETH Z{\" u}rich, Wolfgang-Pauli-strasse 27, CH-8093, Z{\" u}rich, Switzerland}
\author{John D.~Silverman}
\affiliation{Kavli Institute for the Physics and Mathematics of the Universe, the University of Tokyo, Kashiwanoha, Kashiwa, Chiba 277-8583, Japan (Kavli IPMU, WPI)}
\author{David Sanders}
\affiliation{Institute for Astronomy, University of Hawaii, 2680 Woodlawn Drive, Honolulu, HI 96822, USA}
\author{Jeyhan Kartaltepe}
\affiliation{School of Physics and Astronomy, Rochester Institute of Technology, 84 Lomb Memorial Drive, Rochester, NY 14623, USA}
\author{Emanuele Daddi}
\affiliation{Laboratoire AIM-Paris-Saclay, CEA/DSM-CNRS-Universit{\' e} Paris Diderot, Irfu/Service d'Astrophysique, CEA-Saclay, Service d'Astrophysique, F-91191 Gif-sur-Yvette, France}
\author{Alvio Renzini}
\affiliation{INAF Osservatorio Astronomico di Padova, vicolo dell'Osservatorio 5, I-35122 Padova, Italy}
\author{Giulia Rodighiero}
\affiliation{Dipartimento di Fisica e Astronomia, Universit{\' a} di Padova, vicolo dell'Osservatorio, 2, I-35122 Padova, Italy}
\author{Annagrazia Puglisi}
\affiliation{Laboratoire AIM-Paris-Saclay, CEA/DSM-CNRS-Universit{\' e} Paris Diderot, Irfu/Service d'Astrophysique, CEA-Saclay, Service d'Astrophysique, F-91191 Gif-sur-Yvette, France}
\author{Francesco Valentino}
\affiliation{Dark Cosmology Centre, Niels Bohr Institute, University of Copenhagen, Juliane Maries Vej 30, DK-2100 Copenhagen, Denmark}
\author{St{\'e}phanie Juneau}
\affiliation{National Optical Astronomy Observatory, 950 North Cherry Avenue, Tucson, AZ 85719, USA}
\author{Nobuo Arimoto}
\affiliation{Astronomy Program, Department of Physics and Astronomy, Seoul National University, 599 Gwanak-ro, Gwanaku-gu, Seoul 151-742, Korea}
\author{Tohru Nagao}
\affiliation{Graduate School of Science and Engineering, Ehime University, 2-5 Bunkyo-cho, Matsuyama 790-8577, Japan}
\author{Olivier Ilbert}
\affiliation{Aix Marseille Universit{\'e}, CNRS, LAM - Laboratoire d'Astrophysique de Marseille, 38 rue F. Joliot-Curie, F-13388 Marseille, France}
\author{Olivier Le F{\`e}vre}
\affiliation{Aix Marseille Universit{\'e}, CNRS, LAM - Laboratoire d'Astrophysique de Marseille, 38 rue F. Joliot-Curie, F-13388 Marseille, France}
\author{Anton.~M.~Koekemoer}
\affiliation{Space Telescope Science Institute, 3700 San Martin Drive, Baltimore, MD 21218, USA}

\correspondingauthor{Daichi Kashino}
\email{kashinod@phys.ethz.ch}

% Abstract
\begin{abstract}

We present a new data release from the Fiber Multi-Object Spectrograph (FMOS)-COSMOS survey, which contains the measurements of spectroscopic redshift and flux of rest-frame optical emission lines (\Ha, \NII, \SII, \Hb, \OIII) for 1931 galaxies out of a total of 5484 objects observed over the 1.7~deg$^2$ COSMOS field.  We obtained $H$-band and $J$-band medium-resolution ($R\sim3000$) spectra with FMOS mounted on the Subaru telescope, which offers an in-fiber line flux sensitivity limit of $\sim 1 \times 10^{-17}~\mathrm{erg~s^{-1}~cm^{-2}}$ for an on-source exposure time of five hours.  The full sample contains the main population of star-forming galaxies at $z\sim1.6$ over the stellar mass range $10^{9.5}\lesssim M_\ast/M_\odot \lesssim 10^{11.5}$, as well as other subsamples of infrared-luminous galaxies detected by {\it Spitzer} and {\it Herschel} at the same and lower ($z\sim0.9$) redshifts and X-ray emitting galaxies detected by {\it Chandra}.  This paper presents an overview of our spectral analyses, a description of the sample characteristics, and a summary of the basic properties of emission-line galaxies.  We use the larger sample to re-define the stellar mass--star formation rate relation based on the dust-corrected \Ha\ luminosity, and find that the individual galaxies are better fit with a parametrization including a bending feature at $M_\ast\approx10^{10.2}~M_\odot$, and that the intrinsic scatter increases with $M_\ast$ from 0.19 to $0.37~\mathrm{dex}$.  We also confirm with higher confidence that the massive ($M_\ast\gtrsim10^{10.5}~M_\odot$) galaxies are chemically mature as much as local galaxies with the same stellar masses, and that the massive galaxies have lower \SII/\Ha\ ratios for their \OIII/\Hb, as compared to local galaxies, which is indicative of enhancement in ionization parameter.

 \end{abstract}

\section{Introduction}

Over the last decade, numerous rest-frame optical spectral data of galaxies at $1\lesssim z\lesssim 3$ have been delivered by near-infrared spectrographs installed on 8--10-m class telescopes \citep[e.g.,][]{2014ApJ...795..165S,2015ApJS..218...15K,2015ApJ...799..209W,2016MNRAS.456.1195H}.  These datasets have revolutionized our understanding of the formation and evolution of galaxies across the so-called `cosmic noon' epoch that marks the peak and the subsequent transition to the declining phase of the cosmic star formation history.   
Before the data flood by such large near-infrared surveys, however, the relatively narrow redshift range of $1.4<z<1.7$ had long been dubbed the `redshift desert' since all strong spectral features in the rest-frame optical such as \Ha, \OIII, \Hb, and \OII\ are redshifted into the infrared, while strong rest-frame UV features such as C\,{\sc iv}/S\,{\sc ii} absorption lines, Lyman break, and Ly$\alpha$ emission line, are still too blue, thus both being out of reach of conventional optical spectrographs.  This redshift interval had thus remained as the last gap to be explored by dedicated spectroscopic surveys even after recent deep optical spectroscopic surveys such as VIMOS Ultra-Deep Survey (VUDS; see Figure 13 of \citealt{2015A&A...576A..79L}).

To fill in this redshift gap, we have carried out a large spectroscopic campaign, the FMOS-COSMOS survey, first with the low-resolution mode ($R\sim600$) over 2010 November --2012 February and then in the high-resolution mode ($R\sim3000$) over 2012 March--2016 April.  The Fiber Multi-Object Spectrograph (FMOS) is a near-infrared instrument mounted on the Subaru telescope and uniquely characterized by its wide field-of-view (FoV; 30~arcmin in diameter) and high multiplicity (400 fibers), making it one of the ideal instruments to conduct a large spectroscopic survey to detect the rest-frame optical emission lines (e.g., \Hb, \OIII, \Ha, \NII, \SII) at the redshift desert.  
We refer the reader to \citet{2015ApJS..220...12S} for the high-resolution survey design and some early results, and to Kartaltepe et al.,~in~prep for the details of the low-resolution survey.  Spectral datasets obtained through the early runs of the FMOS-COSMOS survey have allowed us to investigate various aspects of star-forming galaxies in the $1.43\le z\le 1.74$ redshift range, including their dust extinction and the evolution of a so-called main sequence of star-forming galaxies \citep{2013ApJ...777L...8K,2014MNRAS.443...19R}, the evolution of the gas-phase metallicity and the stellar mass--metallicity relation \citep{2014ApJ...792...75Z,2017ApJ...835...88K}, the excitation/ionization conditions of main-sequence galaxies \citep{2017ApJ...835...88K}, the properties of far-IR luminous galaxies \citep{2015ApJ...806L..35K}, heavily dust-obscured starburst galaxies \citep{2017ApJ...838L..18P}, and Type-I active galactic nuclei (AGNs) \citep{2013ApJ...771...64M,2018arXiv181007445S}, the spatial clustering of host dark matter halos \citep{2017ApJ...843..138K}, and the number counts of \Ha-emitting galaxies \citep{2017MNRAS.472.4878V}.  Complementary efforts for the follow-up measurement of the [O\,{\sc ii}]$\lambda\lambda3726,3729$ emission lines with Keck/DEIMOS have constrained the electron density \citep{2017MNRAS.465.3220K} and the ionization parameter \citep{2018MNRAS.tmp..962K} for a subset of the FMOS-COSMOS galaxies.  Furthermore, high-resolution molecular line intensity and kinematic mapping have been obtained with ALMA for an FMOS sample of starburst galaxies, which have revealed their high efficiency of converting gas into stars \citep{2015ApJ...812L..23S,2018arXiv181001596S}.   Our ALMA follow up observations also discovered a very unique system, where pair of two galaxies are colliding, and revealed their high gas mass  and highly enhanced star formation efficiency \citep{2018arXiv181001595S}.

In this paper, we present the final catalog of the full sample from the FMOS high-resolution observations over the COSMOS field, which includes measurements of spectroscopic redshifts and fluxes of strong emission lines.  This catalog includes observations done after February 2014 that were not reported in our previous papers.  Based on the latest catalog, we present the basic characteristics of emission-line galaxies, evaluate the possible biases of the FMOS sample with an \Ha\ detection, and then revisit with substantially improved statistics the properties of star-forming galaxies at $z\sim1.6$, including dust extinction, the stellar mass--star formation rate (SFR) relation, and the properties of the interstellar medium (ISM) using the emission-line diagnostics.  

The paper is organized as follows.   
In Sections \ref{sec:FMOS-COSMOS} and \ref{sec:samples} we give an overview of the survey and galaxy samples in the FMOS-COSMOS survey.  
In Section \ref{sec:calib} we describe spectral analyses, emission-line flux measurements, flux calibration, and aperture correction.
In Section \ref{sec:detections} we summarize detections of the emission lines and spectroscopic redshift estimates.
In Sections \ref{sec:lineprop} and \ref{sec:assessment} we present the basic measurements of the emission lines, and assess the quality of the redshift and flux measurements.    
In Section \ref{sec:re_L16} we re-evaluate the characteristics of our FMOS sample relative to the current COSMOS photometric catalog ({\it COSMOS2015}; \citealt{2016ApJS..224...24L}).  
In Section \ref{sec:lephare} we describe our spectral energy distribution (SED) fitting procedure for the stellar mass estimation, and drive SFRs from the rest-frame UV emission and the observed \Ha\ fluxes, with correction for dust extinction.
In Section \ref{sec:MS} we measure the relation between stellar mass and SFR at $z\sim1.6$ and discuss the behavior and intrinsic scatter of the relation.
In Section \ref{sec:lineratio} we revisit  the ionization/excitation conditions of the ionized nebulae by using key emission-line ratio diagnostics, and re-define the $M_\ast$--\NII/\Ha\ relation.
In Section \ref{sec:Ha_vs_OIII} we compare between the \Ha- and \OIII-emitter samples, and discuss possible biases induced by the use of the \OIII\ line as a galaxy tracer.
We give a summary of this paper in Section \ref{sec:summary}.
This paper and the catalog use a standard flat cosmology ($h=0.7,~\Omega_\Lambda=0.7,~\Omega_\mathrm{M}=0.3)$, AB magnitudes, and a \citet{2003PASP..115..763C} initial mass function (IMF).

\section{The FMOS-COSMOS observations \label{sec:FMOS-COSMOS}}

\begin{deluxetable*}{lcccc}
\tablecaption{Summary of Subaru/FMOS HR observations (2012 March -- 2014 February) \label{tb:observations1}}
\tablehead{\colhead{Date (Local Time)}&
		\colhead{Program ID}&
		\colhead{Pointing}&
		\colhead{Grating}&
		\colhead{Total exp time (hr)}}
\startdata
2012-03-12 & UH-B3 & HR4 & $H$-long & 5 \\
2012-03-13 & S12A-096 & HR1 & $H$-long & 5 \\
2012-03-14 & S12A-096 & HR2 & $H$-long & 4.5 \\
2012-03-15 & S12A-096 & HR1 & $H$-long & 5 \\
2012-03-16 & S12A-096 & HR3 & $H$-long & 4 \\
2012-03-17 & S12A-096 & HR1 & $H$-short & 4 \\
2012-03-18 & UH-B5 & HR1 & $J$-long & 4.5 \\
2012-12-28 & UH-18A & HR2 & J-long & 3.5 \\
2013-01-18 & S12B-045I & HR3 & H-long & 3 \\
2013-01-19 & S12B-045I & HR4 & H-long & 3.5 \\
2013-01-20 & UH-18A & HR3 & J-long & 4.5 \\
2013-01-21 & UH-18A & HR4 & J-long & 3.5 \\
2013-12-28 & S12B-045I & HR2 & H-long & 4.25 \\
2014-01-21 & UH-11A & EXT1 & H-long & 2.25 \\
2014-01-23 & UH-11A & EXT2 & H-long & 2 \\
2014-01-24 & S12B-045I & HR3 & H-long & 1.5 \\
2014-01-25 & S12B-045I & HR1 & H-long & 5.25 \\
2014-01-26 & S12B-045I & HR4 & H-long & 5 \\
2014-02-07 & S12B-045I & HR1 & J-long & 4.5 \\
2014-02-08\tablenotemark{a}  & S12B-045I & HR4 & J-long & 5.5 \\
2014-02-09\tablenotemark{a}  & S12B-045I & HR4 & J-long & 5 \\
2014-02-10 & UH-38A & EXT3 & H-long & 5.5 \\
\enddata
\tablenotetext{a}{These two $J$-long observations have been conducted with the same fiber allocation design (i.e., the same galaxies were observed in total 10.5 hours in the two nights.)}
\end{deluxetable*}

\begin{deluxetable*}{lcccc}
\tablecaption{Summary of Subaru/FMOS HR observations (2014 March -- 2016 April) \label{tb:observations2}}
\tablehead{\colhead{Date (Local Time)}&
		\colhead{Program ID}&
		\colhead{Pointing}&
		\colhead{Grating}&
		\colhead{Total exp time (hr)}}
\startdata
2014-03-06 & UH-38A & EXT1 & J-long & 5.5 \\
2014-12-02\tablenotemark{a}  & UH-25A & HR4E & H-long & 2.25 \\
2015-02-08\tablenotemark{a}  & S15A-134I & HR7 & H-long & 4.5 \\
2015-02-11\tablenotemark{a}  & UH-22A & HR7 & H-long & 5 \\
2015-02-12\tablenotemark{a}  & UH-22A & HR6 & H-long & 3.5 \\
2015-04-10\tablenotemark{a}  & UH-22A & HR5 & H-long & 4 \\
2015-04-11\tablenotemark{a}  & UH-22A & HR5 & H-long & 1.5 \\
2016-01-15 & S15A-134I & HR8E & H-long & 4.5 \\
2016-01-16 & S15A-134I & HR4E & H-long & 4.5 \\
2016-01-17 & S15A-134I & HR1E & H-long & 4.5 \\
2016-01-18 & UH-24A & HRC0 & H-long & 5 \\
2016-01-19 & UH-24A & HR6 & H-long & 5 \\
2016-01-20 & UH-24A & HR7 & J-long & 5 \\
2016-03-24 & UH-11A & HR1 & J-long & 3.5 \\
2016-03-26 & S16A-054I & HR2 & J-long & 4.5 \\
2016-03-27 & S16A-054I & HR4 & J-long & 4.5 \\
2016-03-29 & S16A-054I & HR3 & J-long & 4 \\
2016-03-30 & S16A-054I & HR7E & H-long & 4 \\
2016-04-19 & UH-11A & HR1E & J-long & 3.25 \\
2016-04-20 & UH-11A & HR6E & J-long & 3.5 \\
2016-04-21 - 1st half & S16A-054I & HR1 & J-long & 3.5  (3.0 in IRS2) \\
2016-04-22 - 1st half & S16A-054I & HR3 & J-long & 3.5 \\
2016-04-23 - 1st half & S16A-054I & HR2 & J-long & 3.25 \\
2016-04-24 - 1st half & S16A-054I & HR8E & J-long & 3 \\
\enddata
\tablenotetext{a}{Observations from December 2014 to April 2015 have been conducted using only a single spectrograph IRS1.}
\end{deluxetable*}

Here we present a summary of our all FMOS observing runs with the high-resolution mode.  The survey design, observations and data analysis have been described in our previous papers \citep[e.g.,][]{2015ApJ...812L..23S}.

Tables \ref{tb:observations1} and \ref{tb:observations2} summarize all observing runs in the high-resolution (HR) mode from March 2012 to April 2016, with Table \ref{tb:observations1} referring to runs having produced the data used in our previous papers, and Table \ref{tb:observations2} listing the observations afterwards.  Observing runs with a program ID starting with `S' were conducted within the Subaru Japan time (PI John Silverman), while runs with a program ID with `UH' were carried out through the time slots allocated to the University of Hawaii (PI David Sanders).  Although the intended exposure time was five hours for all runs, in some runs it was reduced due to the observing conditions.  We also note that observations from December 2014 to April 2015 were conducted using only a single FMOS spectrograph (IRS1) due to instrumental problem with the second spectrograph (IRS2), thus the number of targets per run was correspondingly reduced by half, while in all other runs $\sim200$ targets were observed simultaneously using the two spectrographs with the cross beam switching mode, in which two fibers are allocated for a single target.

\begin{figure}[htbp] 
   \centering
   \includegraphics[width=3.45in]{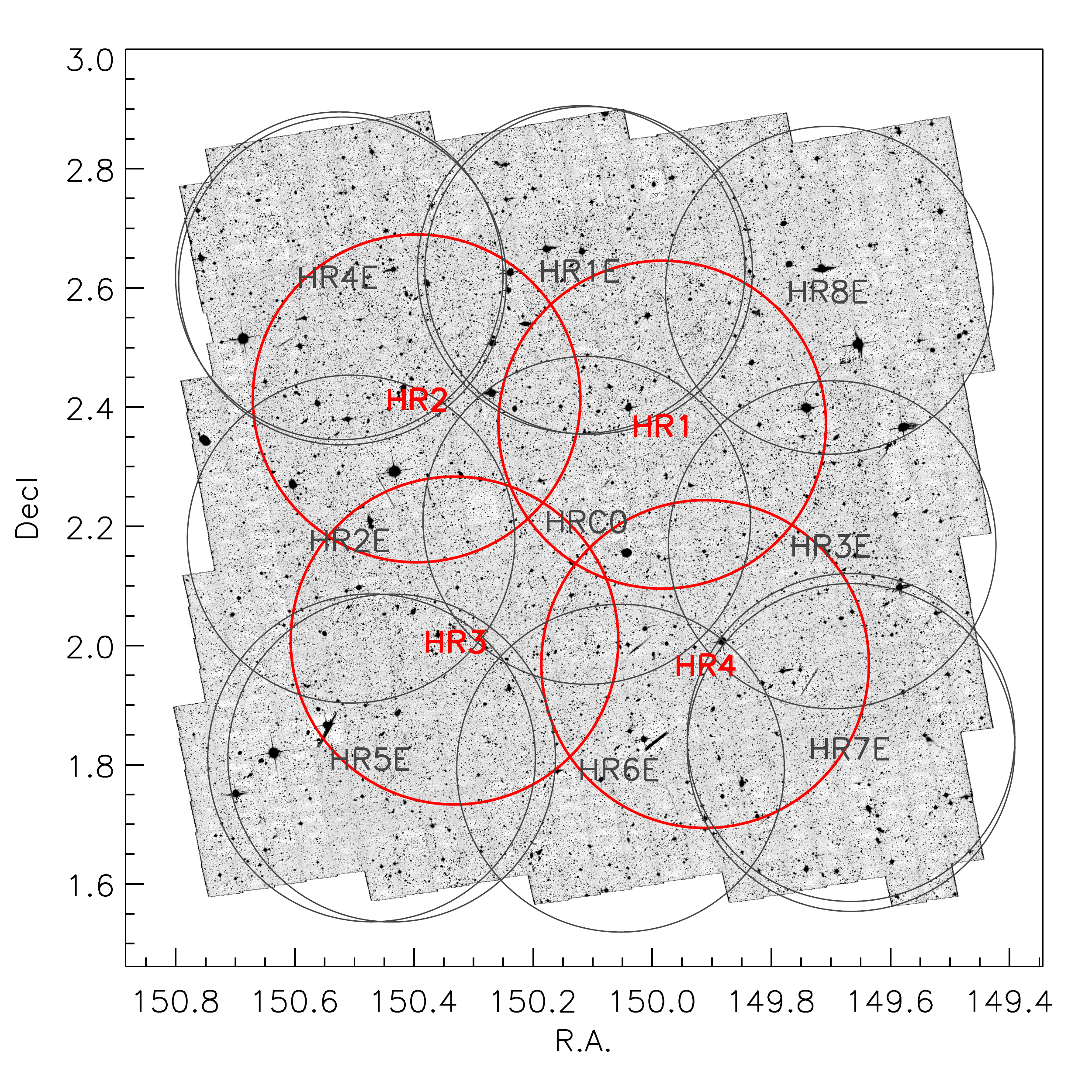} 
   \includegraphics[width=3.45in]{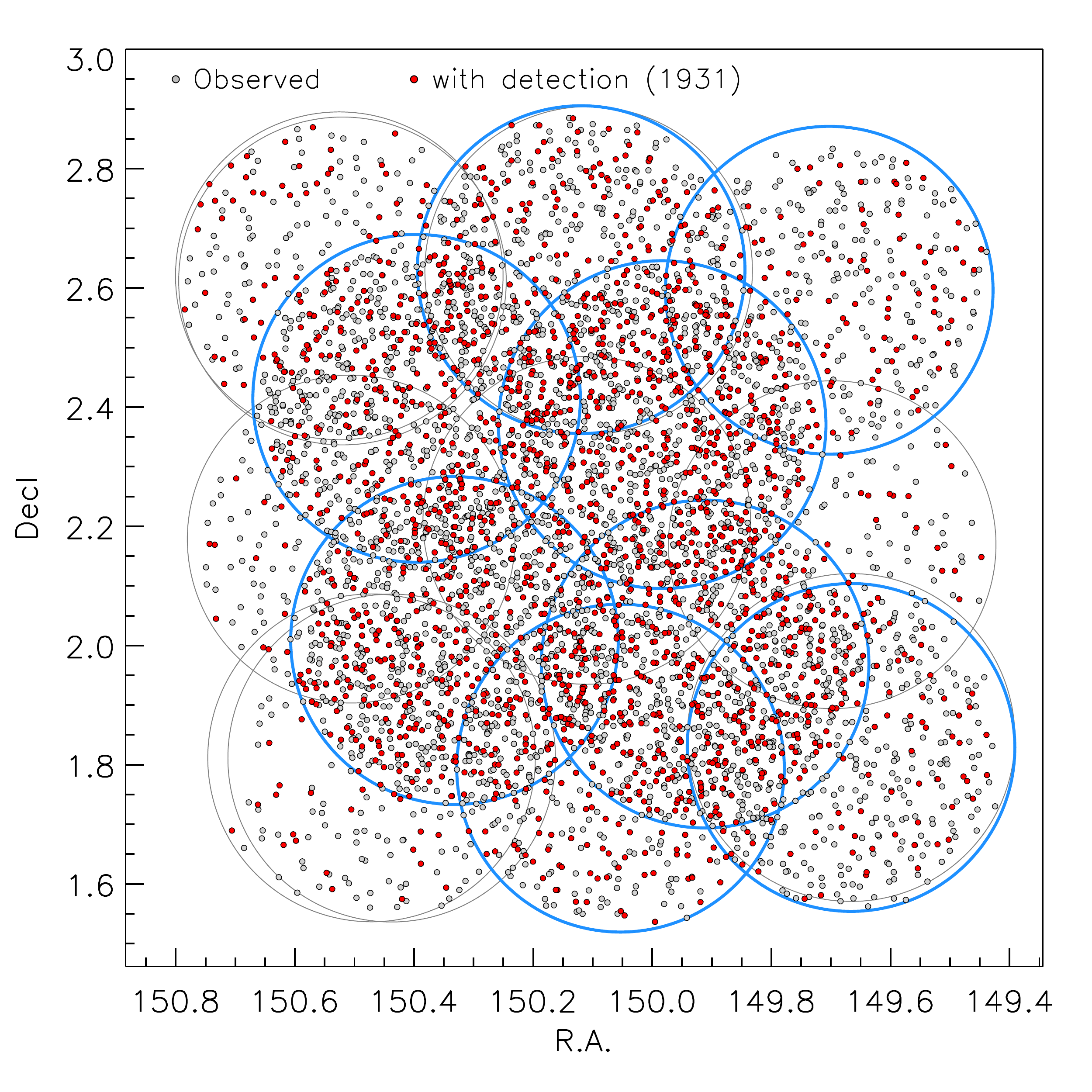}
   \caption{
   Upper panel: the FMOS pawprint overlaid on the {\it HST}/ACS mosaic of the COSMOS field \citep{2007ApJS..172..196K,2010MNRAS.401..371M}.  Large circles show the FoV of each FMOS pointing.  The central area of 0.81~$\mathrm{deg}^2$ covered by four FoVs (HR1--4) are highlighted by red.  
   Lower panel: On-sky distribution of all galaxies in the FMOS-COSMOS catalog (gray circles).  Red circles indicate those with any spectroscopic redshift estimate (1931 objects with $z\mathrm{Flag}\ge1$; see Section \ref{sec:detections}).  The pawprints visited with the $J$-long grating are highlighted by thick blue circles.}
   \label{fig:pawprints}
\end{figure}

Figure \ref{fig:pawprints} shows the complete FMOS-COSMOS pawprint over the {\it Hubble Space Telescope (HST)} Advanced Camera for Surveys (ACS) mosaic image in the COSMOS field (\citealt{2007ApJS..172..196K,2010MNRAS.401..371M}; upper panel) and with the individual objects in the FMOS-COSMOS catalog (lower panel).  Each circle with radius of 16.5 arcmin corresponds to the FMOS FoV and their positions are reported in Table \ref{tb:pawprints}.  $H$-long spectroscopy has been conducted once or more times at all positions, while the $J$-long observations have been conducted only at 8 out of 13 positions due to the reduction of the observing time for bad weather or instrumental troubles.  These eight FoVs are highlighted in the lower panel of Figure \ref{fig:pawprints}.  As clearly shown in the lower panel, the sampling rate is not uniform across the whole survey area due to the difference in the number of pointings and the presence of overlapping regions.  In particular, the central area covered by four FoVs (HR1, 2, 3, and 4) has a higher sampling rate with their larger number of repeat pointings relative to the outer region.  The full FMOS-COSMOS area is $1.70~\mathrm{deg^2}$ and the central area covered by the four FoVs is $0.81~\mathrm{deg^2}$.

\begin{deluxetable}{lcccc}
\tablecaption{Location of the FMOS pawprints\label{tb:pawprints}}
\tablehead{\colhead{Name}&
		\colhead{R.A.}&
		\colhead{Declination}&
		\colhead{$N_\mathrm{visits}$}&
		\colhead{$N_\mathrm{visits}$} \\
		\colhead{}&
		\colhead{(J2000)}&
		\colhead{(J2000)}&
		\colhead{$H$-long}&
		\colhead{$J$-long}}
\startdata
HR1 &    09:59:56.0 &   +02:22:14 & 3~(+1)\tablenotemark{a} & 4 \ \\
HR2 &    10:01:35.0 &   +02:24:52 & 2 & 3 \\
HR3 &    10:01:19.7 &   +02:00:29 & 3 & 3 \\
HR4 &    09:59:38.7 &   +01:58:08 & 3 & 3\tablenotemark{b} \\
HR1E &    10:00:28.6 &   +02:37:49 & 1 & 1 \\
HR2E &    10:02: 1.4 &   +02:10:42 & 1 & 0\\
HR3E &    09:58:48.2 &   +02:10:21 & 1 & 0\\
% ex5 &    10:02: 4.8 &   +02:36:41 & * & *\\
HR4E &    10:02: 6.1 &   +02:37:12 & 2\tablenotemark{c} & 0\\
HR5E &    10:01:51.1 &   +01:48:41 & 2\tablenotemark{c} & 0\\
HR6E &    10:00:12.8 &   +01:47:39 & 2\tablenotemark{c} & 1\\
HR7E &    09:58:28.6 &   +01:49:24 & 3\tablenotemark{c} & 1\\
HR8E &    09:58:38.1 &   +02:35:45 & 1 & 1\\
HRC0 &    10:00:26.4 &   +02:12:36 & 1 & 0\\
\hline
Full area\tablenotemark{d} & $1.70~\mathrm{deg^2}$ \\
HR1--4\tablenotemark{e} & $0.81~\mathrm{deg^2}$ \\
\enddata
\tablenotetext{a}{`+1' denotes an additional $H$-short observation.}
\tablenotetext{b}{Two of the three $J$-long observations in HR4 have conducted with the same fiber allocation (i.e., observed the same galaxies in total $10.5$ hours in two nights; see Table \ref{tb:observations1}).}
\tablenotetext{c}{Observations from 2014 Dec to 2015 Apr have been conducted with only a single spectrograph IRS1 (see Table \ref{tb:observations2}).}
\tablenotetext{d}{Area of the full FMOS-COSMOS survey field.}
\tablenotetext{e}{Area covered by the central four FMOS pawprints (HR1--4).}
\end{deluxetable}

% ====================================================
\section{Galaxies in the FMOS-COSMOS catalog \label{sec:samples}}

\subsection{Star-forming galaxies at $z\sim1.6$ \label{sec:Primary-HL}}

Our main galaxy sample is based on the COSMOS photometric catalogs \citep{2007ApJS..172...99C,2010ApJ...708..202M,2012A&A...544A.156M,2010ApJ...709..644I,2013A&A...556A..55I} that include the Ultra-VISTA/VIRCam photometry.  For observations after February 2015, we used the updated photometric catalog from \citet{2015A&A...579A...2I}.  For each galaxy in these catalogs, the global properties, such as photometric redshift, stellar mass, SFR, and the level of extinction, are estimated from SED fits to the broad- and intermediate-band photometry using LePhare \citep{2002MNRAS.329..355A,2006A&A...457..841I}.  We refer the reader to \citealt{2010ApJ...709..644I,2013A&A...556A..55I,2015A&A...579A...2I} for further details.  For the target selection, we computed the predicted flux of the \Ha\ emission line from the intrinsic SFR and extinction estimated from our own SED fitting adopting a constant star formation history \citep[see][]{2015ApJS..220...12S}.

For the FMOS $H$-long spectroscopy, we preferentially selected galaxies that satisfy the criteria listed below.
\begin{enumerate}
\item $K_\mathrm{S} \le 23.5$, a magnitude limit on the Ultra-VISTA $K_\mathrm{S}$-band photometry (auto magnitude).
\item $1.46 \le z_\mathrm{phot} \le 1.72$, a range for which \Ha\ falls within the FMOS $H$-long spectral window.
\item $M_\ast \ge 10^{9.77}~M_\odot$ (for a Chabrier IMF)
\item Predicted total ({\it not} in-fiber) \Ha\ flux $F^\mathrm{pred}_\mathrm{H\alpha} \ge 1 \times 10^{-16}~\mathrm{erg~s^{-1}~cm^{-2}}$.
\end{enumerate}
We refer to those satisfying all the above criteria as {\it Primary} objects.  From the COSMOS photometric catalog, 3876 objects are identified to meet the above criteria (the {\it Primary-parent} sample), and 1582 objects were observed in the $H$-long mode (the {\it Primary-HL} sample).  

\begin{figure*}[t] 
   \centering
   \includegraphics[width=6.0in]{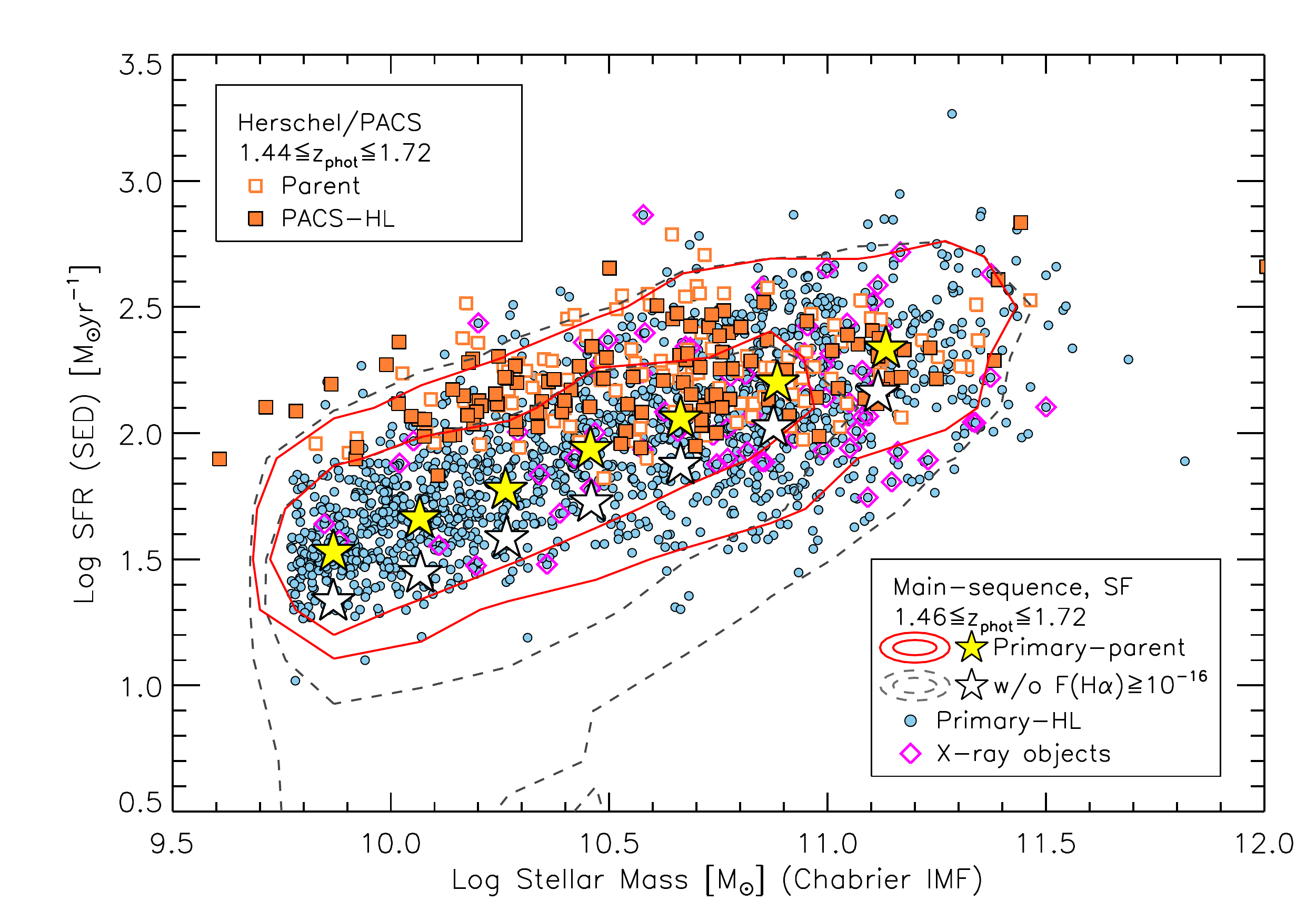}
   \caption{
   $M_\ast$ vs. SFR (from SED fits) for the target samples at $z\sim1.6$ in the FMOS-COSMOS survey.  Red solid and black dashed contours show the distribution (containing 68 and 90\%) of the parent galaxies limited with (i.e., Primary-parent sample) and without the threshold $F_\mathrm{H\alpha}^\mathrm{pred} \ge 1\times10^{-16}~\mathrm{erg~s^{-1}~cm^{-2}}$.  Correspondingly, yellow and white stars indicate the median SFRs in bins of $M_\ast$, respectively, for the parent galaxies with and without the limit on the predicted \Ha\ flux.  Objects in the Primary-HL sample are indicated by blue circles, with X-ray detected objects marked by magenta diamonds.  Orange empty and filled squares indicate the PACS-parent and PACS-HL samples (Section \ref{sec:PEP}).}
   \label{fig:M_vs_SFR_subsamples}
\end{figure*}

Figure \ref{fig:M_vs_SFR_subsamples} shows the SFR as a function of $M_\ast$ for the Primary-parent sample (red contours), and the Primary-HL sample (blue circles).  The observed objects trace the so-called main sequence \citep[e.g.,][]{2007ApJ...660L..43N} of star-forming galaxies over two orders of magnitudes in stellar mass.  However, the limit on the predicted \Ha\ flux removed a substantial fraction (60~\%) of potential targets selected only with the $K_\mathrm{S}$ and $M_\ast$ criteria (shown by black dashed contours).  In Figure \ref{fig:M_vs_SFR_subsamples}, we indicate median SFRs in bins of $M_\ast$ separately for the parent galaxies limited with and without the limit on the predicted \Ha\ flux.  It is shown that the limit on $F^\mathrm{pred}_\mathrm{H\alpha}$ results in the observed sample being biased $\sim0.2~\mathrm{dex}$ higher in the average SFR, at all stellar masses.  We found that the Primary-HL sample includes 70 objects detected by {\it Chandra} X-ray observations (see Section \ref{sec:Chandra}) by checking counterparts.  These X-ray-detected objects are excluded for studies on the properties of a pure star-forming population.  

\begin{figure*}[t] 
   \centering
   \includegraphics[width=7in]{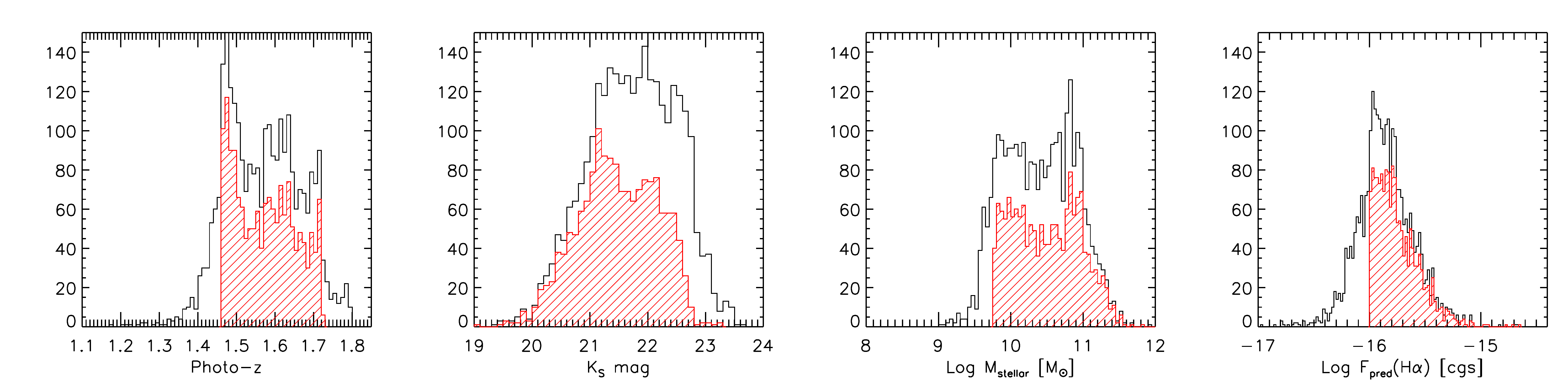}
   \caption{
From left to right, distributions of $z_\mathrm{phot}$, $K_\mathrm{S}$ magnitude, $M_\ast$, and predicted $F_\mathrm{H\alpha}$ for the Primary-HL (red hatched histograms) and the Secondary-HL (plus Primary-HL) sample (empty histograms).}
   \label{fig:tar_comb_hists}
\end{figure*}

In addition to the Primary sample, the FMOS-COSMOS catalog contains a substantial number of star-forming galaxies at $z\sim1.6$ not satisfying all the criteria described above.  This is because the criteria were loosened down to $M_\ast \ge 10^{9.57}~M_\odot$ and/or $F^\mathrm{pred}_\mathrm{H\alpha} \ge 4 \times 10^{-17}~\mathrm{erg~s^{-1}~cm^{-2}}$ for a part of runs, and we also allocated substantial number of fibers through the program to those at $z\sim1.6$ identified in the photometric catalog, but not satisfying all the criteria for the Primary objects.  We refer to these objects observed with the $H$-long grating as the {\it Secondary-HL} sample, which contains 1242 objects.  In Figure \ref{fig:tar_comb_hists}, we show the distributions of galaxy properties for both the Primary-HL and the Primary+Secondary-HL objects.  The Secondary-HL sample includes objects with lower or higher $z_\mathrm{phot}$ and/or lower $M_\ast$ outside the limits, while the majority are those with $F^\mathrm{pred}_\mathrm{H\alpha}$ lower than the threshold.

\begin{figure}[t] 
   \centering
   \includegraphics[width=3.5in]{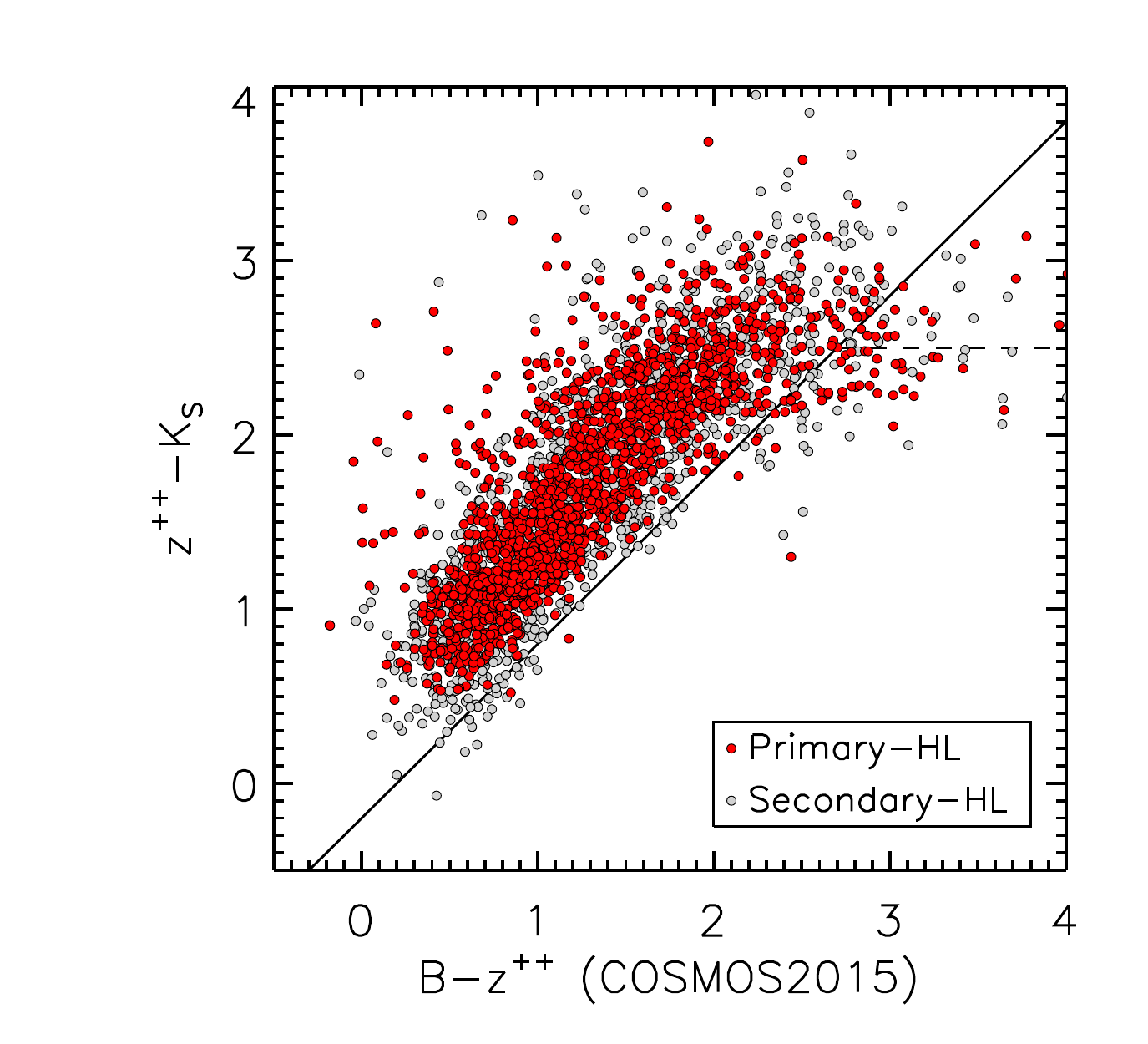}
   \caption{
  Primary-HL (red circles) and Secondary-HL (gray circles) samples in the BzK diagram.  The solid and dashed lines indicate the boundaries for distinguishing $z>1.4$ star-forming, $z>1.4$ quiescent, and $z<1.4$ galaxies, defined by \citet{2004ApJ...617..746D}.
}
   \label{fig:BzK}
\end{figure}

In Figure \ref{fig:BzK}, we show the Primary-HL and Secondary-HL objects in the ($B-z$) vs. ($z-K$) diagram.  These colors are based on the photometric measurements (Subaru $B$ and $z^{++}$, and UltraVISTA $K_\mathrm{S}$) given in the COSMOS2015 catalog \citep{2016ApJS..224...24L}.  It is demonstrated that the majority (95\%) of the Primary+Secondary-HL sample match the so-called s$BzK$ selection \citep{2004ApJ...617..746D}.

\subsection{Far-IR sources from the Herschel PACS Evolutionary Probe (PEP) Survey \label{sec:PEP}}

{\it Herschel}-PACS observations cover the COSMOS field at 100~$\mu$m and 160~$\mu$m, down to a $5\sigma$ detection limits of $\sim 8~\mathrm{mJy}$ and $\sim 17~\mathrm{mJy}$, respectively \citep{2011A&A...532A..90L}.  These limits correspond to a SFR of roughly $100~M_\odot~\mathrm{yr^{-1}}$ at $z\sim1.6$.  We allocated fibers to these FIR-luminous objects for particular studies of starburst and dust-rich galaxies \citep[e.g.,][]{2015ApJ...806L..35K,2017ApJ...838L..18P} also in view of their follow-up with ALMA \citep{2015ApJ...812L..23S,2018arXiv181001595S,2018arXiv181001596S}.  The objects were selected by cross-matching between the PACS Evolutionary Probe (PRP) survey catalog and the IRAC-selected catalog of \citet{2010ApJ...709..644I}, and their stellar mass and SFR are derived from SED fits (further detailed in \citealt{2011ApJ...739L..40R}).  For these objects, a higher priority with respect to fiber allocation had to be made since these objects are rare and would not be sufficiently targeted otherwise. 

Our parent sample of the PACS sources contains 231 objects in the range $1.44 \le z_\mathrm{phot} \le 1.72$, and 116 objects were selected for FMOS $H$-long spectroscopy.  We refer to these objects as the PACS-HL sample.  Figure \ref{fig:M_vs_SFR_subsamples} shows the distribution of the {\it Herschel}-PACS sample in the $M_\ast$ vs. SFR plot.  It is shown that these objects are limited to be above an SFR of $\sim 100~M_\odot~\mathrm{yr^{-1}}$.  Further analyses of this subsample are presented in companion papers (\citealt{2017ApJ...838L..18P}; Kartaltepe et al., in prep).

\subsection{{\it Chandra} X-ray sources \label{sec:Chandra}}

We have dedicated a fraction of FMOS fibers to obtain spectra for optical/near-infrared counterparts to X-ray sources from the {\it Chandra} COSMOS Legacy survey \citep{2009ApJS..184..158E,2016ApJ...819...62C}.  The FMOS-COSMOS catalog includes 84 X-ray-selected objects intentionally targeted as compulsory.  However, there are many X-ray sources other than those, which have been targeted as star-forming galaxies (i.e., the Primary/Secondary-HL sample) or infrared galaxies.  We thus performed position matching between the full FMOS-COSMOS catalog and the full {\it Chandra} COSMOS Legacy catalog\footnote{The Chandra catalogs are available here: http://cosmos.astro.caltech.edu/page/xray}.  In total, we found an X-ray counterpart for 742 (including the intended 84 objects) among all FMOS extragalactic objects.  Most of these X-ray-detected objects are probably AGN-hosting galaxies.  These objects are not included in the analyses presented in the rest of this paper, but studies of these X-ray sources are presented in companion papers (\citealt{2018arXiv181007445S}, Kashino et al. in prep.). 

\subsection{Additional infrared galaxies \label{sec:JKsample}}

We also allocated a substantial number of fibers to observe lower redshift  ($0.7\lesssim z \lesssim 1.1$, where \Ha\ falls in the $J$-long grating) infrared galaxies selected from S-COSMOS {\it Spitzer}-MIPS observations \citep{2007ApJS..172...86S} and {\it Herschel} PACS and SPIRE from the PEP \citep{2011A&A...532A..90L} and HerMES \citep{2012MNRAS.424.1614O} surveys, respectively.  We used the photometric redshifts of \citet{2015A&A...579A...2I} and \citet[][for X-ray detected AGN]{2011ApJ...742...61S} for the source selection.  We derived the total IR luminosity, calculated from the best-fit IR template using the SED fitting code LePhare and integrating from 8 to 1000 microns. These luminosities range between $10^{11} \lesssim L_\mathrm{IR}/L_\odot \lesssim 10^{12.5}$, spanning the luminosity regime of LIRG/ULIRG (Luminous and Ultraluminous Infrared Galaxies, see review by \citealt{1996ARA&A..34..749S}).  
Our parent sample includes 1818 objects between $0.66 \le z_\mathrm{phot} \le 1.06$.  Of those, we observed 344 using the $J$-long grating.   Further analysis of this particular sub-sample will be presented in a future paper (Kartaltepe et al., in prep).

\section{Flux measurement and calibration \label{sec:calib}}
\subsection{Emission-line fitting \label{sec:fitting}}

Our procedure for the emission-line fitting makes use of the  IDL package {\sc mpfit} \citep{2009ASPC..411..251M}.  Candidate emission lines were modeled with a Gaussian profile, after subtracting the continuum.  The \Ha\ and \NII\ or \Hb\ and \OIII\ lines were fit simultaneously while fixing the velocity widths to be the same and allowing no relative offset for the line centroids.  The flux ratios of the doublet \NII$\lambda$6584/6548 and \OIII$\lambda$5007/4959 were fixed to be 2.96 and 2.98, respectively \citep{2000MNRAS.312..813S}.

The spectral data processed with the standard reduction pipeline, FIBRE-pac \citep{2012PASJ...64...59I}, are given in units of $\mathrm{\mu Jy}$, which were converted into flux density per unit wavelength, i.e., $\mathrm{erg~s^{-1}~cm^{-2}~\AA^{-1}}$, before fitting.  The observed flux density $F_{\lambda,i}$, where $i$ denotes the pixel index, was fit with weights defined as the inverse of the squared noise spectra output by the pipeline.  The weights $W_i$ were set to zero for pixels impacted by the OH mask or sky residuals (see Figures 11 and 14 of \citealt{2015ApJ...812L..23S}).

We assessed the quality  of the fitting results based on the signal-to-noise (S/N) ratio calculated from the formal errors on the model parameters returned by the {\sc mpfitfun} code.  We emphasize that these S/N ratios do not include the uncertainties on the absolute flux calibration described in later sections.  In addition, we have also estimated the fraction of flux lost by bad pixels (i.e., pixels with $W_i=0$).  For all lines we define the `bad pixel loss' as  the fraction of the contribution occupied by the bad pixels  to the total integral of the Gaussian profile:
\begin{equation}
f_\mathrm{badpix} = \frac{\sum_{\left\{ i | W_i=0\right\}}  P_i}{\sum_i  P_i}
\end{equation}
where $P_i$ is the flux density of the best-fit Gaussian profile at the $i$th pixel (not the observed spectrum).  We disregard any tentative  line detections if $f_\mathrm{badpix}>0.7$.

\begin{figure}[tbp] 
   \centering
   \includegraphics[width=3.5in]{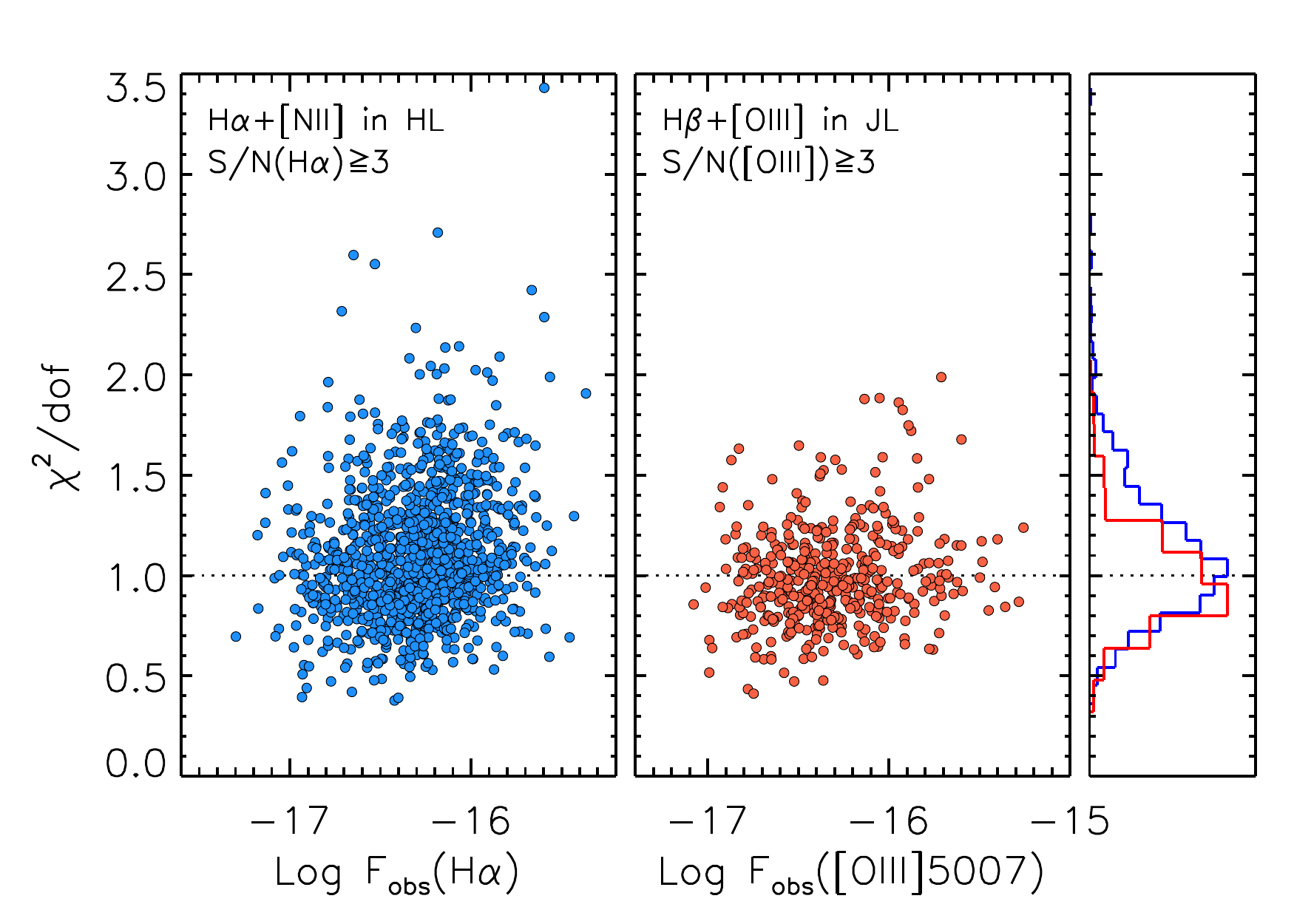}
   \includegraphics[width=3.5in]{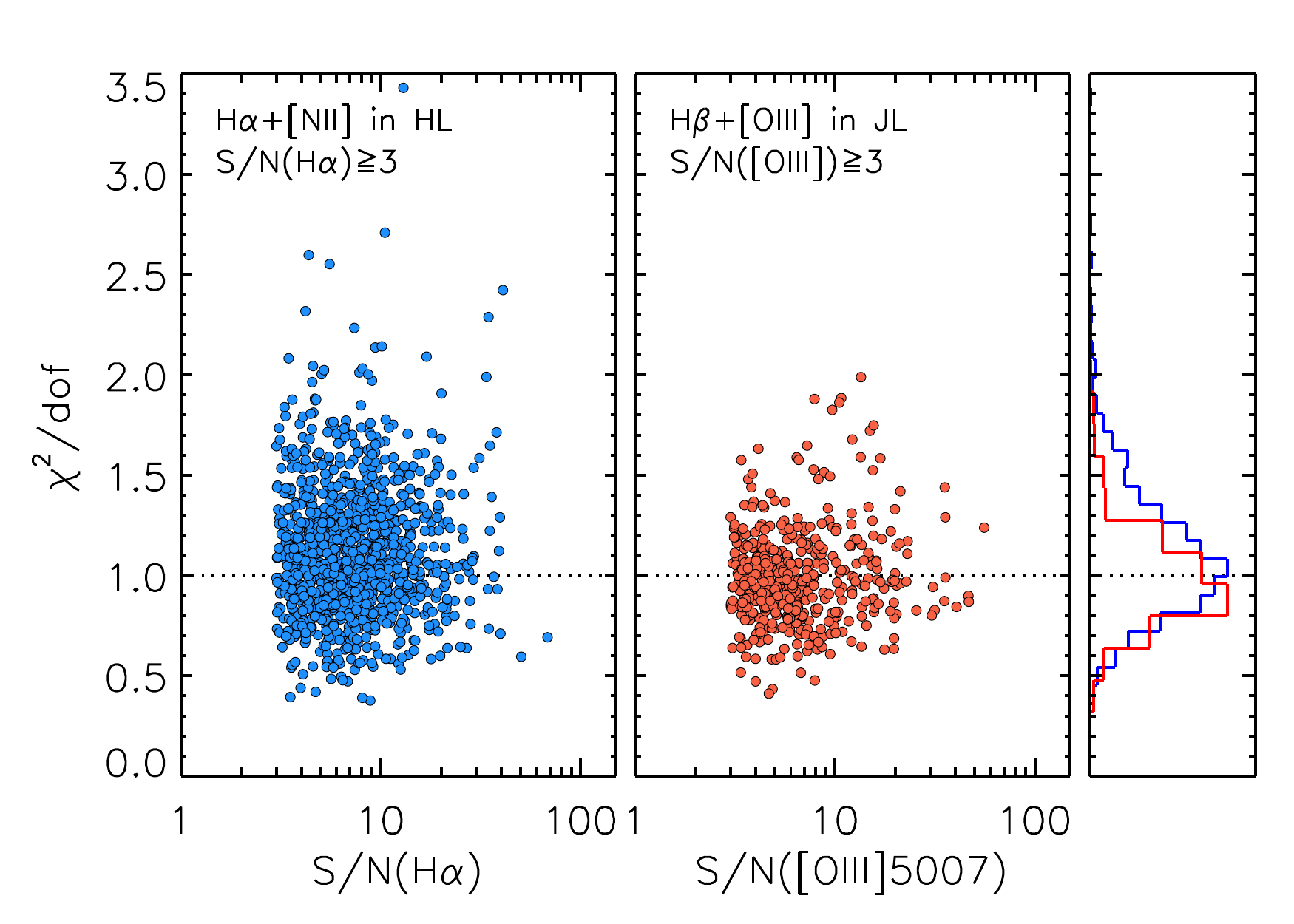} 
   \caption{
   The reduced chi-squared statistic ($\chi^2/\mathrm{dof}$) as a function of observed line flux (upper panel) and S/N (lower panel).  Left and middle panels show the results of fits to \Ha+\NII\ in the $H$-long band, and \Hb+\OIII\ in the $J$-long, respectively.  Right panels show the corresponding normalized distributions of the $\chi^2/\mathrm{dof}$ values. 
}
   \label{fig:fit_chisq}
\end{figure}

The goodness of the line fits is given by the reduced chi-squared statistic, $\chi^2/\mathrm{dof}$ where dof are the degrees of freedom in the fits.  Figure \ref{fig:fit_chisq} shows the resultant $\chi^2/\mathrm{dof}$ values as a function of line strength (upper panel) and of S/N (lower panel), separately for the \Ha+\NII\ in $H$-long and \Hb+\OIII\ in $J$-long.  The distribution of the reduced $\chi^2$ statistics clearly peaks at $\chi^2/\mathrm{dof}\simeq 1$, with no significant trends with either line strength or S/N.

In a relatively few cases, a prominent broad emission-line component was present and we included a secondary, broad component for \Ha\ or \Hb.  Furthermore, we also added a secondary narrow \Ha+\NII\ (or \Hb+\OIII) component with centroid and width different from the primary component, when necessary (e.g., a case that there is a prominent blueshifted component of the \OIII\ line, possibly attributed to an outflow).  Such exceptional handling was applied for only $5\%$ of the whole sample  (108 out of 1931 objects with a line detection).  Most of these objects are X-ray detected and we postpone detailed analysis of these objects to a future paper, while focusing here on the basic properties of normal star-forming galaxies.

\subsection{Upper Limits \label{sec:upperlimits}}

For non-detections of emission lines of interest we estimated upper limits on their in-fiber fluxes if we have a spectroscopic redshift estimate from any other detected lines in the FMOS spectra and the spectral coverage for undetected lines.  The S/N of an emission line depends not only on the flux and the typical noise level of the spectra, but also on the amount of loss due to bad pixels.  These effects have been considered on a case-by-case basis by performing dedicated Monte-Carlo simulations for each spectrum.  

For each object with an estimate of spectroscopic redshift, we created $N_\mathrm{sim}=500$ spectra containing an artificial emission line with a Gaussian profile at a specific observed-frame wavelength of undetected lines based on the $z_\mathrm{spec}$ estimate.  The line width was fixed to a typical FWHM of 300~$\mathrm{km~s^{-1}}$ (Section \ref{sec:Haprop}), and Gaussian noise was added to these artificial spectra based on the processed noise spectrum.  In doing so, we mimicked the impact of the OH lines and the masks.  We then performed a fitting procedure for these artificial spectra with various amplitudes in the same manner as the data, and estimated the 2$\sigma$ upper limit for each un-detected line by linearly fitting the sets of simulated fluxes and the associated S/Ns.

\subsection{Integrated flux density \label{sec:specmag}}

In addition to the line fluxes, we also measured the average flux density within the spectral window for individual objects regardless the presence or absence of a line detection. The average flux density $\left< f_\nu \right>$ and the associated errors were derived by integrating the extracted 1D spectrum of each galaxy as follows.
\begin{eqnarray}
\left< f_\nu \right> = \frac{{\sum_i f_{\nu, i}} W_i R_i}{\sum_i W_i R_i d\lambda} \label{eq:f_nu}\\
\Delta \left< f_\nu \right>= \frac{\sqrt{ \sum_i (N_{\nu,i} W_i R_i d\lambda)^2 }}{\sum_i W_i R_i d\lambda}
\end{eqnarray}
where  $d\lambda=1.25~\mathrm{\AA}$ is the wavelength pixel resolution, $N_i$ is the associated noise spectrum, and $R_i$ is a response curve.

Beside been used to estimate the equivalent widths of detected emission lines, these quantities can also allow for the absolute flux calibration by comparing them with the ground-based $H$ or $J$-band photometry.  For this purpose, we use the fixed $3''$-aperture magnitudes {\tt H(J)\_MAG\_APER3} from the UltraVISTA-DR2 survey \citep{2012A&A...544A.156M} provided in the COSMOS2015 catalog \citep{2016ApJS..224...24L} as reference, applying the recommended offset from aperture to total magnitudes (see Appendix of \citealt{2016ApJS..224...24L}).  For comparison with the reference photometry, we define $R_i$ in the above equations based on the response curve of the VISTA/VIRCam $H$ or $J$-band filters\footnote{The data for the filter response curves are available here: http://www.eso.org/sci/facilities/paranal/instruments/vircam/inst.html}, and flux densities were then converted to (AB) magnitudes.  In the calculation of these equations, we did not exclude the detected emission lines because our primary purpose is to compare these to the ground-based broad-band photometry, which in principle includes the emission line fluxes if exit\footnote{For estimating the emission line equivalent widths, we excluded the emission line components.\ref{sec:Haprop}.}.  We disregard the measurements with $S/N < 5$, and also exclude objects whose A- and/or B-position spectrum (obtained through the ABAB telescope nodding) falls on the detector next to those of flux standard stars since these spectra may be contaminated by leakage from the neighbor bright star spectrum.  Finally, we successfully measured the flux density for 2456 objects observed with the $H$-long grating, and for 1700 objects observed in $J$-long.  

\begin{figure}[t] 
   \centering
   \includegraphics[width=3.5in]{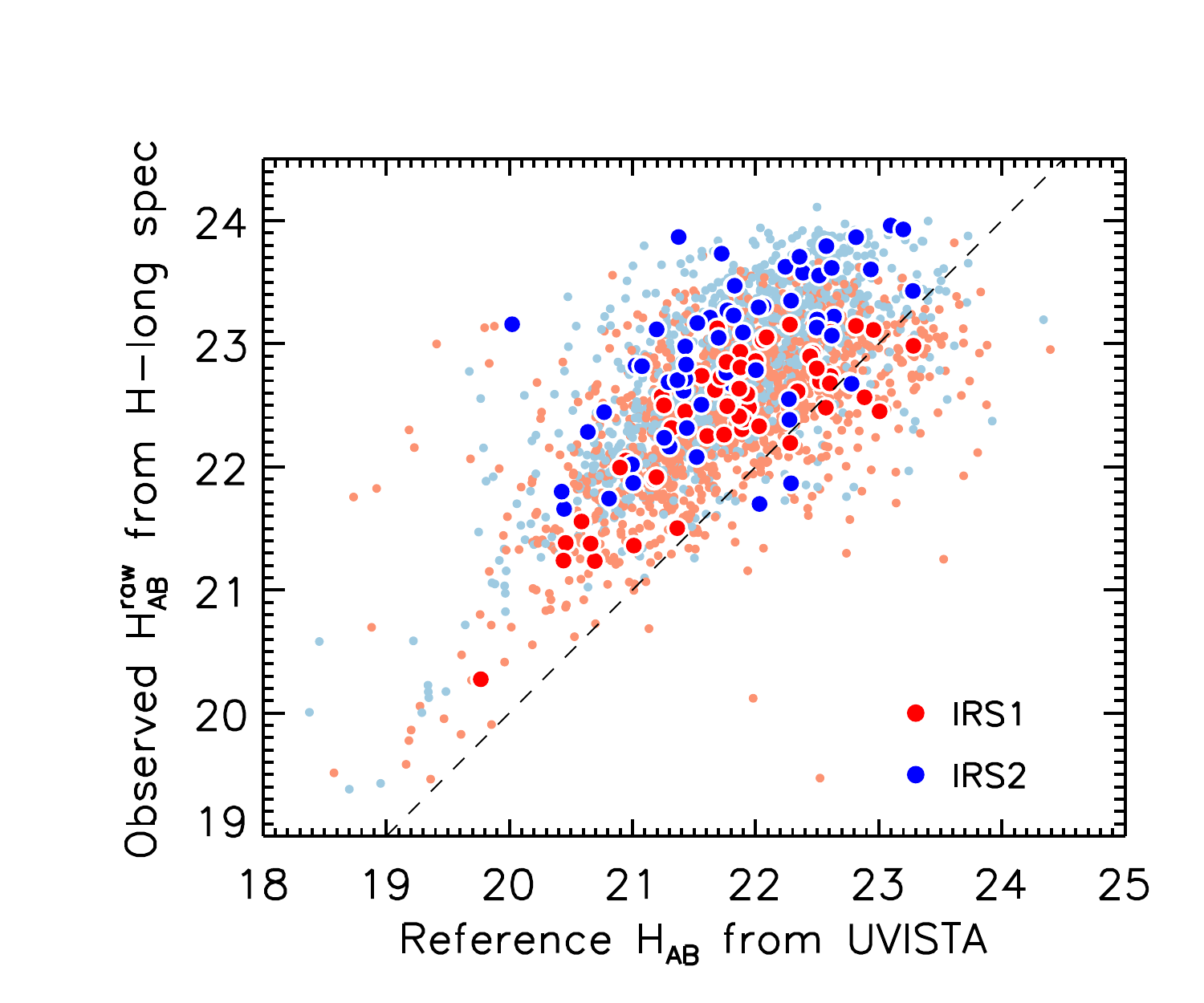}
   \caption{
   Observed `raw' $H_\mathrm{AB}$ from the $H$-long spectra vs. reference $H_\mathrm{AB}$ from the UltraVISTA.  Red and blue points correspond to the measurements with the two spectrographs IRS1 and IRS2, respectively.  Data points from a single observing run (2013-12-28) are highlighted with large symbols.  A global offset of $\sim 1~\mathrm{mag}$ from the one-to-one relation (dashed line) reflects the average aperture loss, while an offset of $\sim0.5~\mathrm{mag}$ between IRS1 (red) and IRS2 (blue) is due to the differential total throughput of these spectrographs.}
   \label{fig:specmag_raw_vs_refmag}
\end{figure}

In Figure \ref{fig:specmag_raw_vs_refmag}, we compared the observed magnitudes $H_\mathrm{AB}$ from the FMOS $H$-long spectra with the  UltraVISTA $H$-band magnitudes, separately for the two spectrographs (IRS1 and IRS2) of FMOS.  Here the observed values were computed from spectra produced by the standard reduction pipeline, and we refer to these as the `raw' magnitude.  Data points from a single observing run (2013-12-28) are highlighted for reference.  It is clear that there is a global offset of $\sim1$ mag in the observed magnitudes relative to the reference UltraVISTA magnitudes.  This reflects the loss flux falling outside the fiber aperture.  In addition, we can also see that an $\sim0.5$ mag systematic offset exists between the two spectrographs.  This offset is due to the difference in the total efficiency of the two spectrographs.  Prior to the aperture correction, we first corrected for this offset between the IRS1 and IRS2, as follows:
\begin{eqnarray}
H_\mathrm{IRS1} = H^\mathrm{raw}_\mathrm{IRS1} + (\left< \Delta H^\mathrm{raw}_\mathrm{IRS2} \right> - \left< \Delta H^\mathrm{raw}_\mathrm{IRS1} \right>)/2, \\
H_\mathrm{IRS2} = H^\mathrm{raw}_\mathrm{IRS2} - (\left< \Delta H^\mathrm{raw}_\mathrm{IRS2} \right> - \left< \Delta H^\mathrm{raw}_\mathrm{IRS1} \right>)/2 
\end{eqnarray}  
where $\left< \Delta H^\mathrm{raw}_{\mathrm{IRS}i} \right>$ is the median offset of the observed magnitude relative to the reference magnitude.  This correction has been done for each observing run independently.  We did the same for the $J$-band observations as well.

\begin{figure}[tbp] 
   \centering
   \includegraphics[width=3.5in]{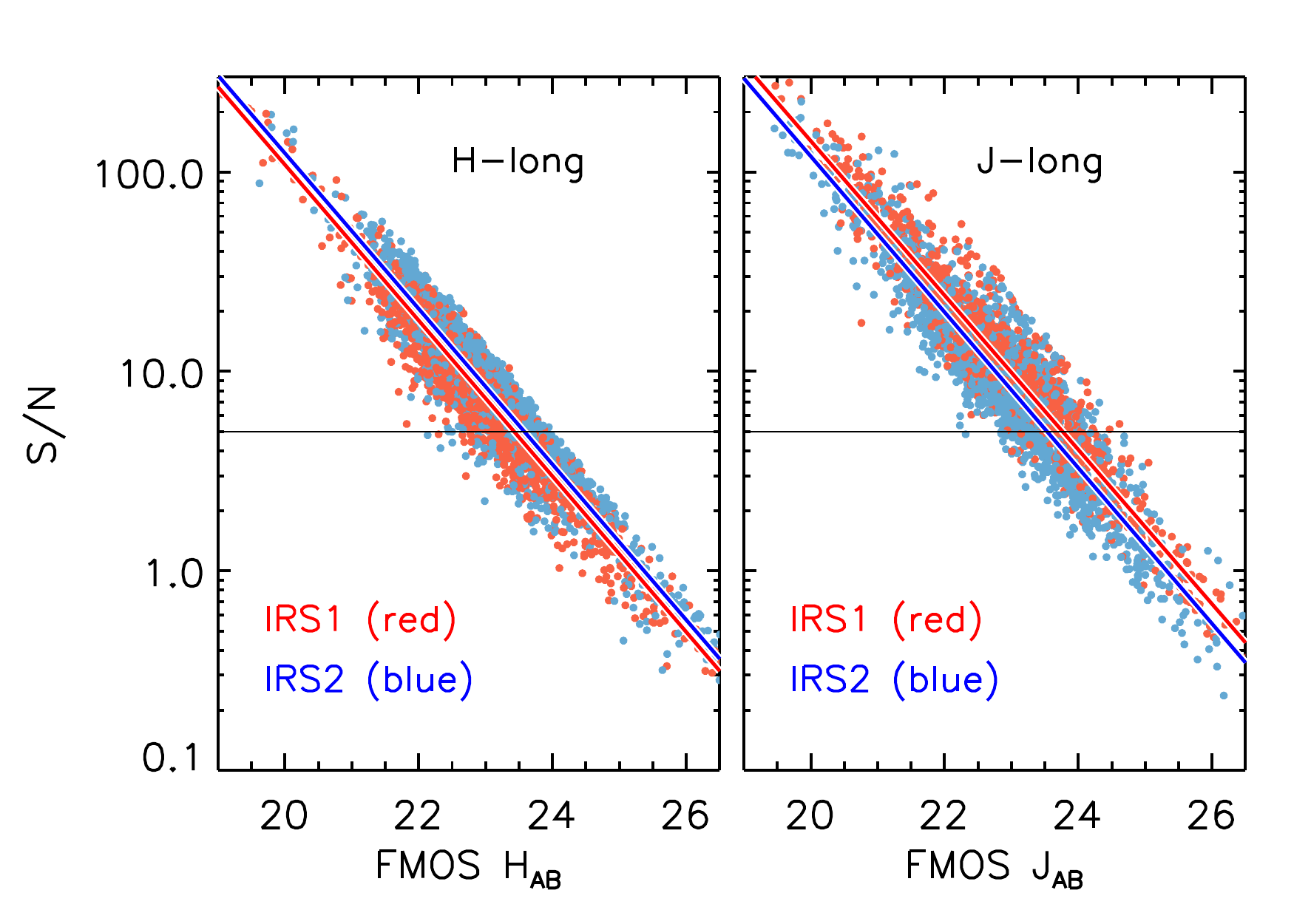} 
   \caption{Observed magnitude from the FMOS $H$-long (left panel) and $J$-long (right panel) spectra versus the estimated S/N.  The data from IRS1 and IRS2 are shown separately with red and blue, respectively.  The color solid lines are linear fits to the data, and horizontal lines indicate the threshold $S/N=5$.}
   \label{fig:mag_vs_SN}
\end{figure}

Figure \ref{fig:mag_vs_SN} shows the observed magnitudes after correcting for the offset between IRS1 and IRS2.  The magnitude from the $H$-long (left panel) and $J$-long (right panel) spectra are shown as a function of S/N ratios, separately for each spectrograph.  The correlations are in good agreement between the two spectrographs, and between the spectral windows.  The threshold $S/N=5$ corresponds to $\approx23.5$ ABmag for both $H$ and $J$.

We emphasize that, in the rest of the paper as well as in our emission-line catalog, the correction for the differential throughput between the two IRSs is applied for all observed quantities, including emission-line fluxes, formal errors and upper limits on line fluxes.  Therefore, catalog users do not need to care about this instrumental issue.  Meanwhile, the fluxes in the catalog denote the in-fiber values, hence the aperture correction should be applied using the correction factors given in the catalog if necessary (see the next subsection for details).

\subsection{Aperture correction \label{sec:apercorr}}

As already mentioned, the emission-line and broad-band fluxes measured from observed FMOS spectra arise from only the regions of each target falling within the $1''\!.2$-diameter aperture of the FMOS fibers.  Therefore, it is necessary to correct for flux falling outside the fiber aperture to obtain the total emission line flux of each galaxy.  The amount of aperture loss depends both on the intrinsic size of each galaxy and the conditions of the observation, which include variable seeing size and fluctuations of the fiber positions (typically $\sim 0''.2$; \citealt{2010PASJ...62.1135K}).  We define three methods for aperture correction.  

First, the aperture correction can be determined by simply comparing the observed $H$ (or $J$) flux density obtained by integrating the FMOS spectra to the reference broad-band magnitude for individual objects.  This method can be utilized for moderately luminous objects for which we have a good estimate of the integrated flux from the FMOS spectra (observed $H_\mathrm{AB}\lesssim22.5$).  This method cannot be applied for objects with poor continuum detection and suffering from the flux leakage from bright objects.  

Second, we can use the average offset of the observed magnitude relative to the reference magnitude for each observing run.  This method can be applied to fainter objects and those with insecure continuum measurement (e.g., impacted by leakage from a bright star) for correcting the emission line fluxes.

Lastly, we determine the aperture correction based on high-resolution imaging data.  In the COSMOS field, we can utilize images taken by the {\it HST}/ACS \citep{2007ApJS..172..196K,2010MNRAS.401..371M} that covers almost entirely the FMOS field and offers high spatial resolution.  The advantage of this method is that we can determine the aperture correction object-by-object taking into account their size property and a specific seeing size of the observing night.  Hereafter we describe in detail this third method (see also \citealt{2013ApJ...777L...8K,2015ApJS..220...12S}).

For each galaxy, the aperture correction is determined from the {\it HST}/ACS $I_\mathrm{F814W}$-band images \citep{2007ApJS..172..196K}.  In doing so, we implicitly assume that the difference between the on-sky spatial distributions of the rest-frame optical continuum (i.e., stellar radiation) and nebular emission is negligible under the typical seeing condition ($\gtrsim 0.5~\mathrm{arcsec}$ in FWHM).  This assumption is reasonable for the majority of the galaxies in our sample, in particular, those at $z>1$ whose typical size is $<1~\mathrm{arcsec}$.  

We performed photometry on the ACS images of the FMOS galaxies using SExtractor version 2.19.5 \citep{1996A&AS..117..393B}.  The flux measurement was performed at the position of the best-matched object in the COSMOS2015 catalog if it exists, otherwise at the position of the fiber pointing, with fixed aperture size.  For the majority of the sample, we use the measurements in the $2^{\prime\prime}$-diameter aperture ({\tt FLUX\_APER2}), but employed a  $3^{\prime\prime}$ aperture ({\tt FLUX\_APER3}) for a small fraction of the sample if the size of the object extends significantly beyond the $2^{\prime\prime}$ aperture, and consequently, the ratio {\tt FLUX\_APER3}/{\tt FLUX\_APER2} is $\gtrsim 1.3$\footnote{In our previous studies \citep{2013ApJ...777L...8K, 2015ApJS..220...12S}, the pseudo-total Kron flux {\tt FLUX\_AUTO} was used as the total $I_\mathrm{F814W}$-band flux, rather than the fixed-aperture flux used in this paper.  Although the conclusions are not affected by the choice, the use of the fixed-aperture gives better reproducibility of photometry as the Kron flux measurement is more sensitive to the configuration to execute the SExtractor photometry.}.  We visually inspected the ACS images to check for the presence of significant contamination by nearby objects, flagging such cases in the catalog.  

Next we smoothed the ACS images by convolving with a Gaussian point-spread function (PSF) for the  effective seeing size.  We then performed aperture photometry with SExtractor to measure the flux in the fixed FMOS fiber aperture {\tt FLUX\_APER\_FIB}, and computed the correction factor as $c_\mathrm{aper} = \textrm{\tt FLUX\_APER2(3)}/\textrm{\tt FLUX\_APER\_FIB}$.  The size of the smoothing Gaussian kernel (i.e., the effective seeing size) was retroactively determined for each observing run to minimize the average offset relative to the reference UltraVISTA broad-band magnitudes \citep{2012A&A...544A.156M} from \citet{2016ApJS..224...24L} (see Section \ref{sec:specmag}).  We note that the effective seeing sizes determined are in broad agreement with the actual seeing conditions during the observing runs ($\sim 0''.5\textrm{--}1''.4$ in FWHM) that were measured from the observed point spread function of the guide stars.

\begin{figure}[t] 
   \centering
   \includegraphics[width=3.5in]{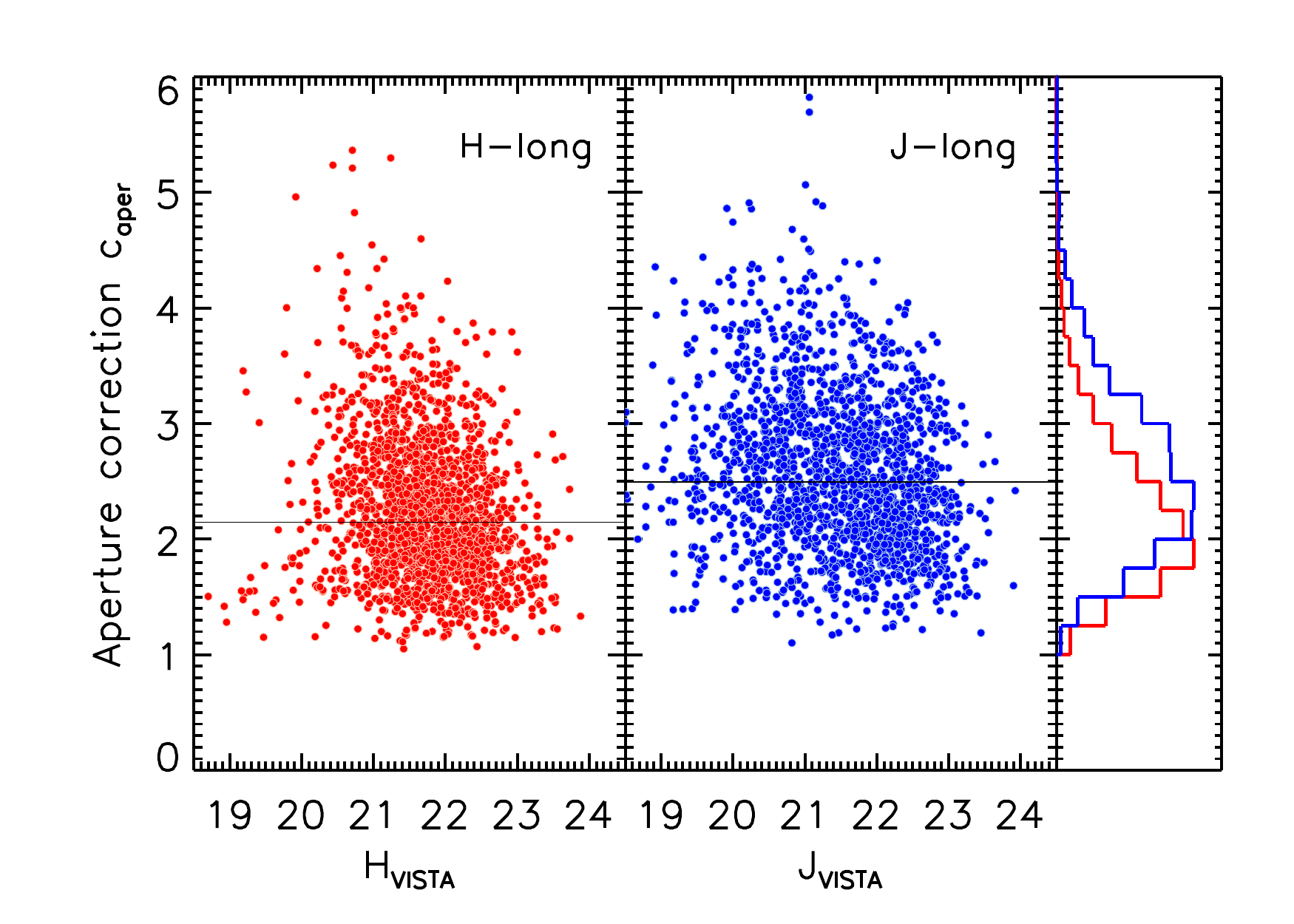} 
   \caption{
   Derived aperture correction factors $c_\mathrm{aper}$ as a function of the reference $H$ or $J$ magnitudes \citep{2016ApJS..224...24L}.  The horizontal solid lines mark the median values.  Histograms show the distribution of $c_\mathrm{aper}$, separately for the $H$- (red) and $J$-long (blue) bands.}
   \label{fig:mag_vs_apercorr}
\end{figure}

Figure \ref{fig:mag_vs_apercorr} shows the derived aperture correction factors as a function of the reference magnitude, separately for the $H$ and $J$ bands.  We excluded insecure estimates of aperture correction, which includes cases where the blending or contamination from other objects are significant.  The aperture correction factors range from $\sim1.2$ to $\sim4.5$, and the median values are 2.1 and 2.5 for the $H$ and $J$ band, respectively.  This small offset between the two bands is due to the fact that seeing is worse for shorter wavelengths under the same condition.  Note that the formal error on the correction factor that comes from the aperture photometry on the ACS image (e.g., {\tt FLUXERR\_APER2}) is small (typically $<5\%$), and thus the scatter seen in Figure \ref{fig:mag_vs_apercorr} is real, reflecting both variations in the intrinsic size of galaxies and the seeing condition of observing nights.

\begin{figure}[tbp] 
   \centering
   \includegraphics[width=3.5in]{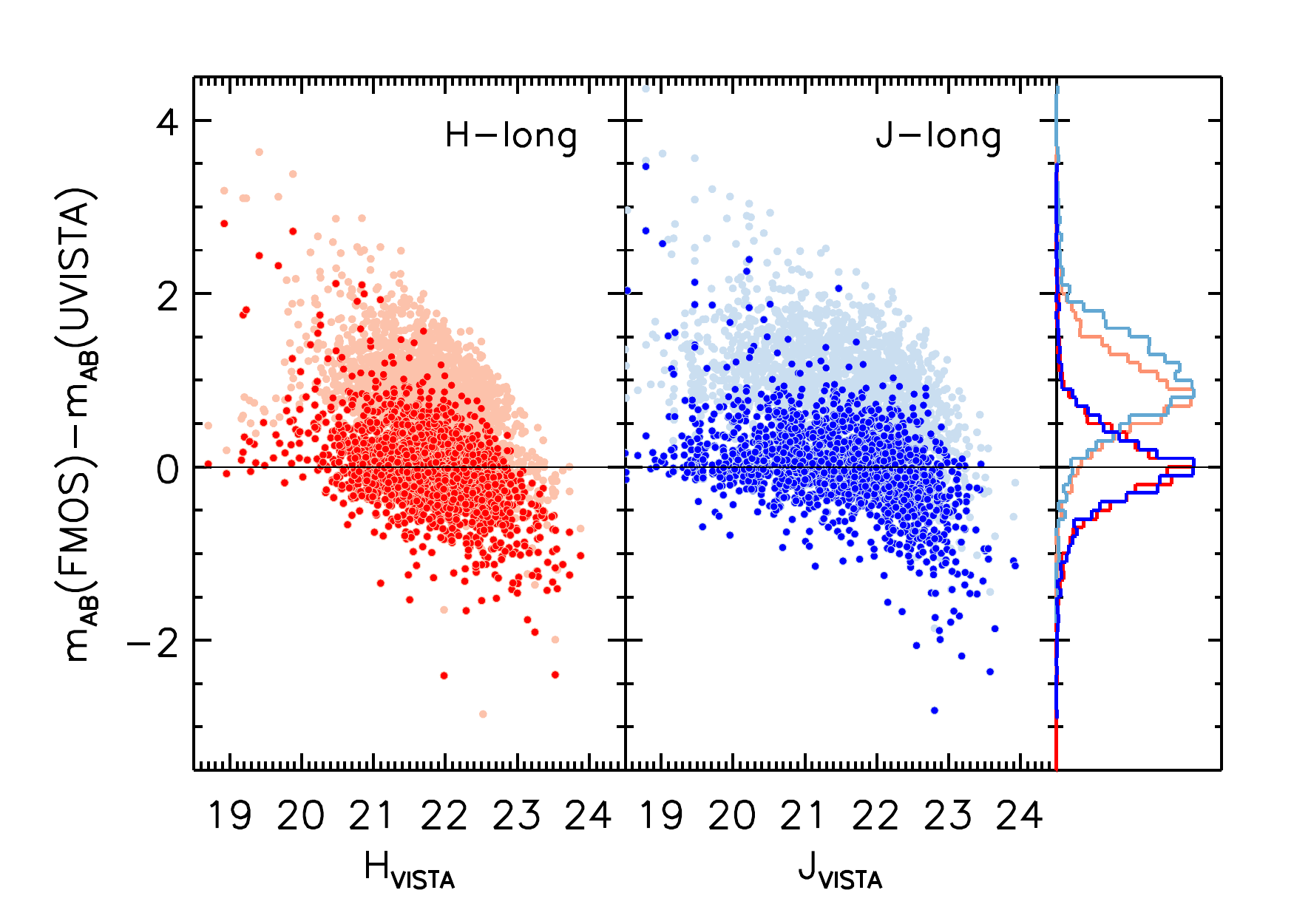} 
   \caption{
   The difference between the observed (FMOS) and reference (UltraVISTA; \citealt{2016ApJS..224...24L}) magnitudes for the $H$- and $J$-bands.  The pale and bright color points correspond respectively to before and after the aperture correction being applied. The histograms show the distribution of the differential magnitudes separately for each band, as well as for before/after the correction.}
   \label{fig:mag_vs_dmag}
\end{figure}

Figure \ref{fig:mag_vs_dmag} shows offsets between the observed and the reference magnitudes before and after correcting for aperture losses, as a function of the reference magnitudes.  The average magnitude offset is mitigated by applying the aperture correction.  After aperture correction, we found that the standard deviation of the magnitude offsets to be 0.42 (0.50) mag, after (before) taking into account the individual measurements errors in both $\left<f_\nu\right>$ of the observed FMOS spectra and the reference magnitude.  There is no significant difference between $H$ and $J$.  Note that this comparison also provides a sense of testing agreement between the first method of aperture correction estimation, described above, that relies on the direct comparison between the observed flux density on the FMOS spectra and the reference magnitude.   

In the catalog, we provide the best estimate of aperture correction for each of all galaxies regardless the presence or absence of spectroscopic redshift estimate.  For 67\% of the sample observed in the $H$-long spectral window and 80\% in $J$-long, the best aperture correction is based on the {\it HST}/ACS image described above.  However, for the remaining objects, the estimates with this method are not robust due to blending, significant contamination from other sources, or any other troubles on pixels of the ACS images.  Otherwise, there is no ACS coverage for some of those falling outside the area (see Figure \ref{fig:pawprints}).  For such cases, we provide as the best aperture correction an alternative estimate based on the second method that uses the average offset of all objects observed together in the same night.  With these aperture correction, the agreement between the aperture-corrected observed flux density and the reference magnitude is slightly worse, with an estimated intrinsic scatter of $\approx0.57~\mathrm{dex}$ for both $H$- and $J$-long, than that based on the ACS image-based aperture correction.  

In the following, we use these best estimates of aperture correction without being aware of which method is used.  Throughout the paper, when any aperture corrected values such as total luminosity and SFRs are shown, the error includes in quadrature a common factor of 1.5 (or 0.17~dex) in addition to the formal error on the observed emission line flux to account for the intrinsic uncertainty of aperture correction.  Lastly, we emphasize that the aperture correction is determined for all the individual objects using the independent observations (i.e., {\it HST}/ACS and Ultra-VISTA photometry) and just average information of the FMOS observations (i.e., mean offset), but not relying on the individual FMOS measurements.  This ensures that the uncertainty of aperture correction is independent of the individual FMOS measurements.

%---------------------------------------------------------------------------------------------
\section{Line detection and redshift estimation \label{sec:detections}}

The full FMOS-COSMOS catalog contains 5247 extragalactic objects that were observed in any of three, $H$-long, $J$-long, or $H$-short bands
\footnote{The full FMOS-COSMOS catalog is available here: \\
\url{http://member.ipmu.jp/fmos-cosmos/fmos-cosmos_catalog_2019.fits}\\ 
For more information, please refer to the README file:\\
\url{http://member.ipmu.jp/fmos-cosmos/fmos-cosmos_catalog_2019.README}}.
The majority of the survey was conducted with the $H$-long grating, collecting spectra of 4052 objects.  The second effort was dedicated to observations in the $J$-long band, including the follow up of objects for which H$\alpha$ was detected in $H$-long to detect other lines (i.e., \Hb\ and \OIII) and observations for lower-redshift objects to detect \Ha.  A single night was used for observation with the $H$-short grating (see Table \ref{tb:observations1}).  In this section, we report spectroscopic redshift measurements and success rates.

\subsection{Spectroscopic redshift measurements \label{sec:redshift_measurements}}

\begin{deluxetable*}{lcccccc}
\tablecaption{Summary of the acquisition of spectra and successful redshift \label{tb:spectra}}
\tablehead{\colhead{Spectra}&
		 \colhead{Wavelength range}&
		 \colhead{$N_\mathrm{obs}$\tablenotemark{}}&
  		 \colhead{$z\mathrm{Flag}=1$}&
 		 \colhead{$=2$}&
		 \colhead{$=3$}&
		 \colhead{$=4$}}
\startdata
Total     & -          &  5247 &  140 &  389 &  507 &  895 \\
\hline
$H$-long  & 1.60--1.80 $\mu$m&  4052 &  117 &  314 &  384 &  694 \\
$H$-short & 1.40--1.60 $\mu$m&   163 &    3 &   12 &   18 &   34 \\
$J$-long  & 1.11--1.35 $\mu$m&  2599 &   77 &  304 &  388 &  807 \\
\hline
HL+HS     & - &   108 &    3 &    9 &   13 &   28 \\
HL+JL     & - &  1441 &   54 &  229 &  266 &  607 \\
HS+JL     & - &    81 &    1 &   11 &   16 &   33 \\
HL+HS+JL  & - &    63 &    1 &    8 &   12 &   28 \\
\enddata
\tablenotetext{a}{The numbers of observed galaxies in specified spectral window(s), i.e., $z\mathrm{Flag}\ge0$.}
\end{deluxetable*}

\begin{deluxetable}{lcccc}
\tablecaption{Summary of the emission-line detection \label{tb:detections}}
\tablehead{\colhead{Line}&
		\colhead{$z_\mathrm{min}$--$z_\mathrm{max}$}&
		\colhead{$1.5\le \mathrm{S/N}<3$}&
		\colhead{$3\le \mathrm{S/N}<5$}&
		\colhead{$\mathrm{S/N}\ge 5$}}
\startdata
\multicolumn{5}{c}{$H$-long} \\
\Ha  & 1.43--1.74 &  111 &  305 &  909 \\
\NII  & 1.43--1.73 &  298 &  274 &  247 \\
\Hb  & 2.32--2.59 &    9 &   13 &   14 \\
\OIII  & 2.21--2.59 &    5 &    8 &   58 \\
\hline
\multicolumn{5}{c}{$H$-short} \\
\Ha  & 1.26--1.46 &    2 &    1 &   21 \\
\NII  & 1.31--1.46 &    2 &    5 &    6 \\
\Hb  & 2.15--2.15 &    1 &    0 &    0 \\
\OIII  & 2.15--2.15 &    0 &    0 &    1 \\
\hline
\multicolumn{5}{c}{$J$-long} \\
\Ha  & 0.70--1.05 &   13 &   50 &  267 \\
\NII  & 0.70--1.04 &   44 &   74 &  134 \\
\Hb  & 1.31--1.74 &  139 &  160 &  100 \\
\OIII  & 1.30--1.69 &   49 &  160 &  296 \\
\enddata
\end{deluxetable}

Out of the full sample, we obtained spectroscopic redshift estimates for 1931 objects.  The determination of spectroscopic redshift is based on the detection of at least a single emission line expected to be either \Ha, \NII, \Hb, or \OIII.  For our initial target selection, galaxies were selected based on the photometric redshift $z_\mathrm{phot}$ so that \Ha+\NII\ and \Hb+\OIII\ are detected in either the $H$-long or $J$-long spectral window.  For the majority of the sample, we identified the detected line as \Ha\ or \OIII\ according to their $z_\mathrm{phot}$.  However, this is not the case for a small number of objects for which we found a clear combination of \Ha+\NII, or \OIII\ doublet (+\Hb) in a spectral window not expected from the $z_\mathrm{phot}$.  For objects observed both in $H$- and $J$-band, we checked whether their independent redshift estimates are consistent. If not, we re-examined the spectra to search for any features that can solve the discrepancy between the spectral windows.  Otherwise, we disregarded line detections of lower S/N.  For objects observed twice or more times, we adopted a spectrum with the highest S/N ratio of the line flux.  For objects with consistent line detections in the two spectral windows (i.e., \Ha+\NII\ in $H$-long, and \Hb+\OIII\ in $J$-long), we regarded a redshift estimate based on higher S/N detection as the best estimate ($z_\mathrm{best}$).  There also objects that were observed twice or more times in the same spectral window.  In particular, the repeat $J$-long observations have been carried out to build up exposure time to detect faint \Hb\ at higher S/N.  In the catalog presented in this paper, however, we adopted a single observation with detections of the highest S/N ratio, instead of stacking spectra taken on different observing runs \footnote{The measurements based on co-added spectra are provided in an ancillary catalog.}.

We assign a quality flag ($z$Flag) to each redshift estimate based on the number of detected lines and the associated S/N as follows (see Section \ref{sec:fitting} for details of the detection criteria).

\begin{itemize}
\item[$z$Flag 0]: No emission line detected.
\item[$z$Flag 1]: Presence of a single emission line detected at $1.5\le S/N<3$.
\item[$z$Flag 2]: One emission line detected at $3\le S/N<5$.
\item[$z$Flag 3]: One emission line detected at $S/N\ge5$.
\item[$z$Flag 4]: One emission line having $S/N\ge5$ and a second line at $S/N\ge3$ that confirms the redshift.
\end{itemize}
The criteria have been slightly modified from those used in \citet{2015ApJS..220...12S} (where $\mathrm{Flag}=4$ if a second line is detected at $S/N\ge1.5$).  Note that objects with $z\mathrm{Flag}=1$ are not used for scientific analyses in the remaining of the paper.

In Table \ref{tb:spectra} we summarize the numbers of observed galaxies and the redshift estimates with the corresponding quality flags.  In the upper three rows, the numbers of galaxies observed with each grating are reported, while the numbers of galaxies observed in two or three bands are reported in the lower four rows.  Table \ref{tb:detections} summarizes the number of galaxies with detections of each of four emission lines.  

\begin{figure}[tbp] 
   \centering
   \includegraphics[width=3.5in]{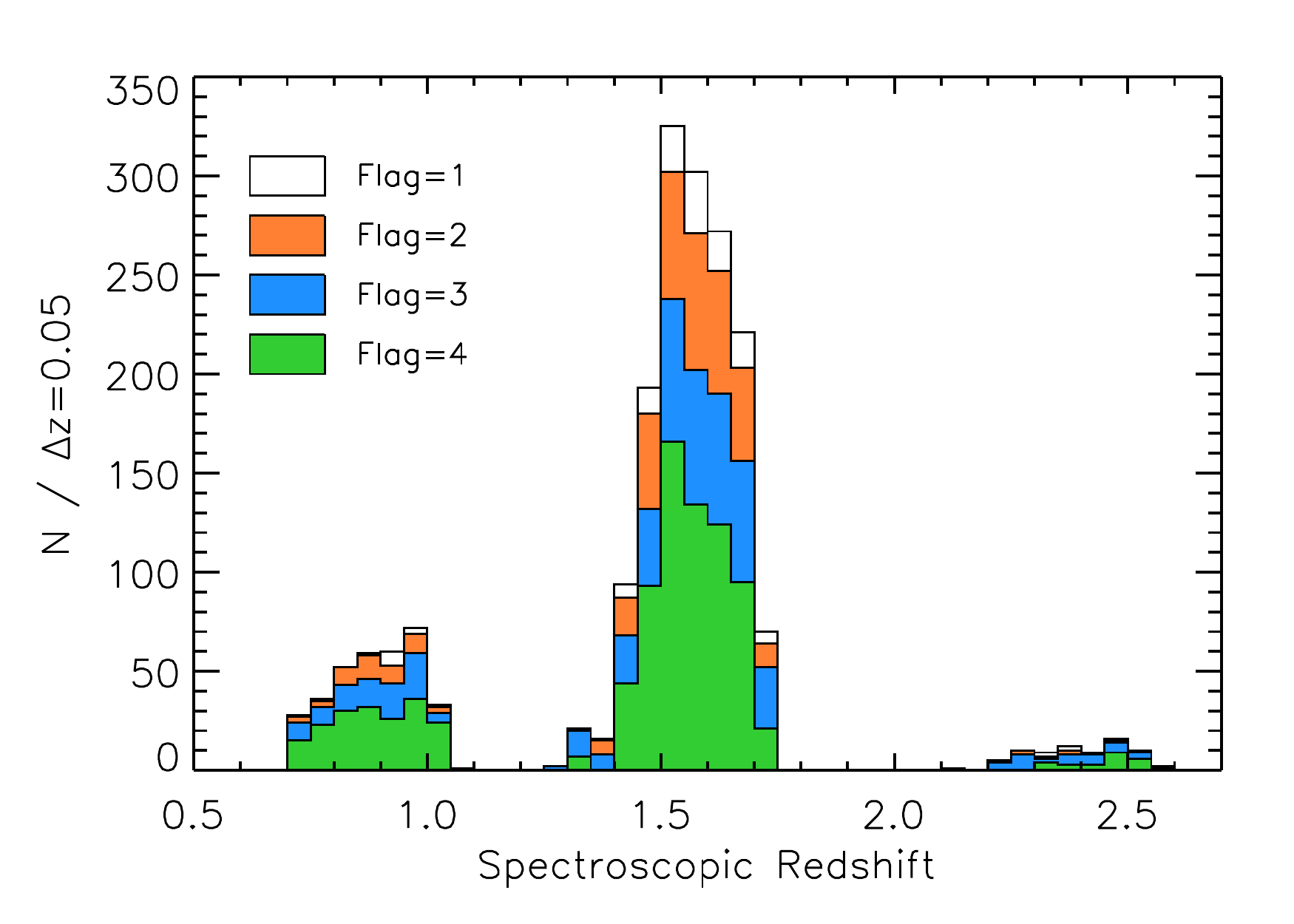} 
   \includegraphics[width=3.5in]{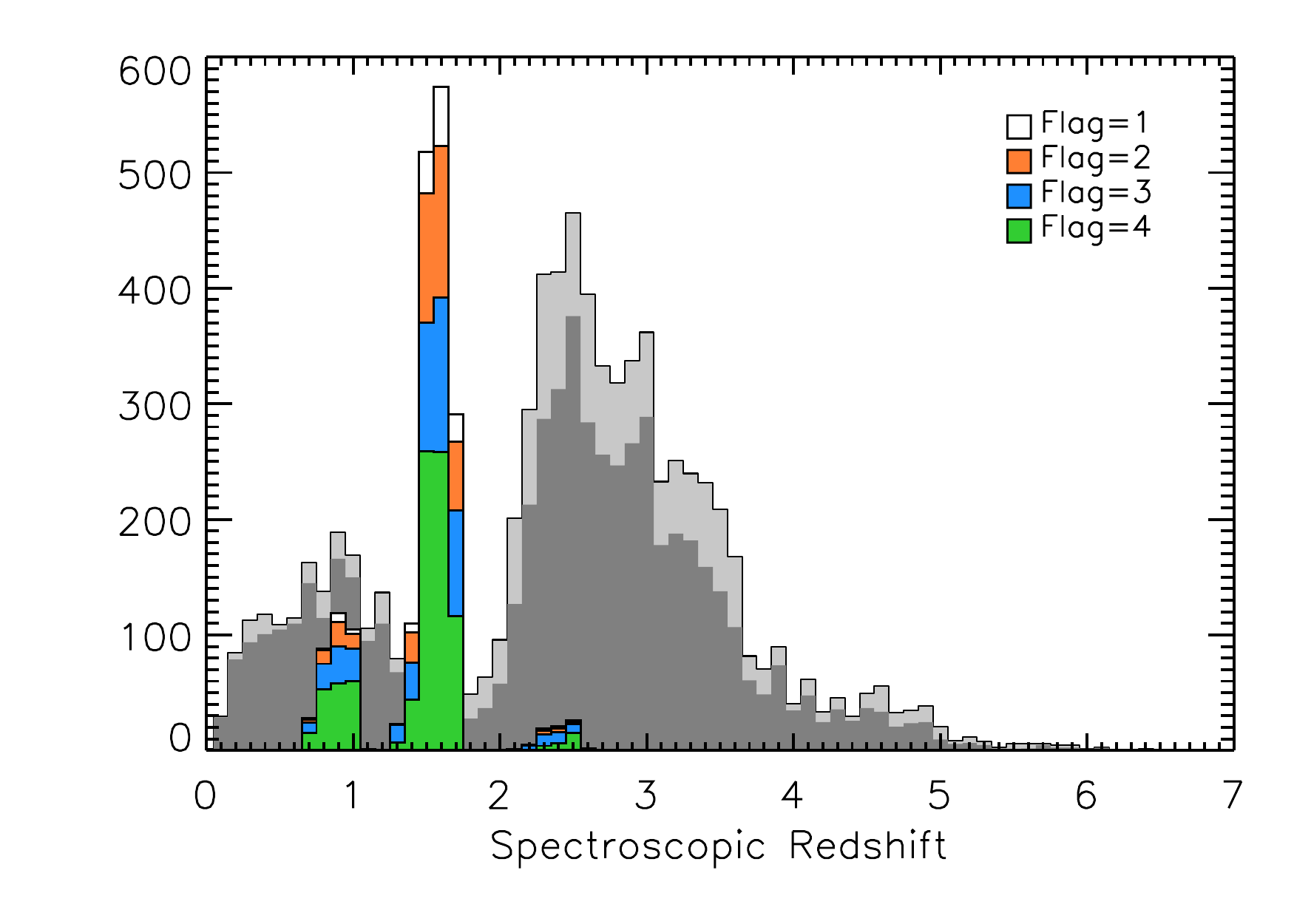} 
   \includegraphics[width=3.5in]{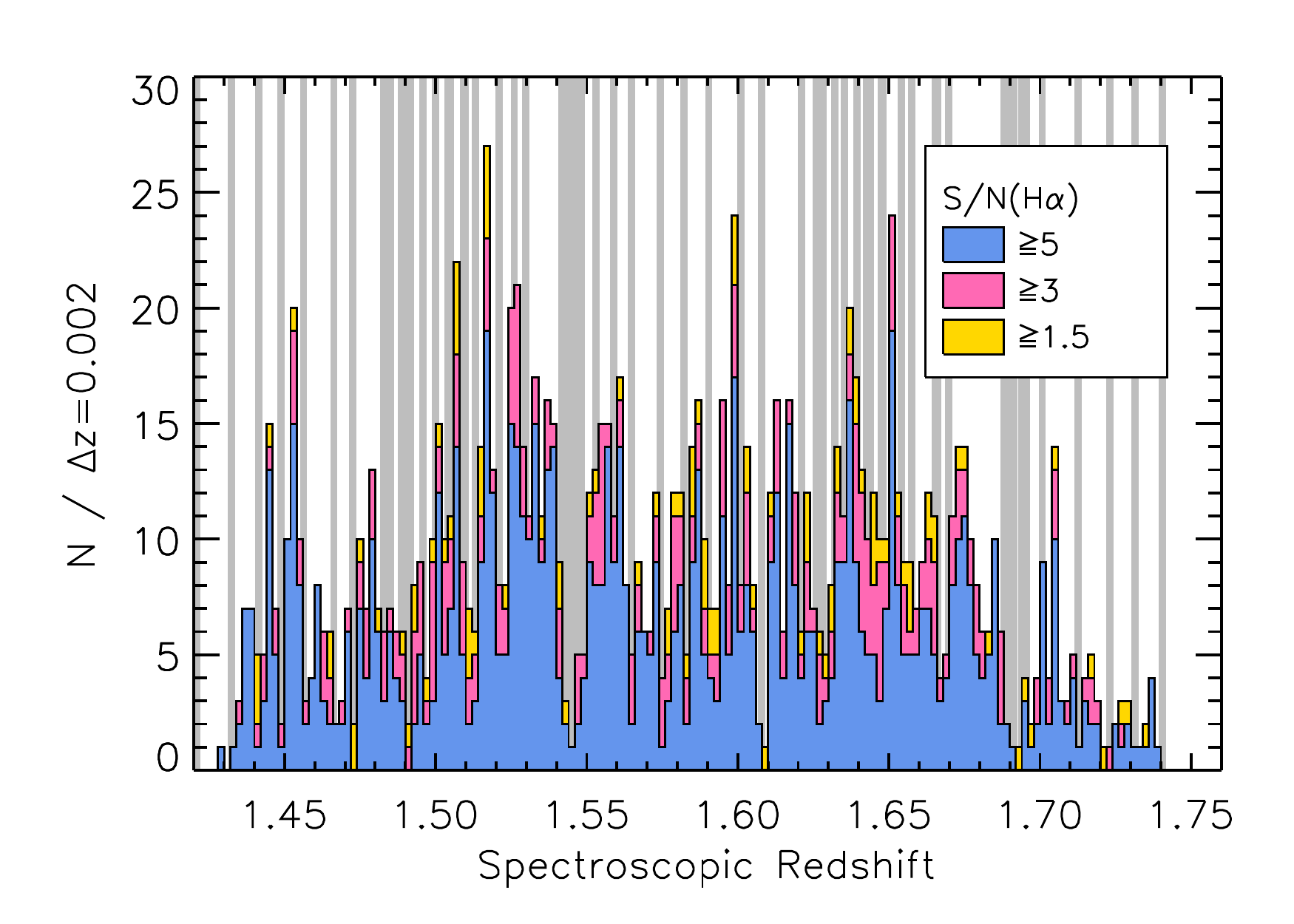} 
   \caption{
   Distribution of spectroscopic redshift measurements for all objects in the full FMOS-COSMOS catalog, split by their quality flags.   The FMOS $z_\mathrm{spec}$ distribution is compared with VUDS \citep[][gray histograms]{2015A&A...576A..79L} in the middle panel.  The bottom panel shows zoom-in of the range $1.42\le z\le 1.76$ with a finer binsize ($\Delta z=0.002$) for objects with an \Ha\ detection in the $H$-long band.  Histograms are color-coded by $\mathrm{S/N}(\mathrm{H\alpha})$ as labeled.  The gray stripes indicate positions of the OH airglow lines, which are converted to redshift with the wavelength of \Ha.}
   \label{fig:histz}
\end{figure}

In the top panel of Figure \ref{fig:histz}, we display the distribution of all galaxies with a spectroscopic redshift estimate split by the quality flag.  There are three redshift ranges, corresponding to possible combinations of the detected emission lines and the spectral ranges, as summarized in Table \ref{tb:detections}.  In the middle panel, we compare the distribution of the FMOS-COSMOS galaxies to the redshift distribution from the VUDS observations \citep{2015A&A...576A..79L}.  It is clear that our FMOS survey constructed a complementary spectroscopic sample that fills up the redshift gap seen in the recent deep optical spectroscopic survey.  In the lower panel of Figure \ref{fig:histz}, we show objects for which \Ha\ is detected in the $H$-long spectra, with the positions of OH lines.  Wavelengths of the OH lines are converted into redshifts based on the wavelength of the \Ha\ emission line as $z_\mathrm{OH}=\lambda_\mathrm{OH}/6564.6\mathrm{\AA}-1$.  It is clear that the number of successful detections of \Ha\ is suppressed near OH contaminating lines.  The OH suppression mask blocks about $30\%$ of the $H$-band.  This reduces the success rate of line detection.

Based on the full sample, we have a 37\% (1931/5247) overall success rate for acquiring a spectroscopic redshift with a quality flag $z\mathrm{Flag}\ge1$, including all galaxies observed in any of the FMOS spectral windows.  We note that given that only $\sim 70\%$ of the $H$-band is available for line detection due to the OH masks, the effective success rate can be evaluated to be $\sim 37/0.7=53\%$.  The full catalog, however, contains various galaxy populations selected by different criteria and many galaxies may satisfy criteria for different selections, i.e., the subsamples overlap each other.  In later subsections, we thus focus our attention separately to each of specific subsamples of galaxies as described in Section \ref{sec:samples}.  In Table \ref{tb:subsamples}, we summarize the successful redshift estimates for each subsample.  

\begin{deluxetable*}{lcccccccc}
\tablecaption{Summary of the \Ha\ detection for the main subsamples \label{tb:subsamples}}
\tablehead{\colhead{}&
		\colhead{}&
		\multicolumn{3}{c}{\Ha\ detection}&
		\multicolumn{4}{c}{Redshift quality flags}\\
		\colhead{Subsample}&
		\colhead{$N_\mathrm{obs}$}&
		\colhead{$1.5\le \mathrm{S/N}<3$}&
		\colhead{$3\le \mathrm{S/N}<5$}&
		\colhead{$\mathrm{S/N}\ge 5$}&
  		 \colhead{$z\mathrm{F}=1$}&
 		 \colhead{$z\mathrm{F}=2$}&
		 \colhead{$z\mathrm{F}=3$}&
		 \colhead{$z\mathrm{F}=4$}}
\startdata
Primary-HL                                 & 1582 & 69 & 168 &  475 & 66 & 162 & 171 & 350 \\
Primary-HL (X-ray removed)      & 1514 & 67 & 161 &  454 & 65 & 155 & 165 & 330 \\
Secondary-HL                            & 1242 & 34 & 91 &  255 & 32 & 96 & 109 & 182 \\
Secondary-HL (X-ray removed) & 1201 & 33 & 87 &  253 & 29 & 91 & 107 & 181 \\
{\it Herschel}/PACS-HL & 116 & 5 & 10 &  38 & 4 & 10 & 10 & 32 \\
Low-$z$ IR galaxies & 344 & 3 & 20 & 149 & 5 & 24 & 35 & 124 \\
 {\it Chandra} X-ray objects & 742 & 12 & 40 &  144 & 18 & 57 & 77 & 129 \\
\hline
\enddata
\end{deluxetable*}

% =========================================================
\subsection{The primary sample of star-forming galaxies at $z\sim1.6$ \label{sec:redshifts_Primary-HL}}

The Primary-HL sample includes galaxies selected from the COSMOS photometric catalog, as described in Section \ref{sec:Primary-HL}.   For these objects, our line identification assumed that the strongest line detected in the $H$-long band is the \Ha\ emission line, although, for some cases, only the \NII\ line was measured and \Ha\ was disregarded due to significant contamination on \Ha.  For other cases with no detections in the $H$-long window, the strongest line detection in the $J$-long spectra was assumed to be the \OIII$\lambda$5007 line.  We observed 1582 galaxies that satisfy the criteria given in Section \ref{sec:Primary-HL} with the $H$-long grating, and successfully obtained redshift estimates with $z\mathrm{Flag} \ge 1$ for 749 (47\%) of them.  The measured redshifts range between $1.36\le z_\mathrm{spec} \le 1.74$.  Focusing on the detection of the \Ha\ line in the $H$-long grating, we successfully detected it for 712 (643) at $\ge 1.5\sigma$ ($\ge 3\sigma$).  We note that the remaining 37 objects includes \NII\ detections with the $H$-long grating, and \Hb\ and/or \OIII\ detections with the $J$-long grating.  In addition to the Primary objects, we also observed other 1242 star-forming galaxies at $z\sim1.6$ which do not match all the criteria for the Primary target (the Secondary-HL sample; see Section \ref{sec:Primary-HL}).  In Table \ref{tb:subsamples}, we summarize the number of redshift measurements for the Primary-HL and the Secondary-HL samples, as well as for the subset after removing X-ray detected objects.

\begin{figure}[tbp] 
   \centering
   \includegraphics[width=3.5in]{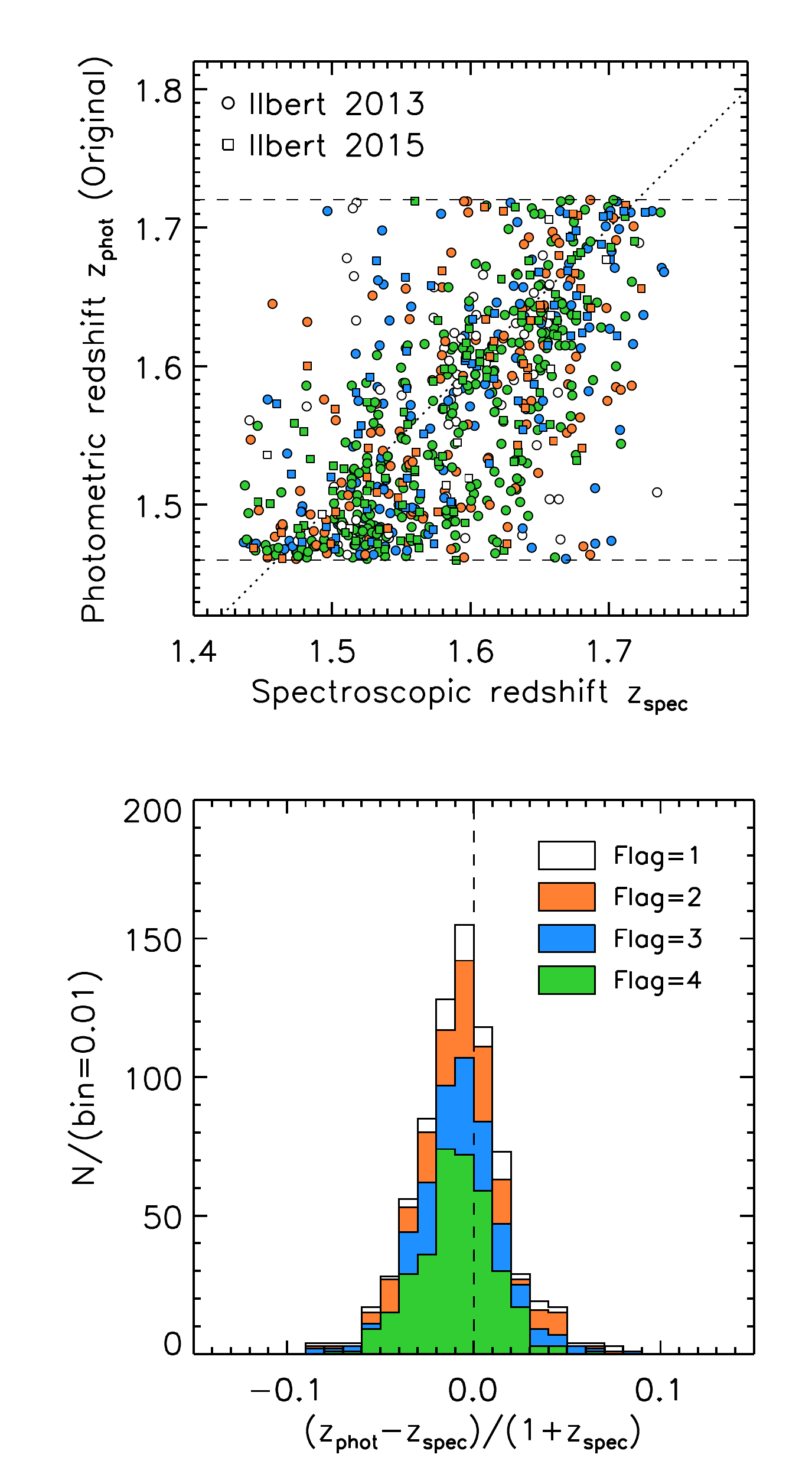} 
   \caption{
   Upper panel: comparison between $z_\mathrm{spec}$ and $z_\mathrm{phot}$ for the Primary-HL sample.  Each point is color-coded by the quality flag of the redshift estimate, as labeled in the lower panel.  Circles indicate the FMOS objects selected based on the photometric redshift from \citet{2013A&A...556A..55I}, while squares indicate the objects based on \citet{2015A&A...579A...2I}.  
   Lower panel: distribution of the differences between the spectroscopic redshifts from FMOS and the photometric redshifts.}
   \label{fig:zs_vs_zp_orig}
\end{figure}

In Figure \ref{fig:zs_vs_zp_orig}, we compare the the spectroscopic redshifts with the photometric redshifts used for the target selection from the photometric catalogs \citep{2013A&A...556A..55I,2015A&A...579A...2I} for the Primary-HL sample.  For those with $z\mathrm{Flag}\ge2$ (i.e., $\ge3\sigma$), the median and the standard deviation $\sigma_\mathrm{std}$ of $(z_\mathrm{phot}-z_\mathrm{spec})/(1+z_\mathrm{spec})$ are $-0.0099$ ($-0.0064$) and $0.028$ (0.024), respectively, after (before) taking into account the effects of limiting the range of photometric redshifts ($1.46 \le z_\mathrm{phot} \le 1.72$).  To account for the edge effects, we adopted a number of sets of the the median offset and $\sigma_\mathrm{std}$ to simulate photometric redshift for each $z_\mathrm{spec}$ measurement, and then determined the plausible values of the intrinsic median and $\sigma_\mathrm{std}$ that can reproduce the observed median offset and $\sigma_\mathrm{std}$ of $(z_\mathrm{phot}-z_\mathrm{spec})/(1+z_\mathrm{spec})$ after applying the limit of $1.46 \le z_\mathrm{phot} \le 1.72$.

\subsection{The Herschel/PACS subsample at $z\sim1.6$}

The PACS-HL sample include 116 objects between $1.44 \le z_\mathrm{phot} \le 1.72$ detected in the {\it Herschel}-PACS observations (Section \ref{sec:PEP}).  We successfully measured spectroscopic redshifts for 56 (43\%) objects with $z\mathrm{Flag}\ge1$, including 32 (28\%) secure measurements ($z\mathrm{Flag}=4$).  These measurements include 43 (3, 2) detection of \Ha\ ($\ge 3\sigma$) in the $H$-long ($H$-short, $J$-long) band, as well as a single higher-$z$ object with a possible detection of the \OIII\ doublet in the $H$-long band ($z_\mathrm{spec}=2.26$).  

\subsection{Lower redshift sample of IR luminous galaxies}

We observed in the $J$-long band 344 lower redshift galaxies selected from the infrared data (see Section \ref{sec:JKsample}), and succeeded to measure spectroscopic redshift with $z\mathrm{Flag}\ge1$ for 188 objects (55\%).  We detected the \Ha\ emission line at $S/N\ge1.5$ ($\ge3.0$) for 172 (169) objects.  We note that 6 objects have detection of \Hb+\OIII\ in the J-long, thus not being within the lower redshift window.

\subsection{Chandra X-ray sample}

We observed in total 742 objects detected in the X-ray from the {\it Chandra} COSMOS Legacy survey \citep{2009ApJS..184..158E,2016ApJ...819...62C}.  Of them, 385 and 533 objects were observed with the $H$-long and $J$-long gratings, while 177 were observed with both of these.  We obtained a redshift estimate for 281 (263) objects with $z\mathrm{Flag} \ge 1$ ($\ge2$).  The entire sample of the X-ray objects include 75 lower redshift ($0.72 \le z_\mathrm{spec} \le 1.1$) objects with a detection of \Ha+\NII\ in the $J$-long band, and 29 (1) higher redshift ($2.1 \le z_\mathrm{spec} \le 2.6$) objects with a detection of \Hb+\OIII\ in the $H$-long ($H$-short).  The remaining majority of the sample are those at intermediate redshift range with detections of \Ha+\NII\ in the $H$-long band, and/or \Hb+\OIII\ in the $J$-long band.

%-------------------------------------------------------------
\section{Basic properties of the emission lines \label{sec:lineprop}}

\subsection{Observed properties of \Ha \label{sec:Haprop}}

\begin{figure}[tbp] 
   \centering
   \includegraphics[width=3.5in]{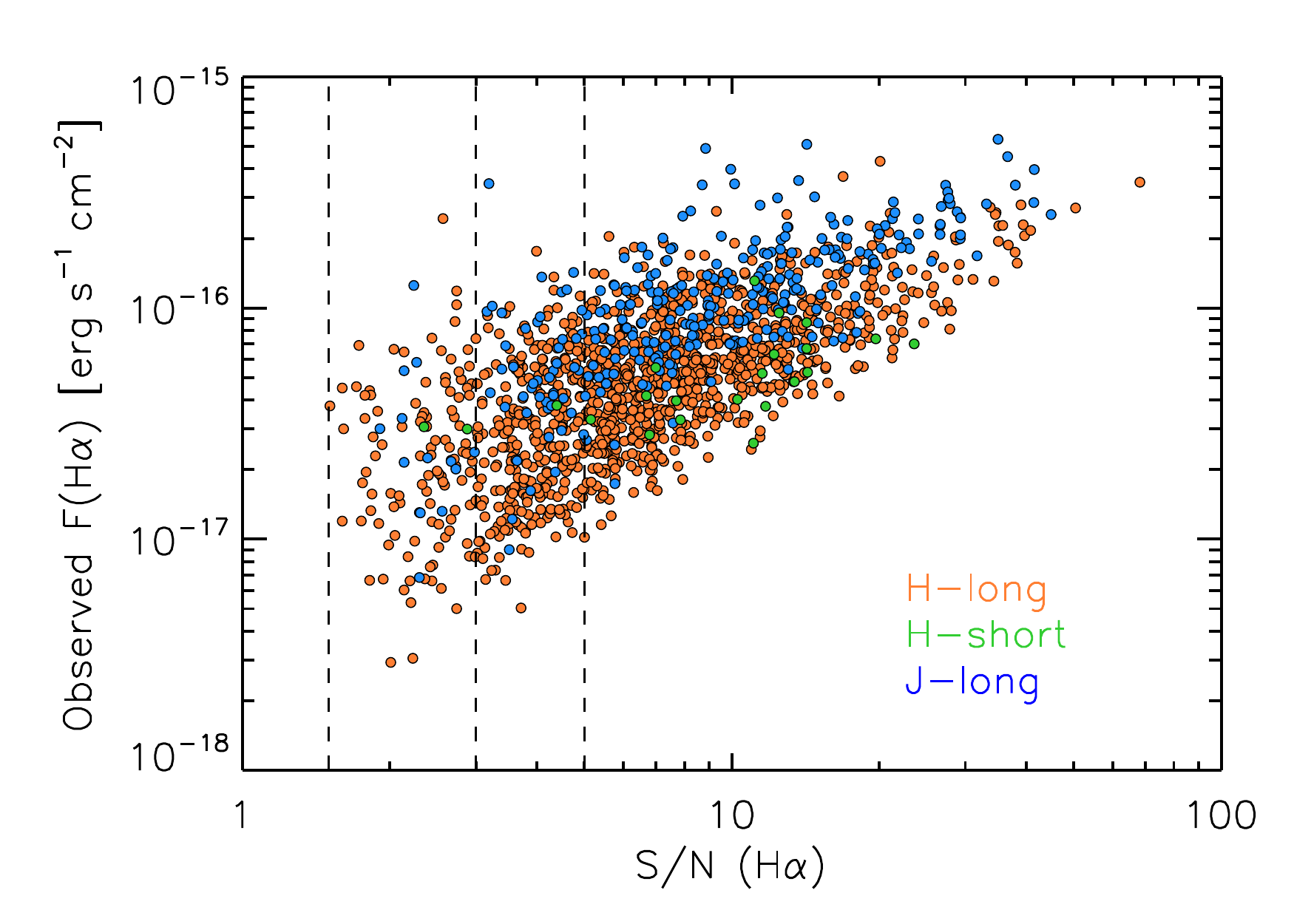} 
   \caption{
   Observed \Ha\ flux (neither corrected for the aperture loss nor extinction) as a function of observed formal S/N for individual galaxies, shown separately for each spectral window as labeled.  Vertical dashed lines indicates $S/N=1.5$ (limit for detection), 3 (limit for $z\mathrm{Flag}=2$), and 5 (limit for $z\mathrm{Flag}=3$).
   }
   \label{fig:flux_SN_Ha}
\end{figure}

In Figure \ref{fig:flux_SN_Ha} we plot the observed in-fiber \Ha\ flux $F_\mathrm{H\alpha}$ (neither corrected for dust extinction nor aperture loss) as a function of associated S/N for each galaxy in our sample, split by the spectral window.  As naturally expected, there is a correlation between $F_\mathrm{H\alpha}$ and S/N, but with large scatter in $F_\mathrm{H\alpha}$ at fixed S/N.  This is mainly due to the presence of `bad pixels' impacted by OH masks and residual sky emission (see Section \ref{sec:fitting}).  The figure indicates that, in the $H$-long band, the best sensitivity achieves $F_\mathrm{H\alpha} \sim 10^{-17}~\mathrm{erg~s^{-1}~cm^{-2}}$ at $S/N=3$, while the average is $\sim 3\times 10^{-17}~\mathrm{erg~s^{-1}~cm^{-2}}$.

\begin{figure}[tbp] 
   \centering
   \includegraphics[width=3.5in]{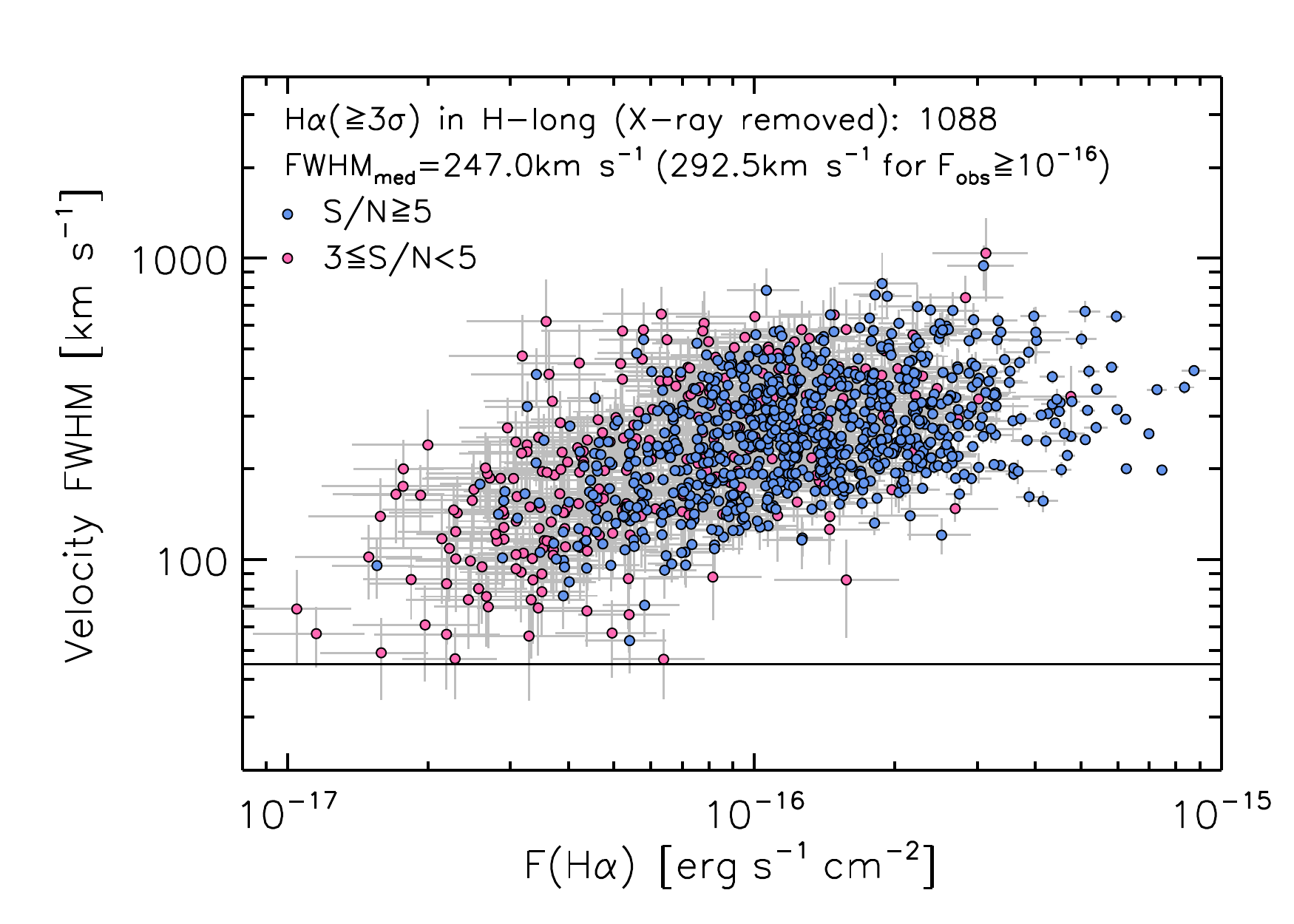} 
   \caption{
   Correlation between aperture-corrected \Ha\ flux (not corrected for dust) and line width (FWHM) in velocity units.  The sample shown here is restricted to those with an \Ha\ detection at $3\le S/N<5$ (magenta) and $S/N\ge5$ (blue) in $H$-long ($1.43 \le z \le 1.74$).  The horizontal line indicates the velocity resolution limit ($45~\mathrm{km~s^{-1}}$).}
   \label{fig:fobsHa_vs_FWHM}
\end{figure}

In Figure \ref{fig:fobsHa_vs_FWHM}, we show the correlation between $F_\mathrm{H\alpha}$ (corrected for aperture, but not for dust) and the full width at half maximum (FWHM) of the \Ha\ line in velocity units for galaxies with an \Ha\ detection ($\ge3.0\sigma$) in the $H$-long band ($1.43\le z_\mathrm{spec} \le1.74$).  The emission line widths are not deconvolved for the instrumental velocity resolution ($\approx 45~\mathrm{km~s^{-1}}$ at $z\sim1.6$).  Although there is a weak correlation between these quantities, the line width becomes nearly constant at $F_\mathrm{H\alpha}\gtrsim1\times10^{-16}~\mathrm{erg~s^{-1}~cm^{-2}}$.  The central 90 percentiles of the observed FWHM is 108--537~$\mathrm{km~s^{-1}}$ with the median at 247~$\mathrm{km~s^{-1}}$.  Limiting to those with $F_\mathrm{H\alpha}\ge1\times10^{-16}~\mathrm{erg~s^{-1}~cm^{-2}}$, the median is 292~$\mathrm{km~s^{-1}}$.  

\begin{figure}[tbp] 
   \centering
   \includegraphics[width=3.5in]{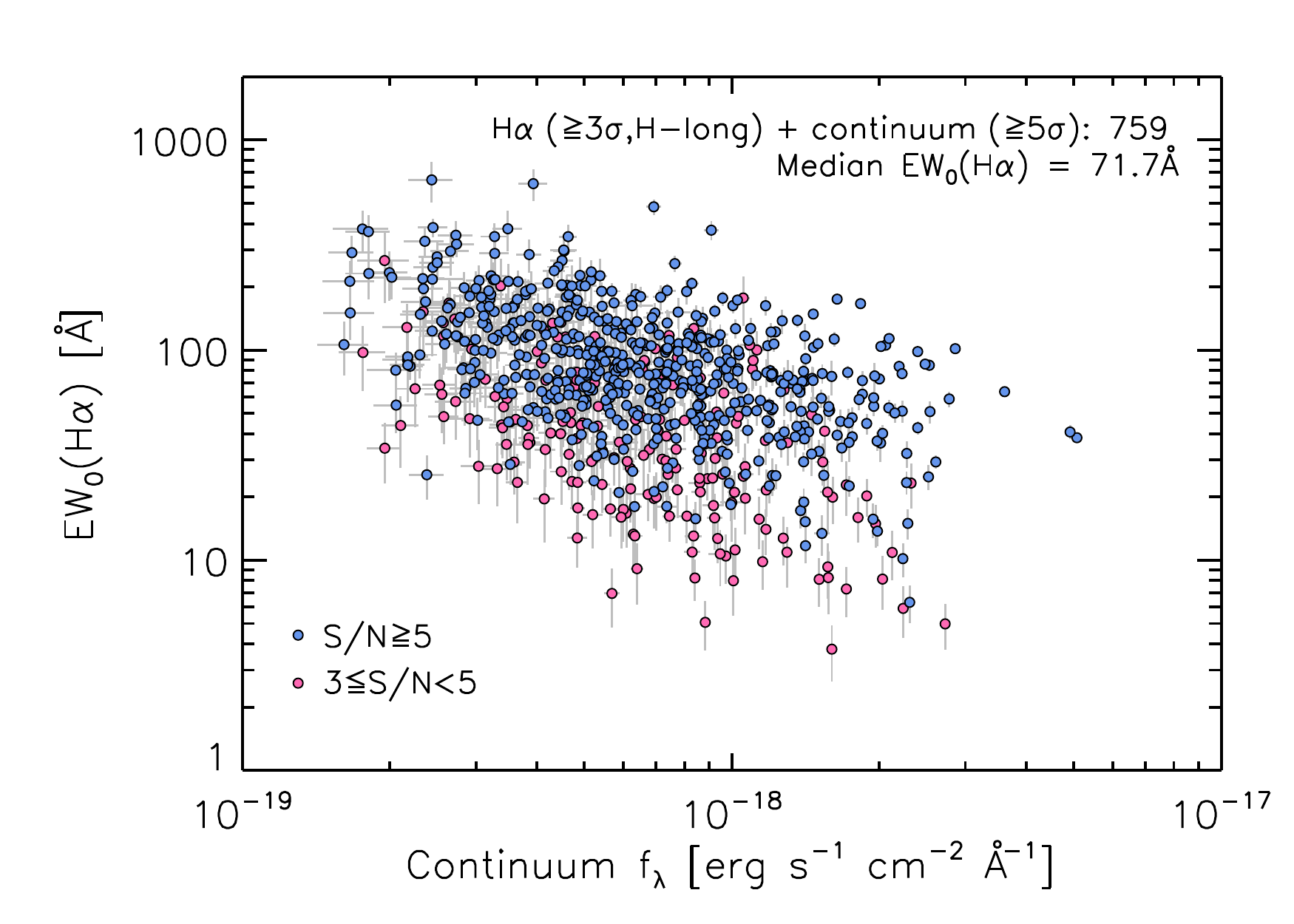} 
   \caption{Rest-frame Equivalent width $EW_0(\mathrm{H\alpha})$ as a function of aperture-corrected, average continuum flux density $\left< f_\mathrm{\lambda,con} \right>$.  Objects shown are limited to have both \Ha\ detection ($\ge3\sigma$) and reliable continuum detection ($\ge5\sigma$).  Symbols are the same as in Figure \ref{fig:fobsHa_vs_FWHM}.  }
   \label{fig:flam_vs_EWHa}
\end{figure}

In Figure \ref{fig:flam_vs_EWHa}, we show the rest-frame equivalent width ($EW_0$) of the \Ha\ emission line as a function of aperture-corrected continuum flux density $\left< f_\mathrm{\lambda,con} \right>$ averaged across the $H$-long spectral window.  The continuum flux density was computed with Equation \ref{eq:f_nu}, excluding the emission line components.  The equivalent widths were not corrected for differential extinction between stellar continuum and nebular emission.  The 759 objects shown here are limited to have a detection of \Ha\ at $\ge 3\sigma$ in $H$-long and a secure measurement of the continuum level ($\ge 5\sigma$).   The observed $EW_0(\textrm{\Ha})$ ranges from $\approx 10$ to $300~\mathrm{\AA}$ with the median $\left< EW_0(\textrm{\Ha}) \right> = 71.7~\mathrm{\AA}$.  The sample shows a clear negative correlation between $\left< f_\mathrm{\lambda,con} \right>$ and $EW_0(\textrm{\Ha})$.  The continuum and \Ha\ flux reflect, respectively, $M_\ast$ and SFR.  Thus, this correlation may be shaped by the facts that specific SFR ($\mathrm{sSFR}=\mathrm{SFR}/M_\ast$) decreases on average with $M_\ast$.

\begin{figure}[tbp] 
   \centering
   \includegraphics[width=3.5in]{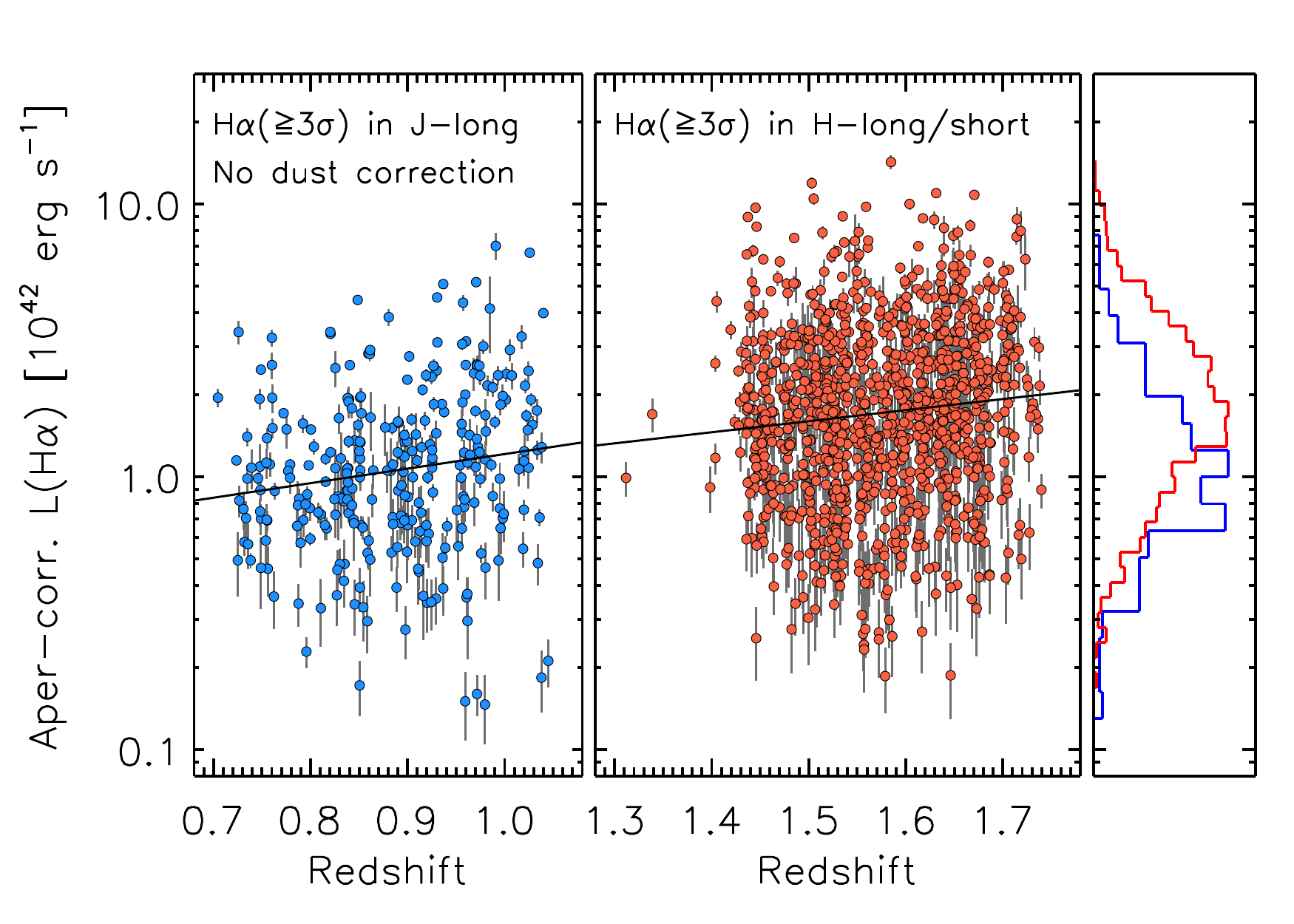} 
   \caption{
   Observed \Ha\ luminosity (corrected for aperture loss,  but not for dust extinction) as a function of redshift in the two redshift ranges, corresponding to the \Ha\ detection in $J$-long (left panel) and $H$-long/short (middle panel).  The solid lines indicate the linear regression, being fitted independently in each redshift range.  Histograms show the normalized distribution of $L_\mathrm{H\alpha}$ for each redshift range as color-coded (right panel).}
   \label{fig:z_vs_LHa}
\end{figure}

In Figure \ref{fig:z_vs_LHa}, we plot the observed \Ha\ luminosity, $L_\mathrm{H\alpha}$ (corrected for aperture loss, but not for dust extinction), as a function of redshift, separately in the two redshift ranges that correspond to where \Ha\ is detected ($J$-long or $H$-long/short).  The observed $L_\mathrm{H\alpha}$ is a weak function of redshift, increasing towards higher redshift, as shown by the linear regression that is derived in each range.  This trend is almost negligible compared to the range spanned by the sample ($\Delta z\approx0.4$ each).   For objects with an \Ha\ detection ($\ge3\sigma$) in the $H$-long spectral window, the central 90th percentiles of $L_\mathrm{H\alpha}$ is $10^{41.7}\textrm{--}10^{42.7}~\mathrm{erg~s^{-1}}$ with median $\left< L_\mathrm{H\alpha}\right> = 10^{42.25}~\mathrm{erg~s^{-1}}$.

\subsection{Sulfur emission lines}

\begin{deluxetable}{lccc}
\tablecaption{Summary of the detections ($\ge3\sigma$) of the \SII$\lambda\lambda$6717,6731 lines \label{tb:SII}}
\tablehead{\colhead{Subsample criteria}&
		\colhead{\SII$\lambda$6717} & 
		\colhead{\SII$\lambda$6731} & 
		\colhead{Both}}
\startdata
Any & 146 & 111 & 55 \\
in $H$-long & 98 & 72 & 30 \\
in $J$-long & 47 & 39& 25 \\
w/ \Ha~($\ge 3\sigma$) in HL & 84& 54&  22 \\
\enddata
\end{deluxetable}

The Sulfur emission lines \SII$\lambda$6717,6731 fall in the $H$-long ($J$-long) spectral window together with \Ha\ at $1.43<z<1.68$ ($0.70<z<1.00$).   For those with a detection of \Ha\ and/or \NII, we fit the \SII\ lines at the fixed spectroscopic redshift determined from \Ha+\NII, as described in \citet{2017ApJ...835...88K}.

\begin{figure}[t] 
   \centering
   \includegraphics[width=3.5in]{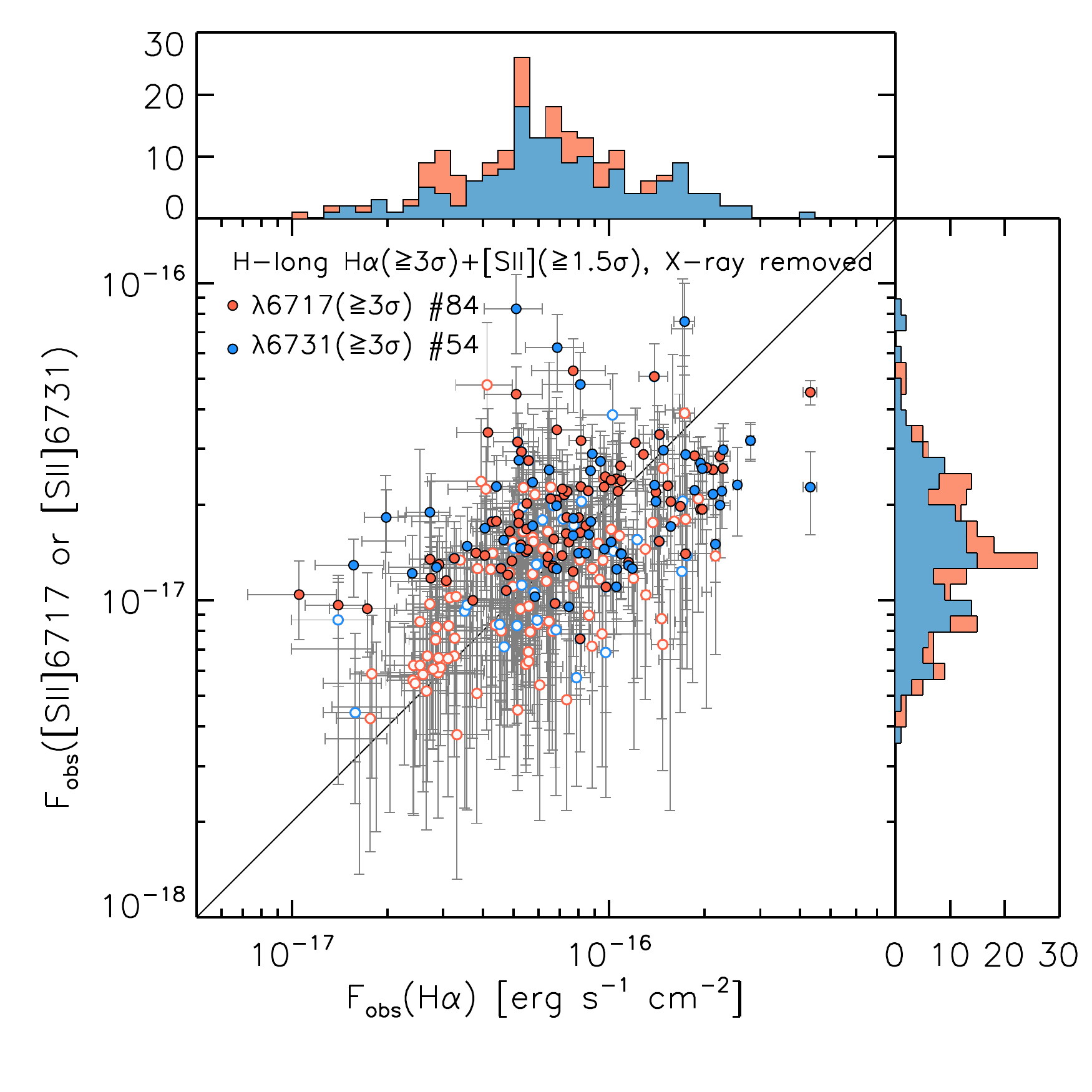}
   \caption{Correlation between observed \Ha\ and \SII\ fluxes detected in the $H$-long band, neither corrected for aperture effects nor extinction.  Red and blue filled (open) circles indicate the \SII$\lambda6717$ and \SII$\lambda6731$ fluxes with $S/N\ge3$ ($1.5\le S/N<3$), respectively. The diagonal solid line indicates the relation of $F_\mathrm{[SII]}=F_\mathrm{H\alpha}/5$.}
   \label{fig:flux_Ha_vs_SII}
\end{figure}

We successfully detected the \SII\ lines for a substantial fraction of the sample.  Table \ref{tb:SII} summarizes detections of the \SII\ lines.  In total, we detected \SII$\lambda6717$ and \SII$\lambda6731$ at $\ge 3\sigma$ for 146 and 111 objects, respectively, with 55 with both detections at $\ge 3\sigma$ (see the top row in Table \ref{tb:SII}).  Limiting those to have an \Ha\ detection ($>3\sigma$) in $H$-long, we detected \SII$\lambda6717$ for 84, \SII$\lambda6731$ for 54, and both of these for 22 objects (all at $\ge3\sigma$).  In Figure \ref{fig:flux_Ha_vs_SII}, we show the observed fluxes of \SII$\lambda$6717 and \SII$\lambda$6731 as a function of observed \Ha\ flux, neither corrected for dust nor aperture loss.  The observed flux of the single \SII\ line is on average $\approx1/5$ times the observed \Ha\ flux, ranging from $F_\mathrm{[SII]}\approx 4\times10^{-18}$ to $8\times10^{-17}~\mathrm{erg~s^{-1}~cm^{-2}}$.

% ===============================================================
\section{Assessment of the redshift and flux measurements \label{sec:assessment}}
\subsection{Redshift accuracy}

\begin{figure}[tbp] 
   \centering
   \includegraphics[width=3.4in]{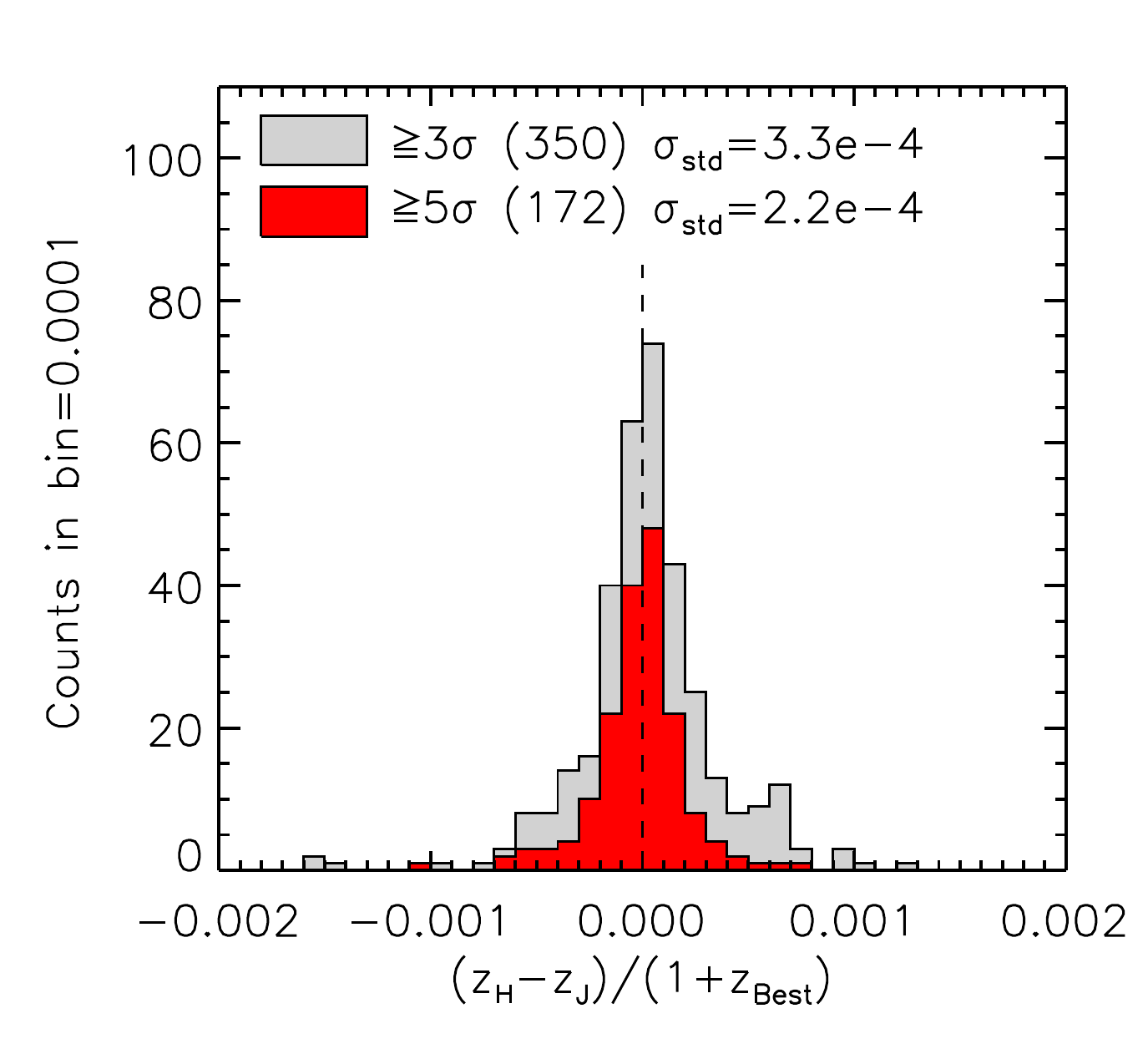} 
   \includegraphics[width=3.4in]{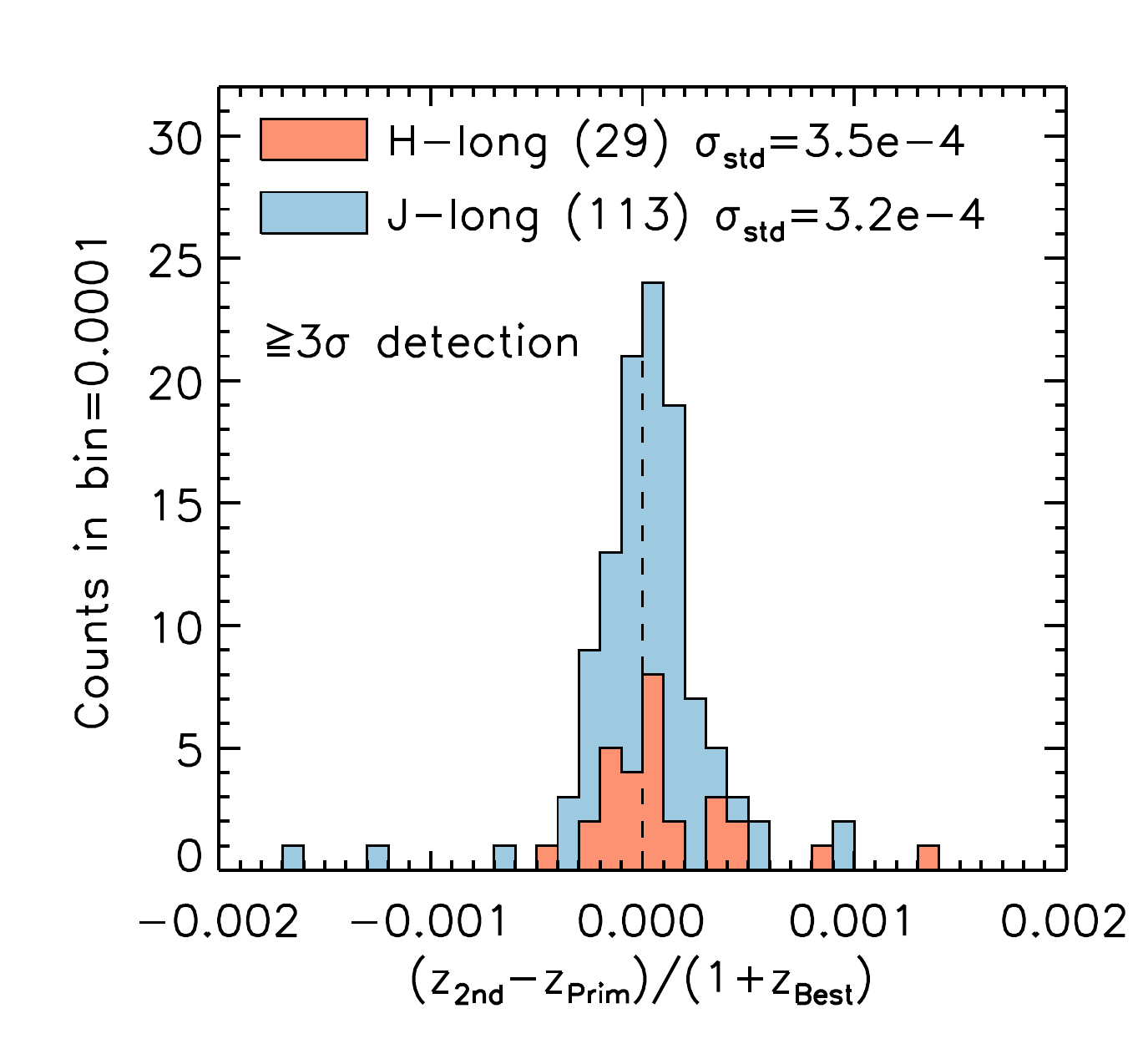} 
   \caption{Upper panel: distribution of $(z_H - z_J)/(1+z_\mathrm{best})$, the difference between spectroscopic redshifts measured from \Ha+\NII\ detected in the $H$-long and those from \Hb+\OIII\ in the $J$-long spectra.  The red histogram represents the subsample with a $\ge5\sigma$ detection in both $H$- and $J$-long spectra.  Lower panel: distribution of $(z_\mathrm{2nd}-z_\mathrm{prim})/(1+z_\mathrm{best})$ (see text).  Red and blue histograms correspond independently to the measurements in the $H$-long and $J$-long, respectively.  Here, the line detections are limited to be $\ge3\sigma$.  In each panel, the values of standard deviation are denoted.}
   \label{fig:hist_dzs}
\end{figure}

To evaluate the accuracy of our redshift estimates, we compared spectroscopic redshifts measured from \Ha+\NII\ detected in the $H$-long spectra and those measured from \Hb+\OIII\ in the $J$-long spectra.  In the top panel of Figure \ref{fig:hist_dzs}, we show the distribution of $(z_H - z_J)/(1+z_\mathrm{best})$ for 350 galaxies with independent line detections in the two spectral windows both at $\ge3\sigma$.  Of these, 172 objects have detections both at $\ge5\sigma$.  Here, the best estimate of redshift $z_\mathrm{best}$ is based on a detection with a higher S/N between the two spectral windows.  The standard deviation $\sigma_\mathrm{std}$ of $dz/(1+z)$ is $3.3\times10^{-4}$ for objects with $\ge3\sigma$ detection ($\sigma_\mathrm{std} = 2.2\times10^{-4}$ for $\ge5\sigma$), with a negligibly small median offset ($2.6\times 10^{-5}$).  The estimated redshift accuracy $\sigma_\mathrm{std}/\sqrt{2}$ is thus to be $\approx70~\mathrm{km~s^{-1}}$.

An alternative check of redshift accuracy can be done using objects that have been observed twice or more times with the same grating on different nights.  For these objects, we have selected the best spectrum to construct the line measurement catalog.  However, the ``secondary'' measurements can be used to evaluate the ``primary'' ones.  In the lower panel of Figure \ref{fig:hist_dzs}, we show the distribution of the difference between the primary ($z_\mathrm{prim}$) and the second-best ($z_\mathrm{2nd}$) redshift measurements, separately for measurements obtained in $H$-long (29 objects) and in $J$-long band (113).  Objects are limited to those with $\ge3\sigma$ detection in the primary and secondary spectra.  The standard deviation $\sigma_\mathrm{std}$ of $(z_\mathrm{2nd}-z_\mathrm{prim})/(1+z_\mathrm{best})$, reported in the figure for both $H$-long ($\sigma_\mathrm{std} = 3.5\times10^{-4}$) and $J$-long ($\sigma_\mathrm{std} = 3.3\times10^{-4}$) observations, is similar to that estimated by comparing the $H$-long and $J$-long measurements.

\subsection{Flux accuracy, using repeat observations \label{sec:comp_repeat}}

\begin{figure}[tbp] 
   \centering
   \includegraphics[width=3.6in]{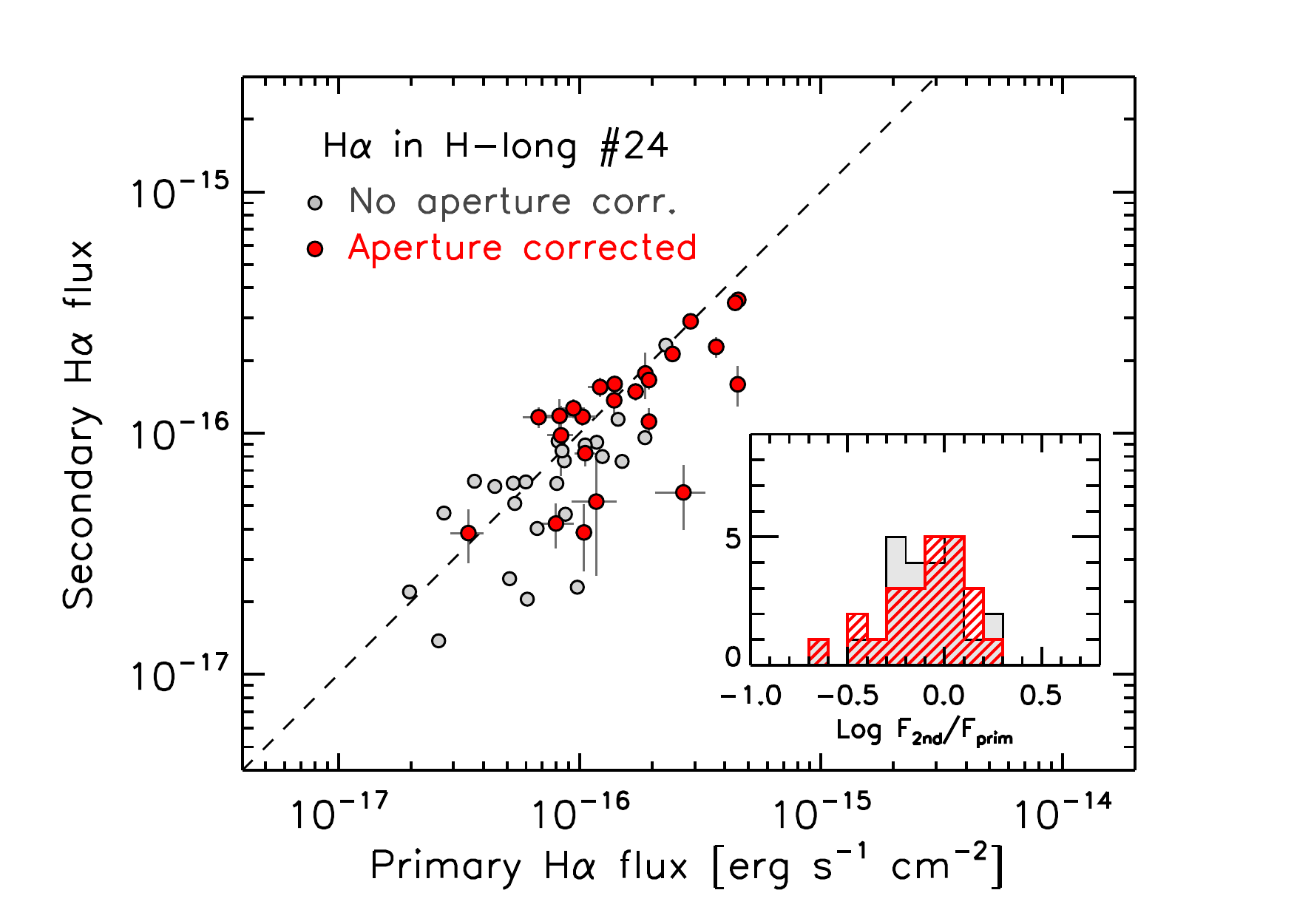} 
   \includegraphics[width=3.6in]{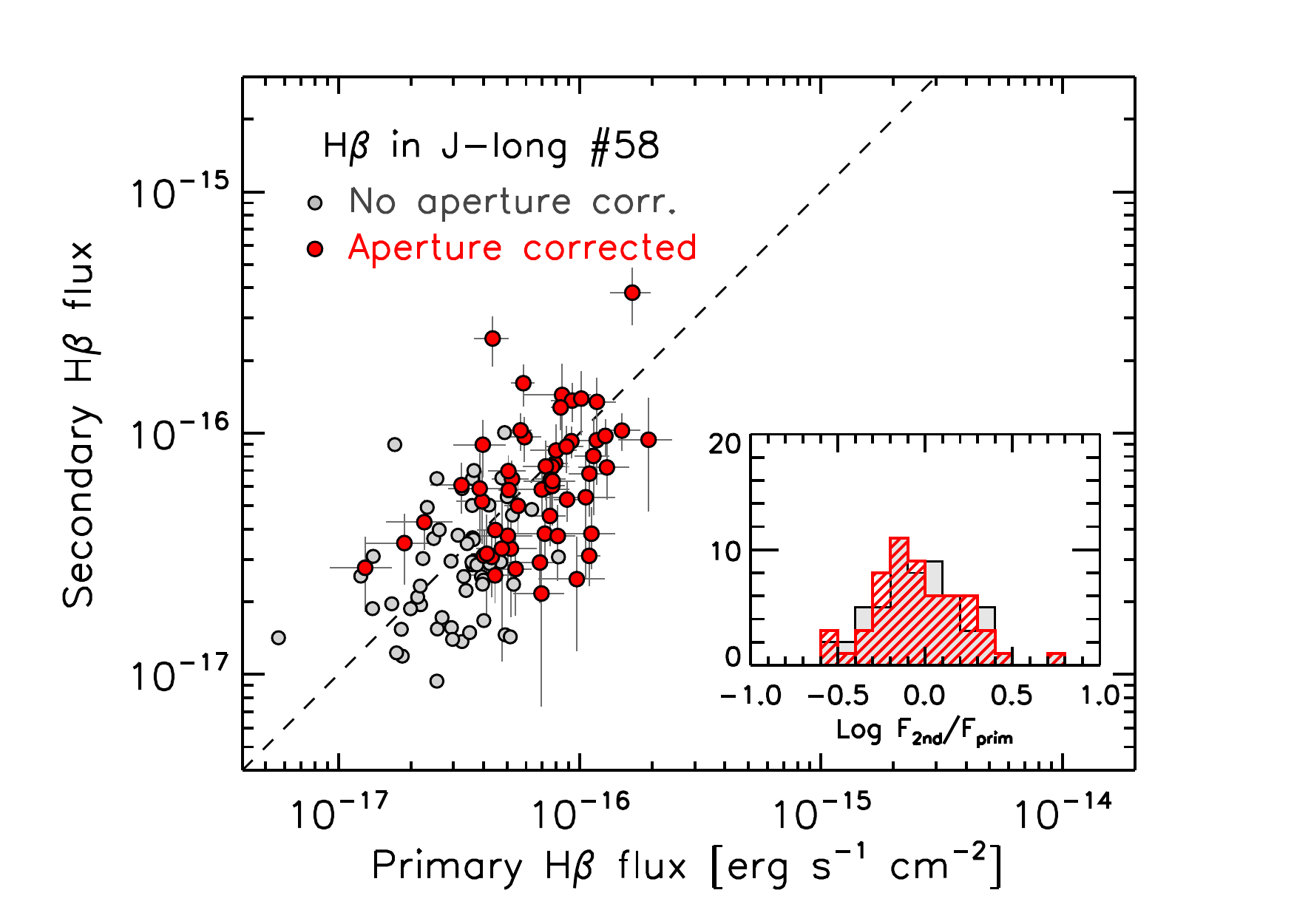} 
   \includegraphics[width=3.6in]{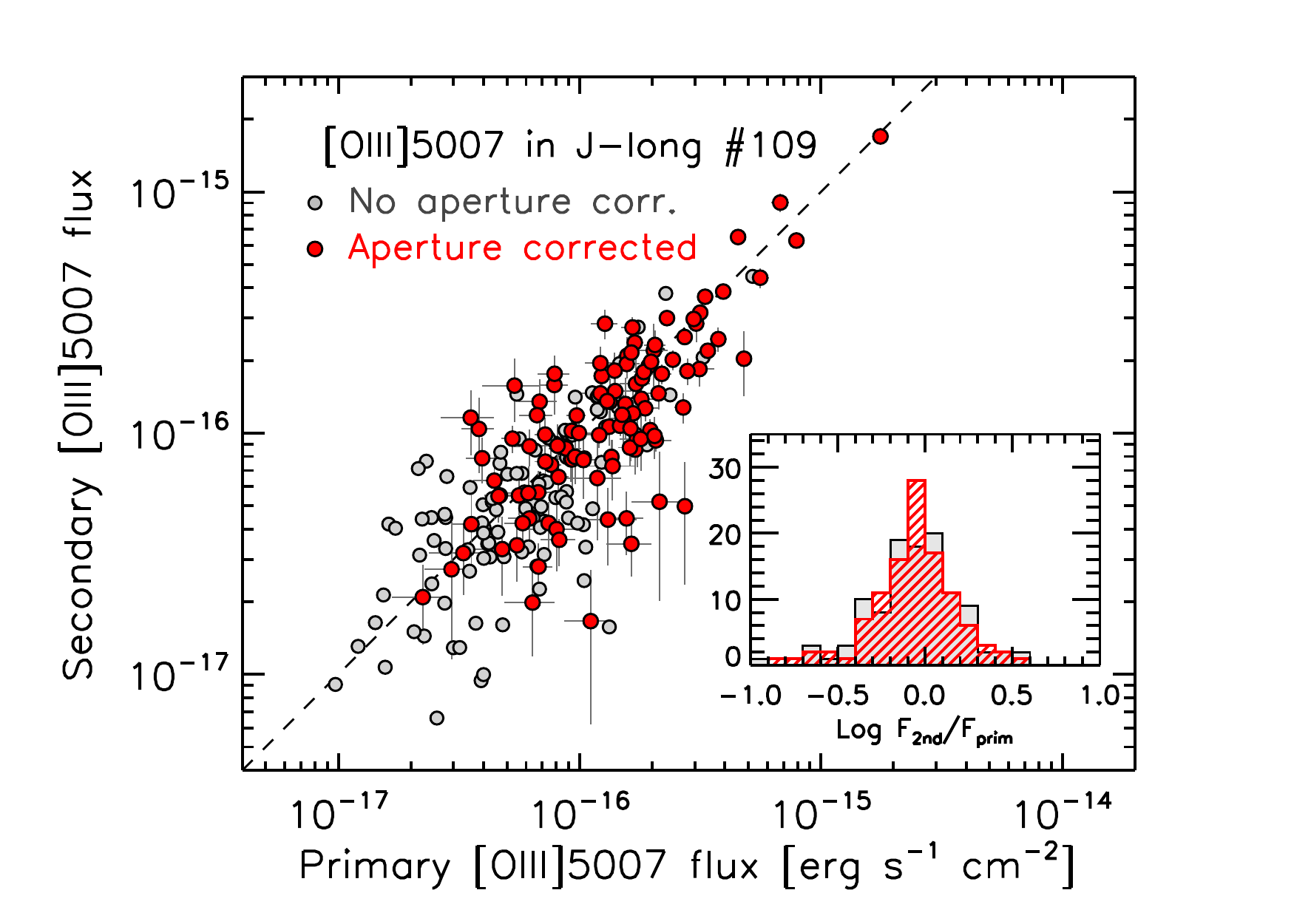}
   \caption{
   Comparison of the best (primary) and the second best (secondary) measurements of the \Ha\ flux in the $H$-long (top panel), and \Hb\ and \OIII$\lambda5007$ fluxes in the $J$-long window (middle and bottom panels) for the repeated objects.  Red and gray circles indicate the observed fluxes with and without aperture correction.  Inset panels show the distribution of the flux ratios $\log(F_\mathrm{second}/F_\mathrm{primary})$ before (gray) and after (red hatched histogram) aperture correction.
   }
   \label{fig:comp_repeat}
\end{figure}

The secondary measurements can be also used to evaluate the accuracy of emission-line flux measurements.  In Figure \ref{fig:comp_repeat}, we compare the secondary and primary measurements of the \Ha\ flux in the $H$-long window (upper panel; 24 objects), and the \Hb\ (middle panel; 58 objects) and \OIII$\lambda$5007 fluxes (lower panel; 109 objects) in the $J$-long window.  Because the two measurements are based on spectra taken under different seeing conditions, the aperture correction needs to be applied for comparison.  We remind that the aperture correction is evaluated once for each object and observing night.  It is shown that the primary and secondary measurements are in good agreement, as well as that aperture correction improves their agreement as shown by histograms in the inset panels.  We found the intrinsic scatter of these correlations to be 0.19, 0.21, and 0.19~dex for \Ha, \Hb, and \OIII, respectively, after taking into account the effects of the individual formal errors of the observed fluxes.  These intrinsic scatters should be attributed to the uncertainties of the aperture corrections, and indeed similar to the estimates made in Section \ref{sec:apercorr} (see Figure \ref{fig:mag_vs_dmag}).

%------------------------------------------------------
\subsection{Comparison with MOSDEF}
Part of our FMOS-COSMOS targets were observed in the MOSFIRE Deep Evolution Field (MOSDEF) survey  \citep{2015ApJS..218...15K}.  The latest public MOSDEF catalog, released on 11 March 2018, contains 616 objects in the COSMOS field.  Cross-matching with the FMOS catalog, we found 45 sources included in both catalogs, and of these, 15 objects have redshift estimates in both surveys.  

Among the matching objects, all 11 FMOS measurements with $z\mathrm{Flag}=4$ and a single $z\mathrm{Flag}=3$ agree with the MOSDEF measurements, which all have a quality flag ({\tt Z\_MOSFIRE\_ZQUAL}) of 7 (based on multiple emission lines at $S/N\ge2$).  The three inconsistent measurements are as follows. An object ($z_\mathrm{FMOS}=1.515$ with $z\mathrm{Flag}=3$) has a \OIII\ detection at $>5\sigma$, with a possible consistent detection of \Ha, in the FMOS spectra, while the MOSDEF measurement is $z=2.1$ with a flag of 7.  The photometric redshift $z_\mathrm{phot}=1.674$ \citep{2016ApJS..224...24L} prefers the FMOS measurement.  For the remaining two (FMOS/MOSDEF estimates (flags) are $z_\mathrm{FMOS/MOSDEF}=1.581/2.555$ (2/6) and $z_\mathrm{FMOS/MOSDEF}=1.584/2.100$ (1/7), respectively), the detections of \Ha\ on the FMOS spectra are not robust, both being significantly affected by the OH mask.  The photometric redshift prefers $z_\mathrm{MOSDEF}$ for the former ($z_\mathrm{phot}=2.612$), while $z_\mathrm{FMOS}$ for the latter ($z_\mathrm{phot}=1.458$).  We note that the redshift range of the MOSDEF survey is $1 \lesssim z \lesssim 3.5$, but having a higher sampling rate at $2<z<2.6$.  Therefore, it is not straightforward to estimate the failure rate in our survey, which could be overestimated.  The small sample size of the matching objects also makes it difficult.  However, we could conclude that, for objects with $z$Flag=3 and 4, the failure rate should be below $10\%$ ($1/13=7.7\%$). 

For these 12 consistent measurements, we found the median offset and standard deviation of $(z_\mathrm{FMOS}-z_\mathrm{MOSDEF})/(1+z_\mathrm{MOSDEF})$ to be $-1.63\times10^{-5}$ ($4.9~\mathrm{km~s^{-1}}$) and $2.57\times10^{-4}$ ($77~\mathrm{km~s^{-1}}$).  This indicates that there is no significant systematic offset in the wavelength calibration of the FMOS survey relative to the MOSDEF survey.

%------------------------------------------------------
\subsection{Comparison with 3D-HST}

A part of the CANDELS-COSMOS field \citep{2011ApJS..197...35G} is covered by the 3D-HST survey \citep{2012ApJS..200...13B}, which is a slitless spectroscopic survey using the {\it HST}/WFC3 G141 grism to obtain near-infrared spectra from 1.10 to $1.65~\mathrm{\mu m}$.  This configuration yields detections of \Ha\ and \OIII\ lines for those matched to the FMOS catalog.  For redshift and flux comparisons with our measurements, we employed the public `linematched' catalogs (ver. 4.1.5) for the COSMOS field \citep{2016ApJS..225...27M}, in which the spectra extracted from the grism images were matched to photometric targets \citep{2014ApJS..214...24S}\footnote{Available here: http://3dhst.research.yale.edu/Home.html}.  Cross-matching the 3D-HST and the FMOS catalogs, we found 78 objects that have redshift measurements from both surveys.  We divided these objects into two classes according to the quality flags in the 3D-HST catalog ({\tt flag1} and {\tt flag2}).  The `good' class contains 67 objects with both $\textrm{\tt flag1}=0$ and $\textrm{\tt flag2}=0$, while the remaining 11 objects are classified to the `warning' class.

\begin{figure}[tbp] 
   \centering
   \includegraphics[width=3.7in]{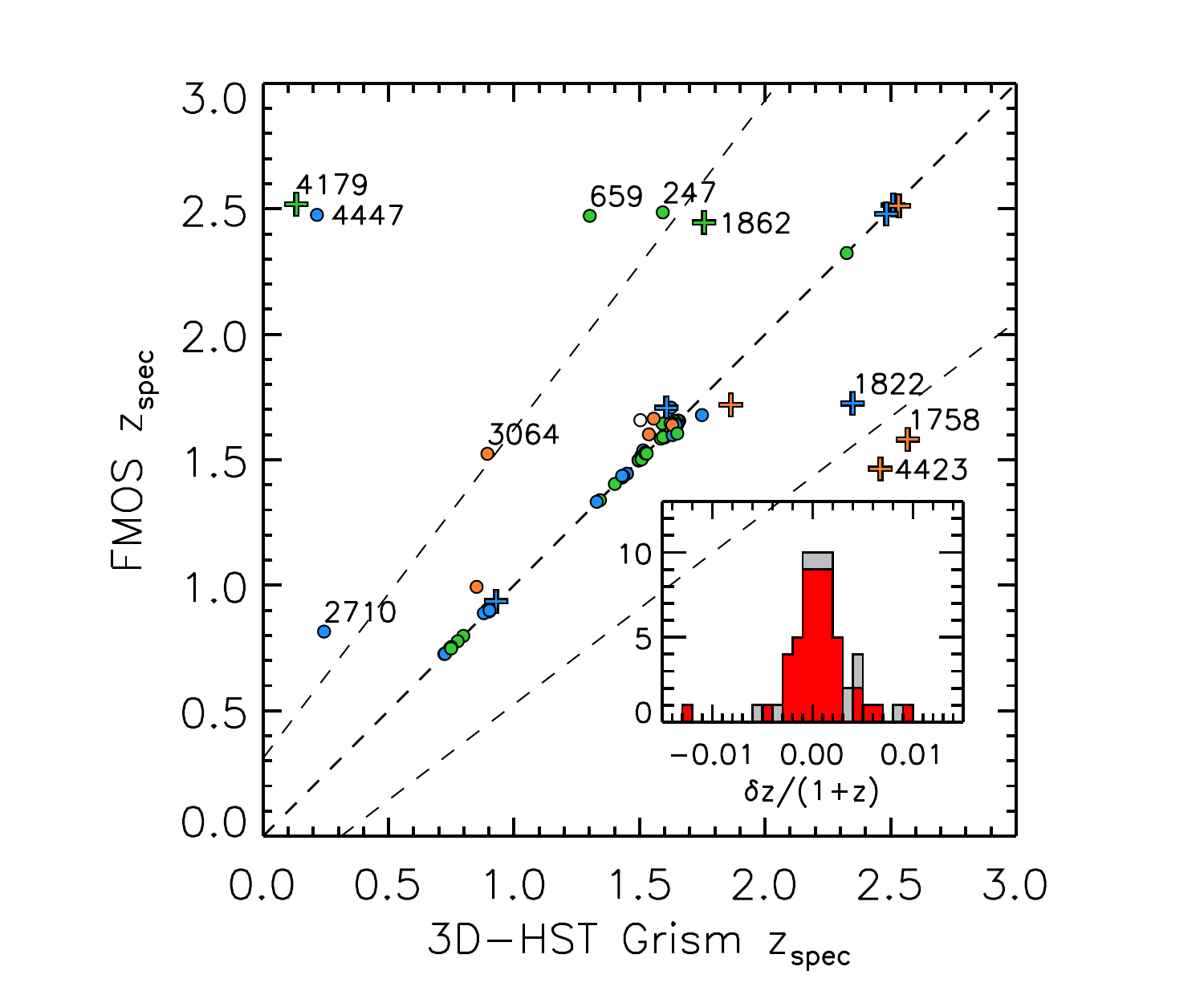}
   \caption{
   Comparison between spectroscopic redshift estimates from FMOS and 3D-HST.  Colors indicate the quality flag of the FMOS estimation: $z\mathrm{Flag}=1~(\mathrm{white})$, $2~(\mathrm{red})$, $3~(\mathrm{blue})$, and $4~(\mathrm{green})$.  Circles and cross symbols correspond to the `good' and `warning' classes, respectively, according  to the  3D-HST flags (see text).  The middle dashed line indicates the  one-to-one relation, while the other two dashed lines correspond to cases that, assuming the 3D-HST measurements are correct, the \Ha\ line is misidentified as \OIII$\lambda$5007 (upper line), or \OIII\ is misidentified as \Ha.  Some objects are labeled (see text).}
   \label{fig:comp_3dHST_zs}
\end{figure}

In Figure \ref{fig:comp_3dHST_zs}, we compare our FMOS redshift estimates to those from 3D-HST for the 78 matching sources.  The colors indicate the quality flags of the FMOS measurements (see Section \ref{sec:redshift_measurements}) and the symbols correspond to the quality classes of the 3D-HST measurements as defined above.  In the inset panel, we show the distribution of $(z_\mathrm{FMOS}-z_\mathrm{3DHST})/(1+z_\mathrm{3DHST})$ for all objects along the diagonal one-to-one line (grey histogram) and for the subsample with FMOS $z\mathrm{Flag}\ge3$ and in the 3D-HST `good' class (red histogram).  We find that average offset of $dz/(1+z)=0.0009$ and a standard deviation of $\sigma_\mathrm{std}=0.0029$ (0.0022 for the `good' sample) after 3-$\sigma$ clipping, which corresponds to $870~\mathrm{km~s^{-1}}$, consistent with the typical accuracy of the redshift determination in 3D-HST \citep{2016ApJS..225...27M}.

We further examine the possible line misidentification for the 10 cases where two spectroscopic redshifts are inconsistent (labeled in Figure \ref{fig:comp_3dHST_zs}).  Of these objects, we found that the FMOS $z_\mathrm{spec}$ is quite robust for three $z\mathrm{Flag}=4$ (IDs 247, 659 4179) and one $z\mathrm{Flag}=3$ objects (ID 4447).  A single $z\mathrm{Flag}=3$ object (ID 2710, $z_\mathrm{spec}=0.815$) has a clear detection of a single line.  If this line is [S\,{\sc iii}]$\lambda9531$ in reality, the corresponding redshift agrees with that from 3D-HST.   For other five objects, our FMOS measurements are not fully robust, including a single $z\mathrm{Flag}=4$ object (ID 1862), whilst four of these are flagged as `warning' in 3D-HST.  We thus would conclude that the possibility of the line misidentification (including fake detections) is equal or less than 6/67=9\%, even down to $z$Flag$\ge2$.  A similar estimate of the possibility ($\lesssim 10\%$) has been obtained from a comparison with the zCOSMOS-Deep survey (\citealt{2007ApJS..172...70L}) for matching objects \citep[see][]{2015ApJS..220...12S}.

\begin{figure}[tbp] 
   \centering
   \includegraphics[width=3.5in]{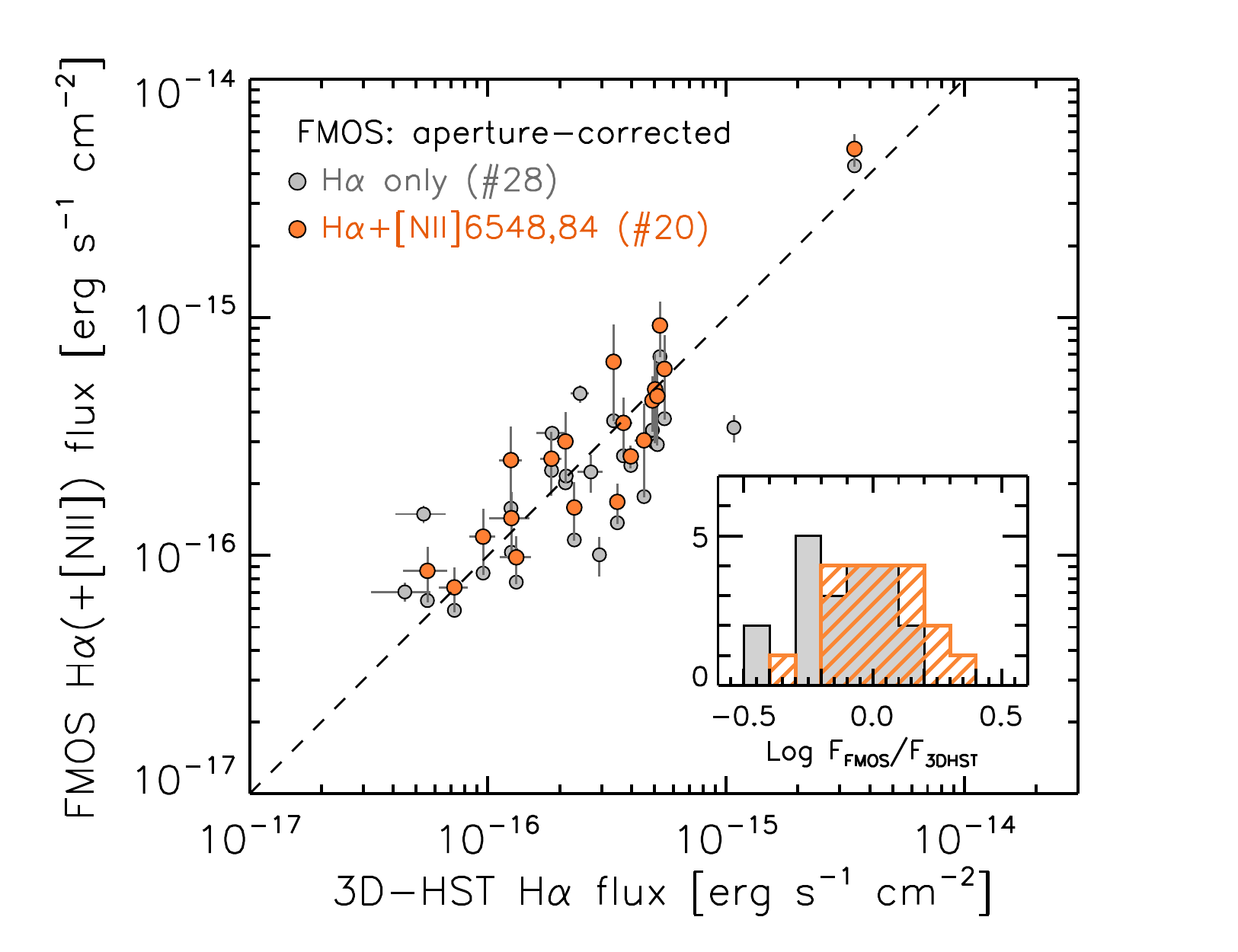}
   \includegraphics[width=3.5in]{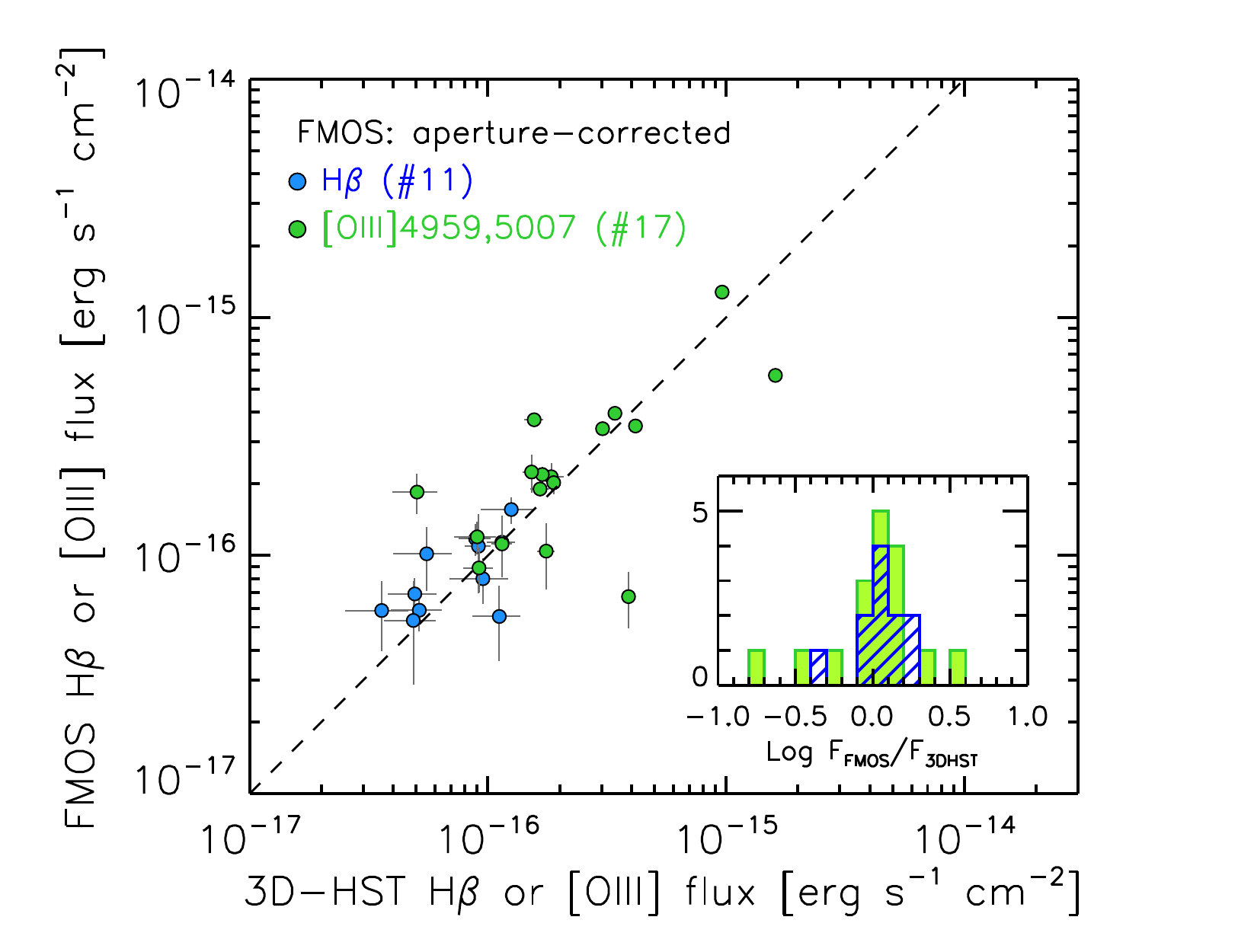}
   \caption{Comparison of the fluxes of \Ha\ (+\NII) (upper panel), and \Hb\ and \OIII\ (lower panel) measured by FMOS and 3D-HST.  The FMOS measurements are corrected for aperture effects.  Gray circles indicate the \Ha\ fluxes measured by FMOS, while orange circles indicate the \Ha+\NII$\lambda6548.,6584$ fluxes to match the 3D-HST measurements, in which the \Ha\ and \NII\ lines are blended.  Blue and green circles indicate \Hb\ and \OIII$\lambda$4959,5007 fluxes, respectively.  Dashed lines indicate a one-to-one relation.  Inset panels indicate distribution of the FMOS-to-3D-HST flux ratios, color-coded as the symbols.}
   \label{fig:comp_3dHST_flux}
\end{figure}

Next we compare our flux measurements to those in 3D-HST for the matching sample.  In contrast to fiber spectroscopy, slitless grism spectroscopy is less affected by aperture losses, and therefore offer an opportunity to check our measurements with aperture correction.  Objects used for this comparison are limited to have a detection of \Ha\ or \OIII\ at $\ge 3\sigma$ in both FMOS and 3D-HST and a consistent redshift estimation ($|dz|/(1+z)\le0.01$).  In the top panel of Figure \ref{fig:comp_3dHST_flux} we compare \Ha\ fluxes measured from FMOS (corrected for aperture loss) with those from 3D-HST for 28 objects.  Here the \Ha\ fluxes from 3D-HST includes the contribution from \NII$\lambda\lambda$6548,6584 because these lines are blended with \Ha\ due to the low spectral resolution ($R\sim100$).  Therefore, we also show the total fluxes of \Ha\ and \NII$\lambda\lambda$6548,6584 for the FMOS measrurements (orange circles).  Eight of the 28 objects have no detection of \NII.  Even with no inclusion of \NII, good agreement is seen between the measurements of both programs,  with a median offset of $-0.09~\mathrm{dex}$ and an rms scatter of $0.23~\mathrm{dex}$.  As naturally expected, the inclusion of \NII\ lines further improves the agreement, resulting in an offset of $0.02~\mathrm{dex}$ and a scatter of $0.17~\mathrm{dex}$.

The bottom panel of Figure \ref{fig:comp_3dHST_flux} shows the comparisons of observed \Hb\ and \OIII\ fluxes.  We show the total fluxes of \OIII$\lambda\lambda$4959,5007 for the FMOS measurements because 3D-HST does not resolve the \OIII\ doublet.  Similarly to \Ha, there is good agreement between flux measurements for these lines with little average offset ($<0.1~\mathrm{dex}$) and small scatter ($<0.2~\mathrm{dex}$). 

The agreement of our flux measurements with the slitless measurements from 3D-HST indicates the success of our absolute flux calibration including aperture correction.  The scatter found in the comparisons ($\sim 0.2~\mathrm{dex}$) is equivalent to those found in comparisons using repeat observation (Section \ref{sec:comp_repeat}), as well as to the typical uncertainty in the aperture correction (Section \ref{sec:apercorr}).

%------------------------------------------------------------------------------------------------------------
\section{Retroactive evaluation of the FMOS-COSMOS sample \label{sec:re_L16}}

The master catalog of our FMOS survey contains various galaxy populations selected in different ways, as described above.  Even for the primary population of star-forming galaxies at $z\sim1.6$ whose \Ha\ is expected to be detected in the $H$-long spectral window, the quantities used for the selection such as photometric redshift, stellar mass, and predicted \Ha\ fluxes had been updated during the period of the project.  Therefore, it is useful to re-evaluate the FMOS sample using a single latest photometric catalog as a base, in which galaxy properties are derived in a consistent way.  For the retroactive characterization of the sample, we rely on the COSMOS2015 catalog \citep{2016ApJS..224...24L}, which contains an updated version of photometry and photometric redshifts, as well as estimates of stellar mass and SFR, for objects across the full area of the COSMOS field.

\begin{figure}[t] 
   \centering
   \includegraphics[width=3.5in]{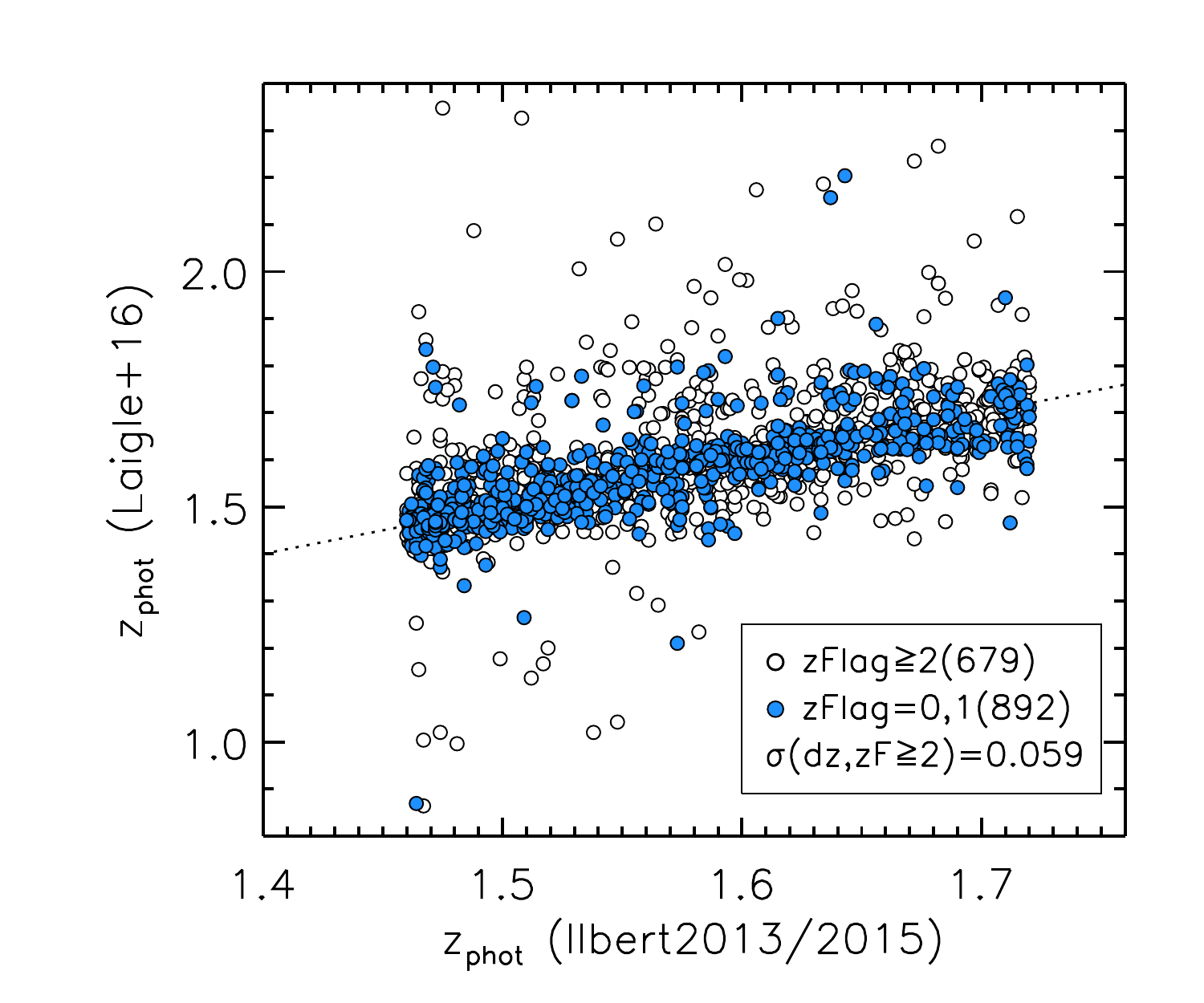}
   \caption{Comparison between the photometric redshifts from our original parent catalog based on \citet{2013A&A...556A..55I,2015A&A...579A...2I} and the COSMOS2015 catalog \citep{2016ApJS..224...24L}.  White and blue circles indicate objects with $z$Flag$=0$ and 1 and those with $z$Flag$\ge2$, respectively.  Dotted line indicates a one-to-one relation.  }
   \label{fig:zp_orig_vs_zp_Laigle}
\end{figure}

In Figure \ref{fig:zp_orig_vs_zp_Laigle}, we compare the photometric redshifts in the COSMOS2015 catalog with those originally used for target selection.  Here, we show FMOS objects that are included in the Primary-HL sample defined in Section \ref{sec:Primary-HL} and are matched in the COSMOS2015 catalog.  It is clear that the photometric redshift estimates from the different versions of the COSMOS photometric catalogs are in good agreement.  The standard deviation of $dz$ for the $z\mathrm{Flag}\ge2$ objects is $\sigma (dz)=0.059$ after 5-$\sigma$ clipping, which is in good agreement with the typical errors of the photometric redshifts relative to the spectroscopic redshifts (see Section \ref{sec:redshifts_Primary-HL}).

\subsection{Sample construction \label{sec:re_L16_sample}}

\begin{deluxetable*}{llcc}
\tablecaption{Summary of the sampling and success rates \label{tb:re_L16}}
\tablehead{\colhead{Sample}&
		\colhead{Selection}&
		\colhead{$N_ \mathrm{galaxies}$}&
		\colhead{Fraction}}
\startdata
Broad L16 & in the FMOS field (1.35~deg$^2$) & & \\
& {\tt \&\&} $\mathrm{TYPE}=0$ (flagged as a galaxy)  & & \\
& {\tt \&\&} $\mathrm{FLAG\_COSMOS}=1$   & & \\
& {\tt \&\&} $\mathrm{FLAG\_HJMCC}=0$  & & \\
& {\tt \&\&} $\mathrm{FLAG\_PETER}=0$  & & \\
& {\tt \&\&} $K_\mathrm{S} \le 24.0$ {\tt \&\&} $1.3 \le z_\mathrm{phot}\le 1.9$  & 39435 & \\
\hline
$\cap$ FMOS HL & {\tt \&\&} Observed in the $H$-long & 2878 & 7.3\% (2878/39435) \\
$\cap$ FMOS HL + \Ha& {\tt \&\&} \Ha\ detection ($3\sigma$) in the $H$-long & 1014 & 35.2\% (1014/2878) \\
\hline\hline
Selected L16 & Criteria for Parent L16 & & \\
& {\tt \&\&} $K_\mathrm{S}\le 23.0$  {\tt \&\&} $1.43 \le z_\mathrm{phot} \le 1.74$ &  &  \\
& {\tt \&\&} $\log M_\ast/M_\odot \ge 9.6$ &  \\
& {\tt \&\&} $F^\mathrm{pred}_\mathrm{H\alpha} \ge 1\times10^{-16}~\mathrm{erg~s^{-1}~cm^{-2}}$ & 3714 &  \\
\hline
$\cap$ FMOS HL& {\tt \&\&} Observed in the $H$-long & 1209 & 32.6\% (1209/3714) \\
$\cap$ FMOS HL + \Ha& {\tt \&\&} \Ha\ detection ($3\sigma$) in the $H$-long & 628 & 51.9\% (628/1209) \\
\enddata
\end{deluxetable*}

In this section, we focus our attention on the star-forming population at $z\sim1.6$, in particular, with a detection of \Ha\ at $\ge 3\sigma$.  Therefore, we limit this discussion to those that were observed in the $H$-long band.  We excluded X-ray objects identified in the {\it Chandra} COSMOS Legacy catalog (see Section \ref{sec:Chandra}).  

We first construct a broad sample from COSMOS2015, named {\it Broad-L16}, to be sufficiently deep relative to the FMOS sample.  We limit the sample to be flagged as a galaxy ({\tt TYPE}=0), being within the strictly defined 2~deg$^2$ COSMOS field ({\tt FLAG\_COSMOS}$=1$), inside the UltraVISTA field ({\tt FLAG\_HJMCC}$=0$), inside the good area (not masked area) of the optical broad-band data ({\tt FLAG\_PETER}$=0$), and inside the FMOS area covered by all the pawprints (see Figure \ref{fig:pawprints}).  As a consequence, the effective area used for this evaluation is $1.35~\mathrm{deg^2}$ , after removing the masked regions\footnote{The DS9-format region files for the outlines and masked regions are available online: http://cosmos.astro.caltech.edu/page/photom.}  For details of these flags, we refer the reader to \citet{2016ApJS..224...24L}.  

We further impose a limit on the photometric redshift $1.3 \le z_\mathrm{phot} \le 1.9$ and the UltraVISTA $K_\mathrm{S}$-band magnitude $K_\mathrm{S}\le 24.0$, where we use the $3^{\prime\prime}$-aperture magnitude ({\tt KS\_MAG\_APER3}).  This limiting magnitude corresponds to the $3\sigma$ limit in the UltraVISTA deep layer, and $\ge5\sigma$ for the Ultra-Deep layer.  Finally, we find 39,435 galaxies satisfying these criteria.  We then performed the position matching between the $H$-long sample and the COSMOS2015 catalog with a maximum position error of 1.0 arcsec, yielding the matched sample that consists of 2878 objects (Broad-L16 $\cap$ FMOS-HL) \footnote{We note that, from matching the full FMOS-COSMOS catalog to the full COSMOS2015 catalog, we find best-matched counterpart for 5157 extragalactic objects.  The public FMOS-COSMOS catalog contains the best-matched ID for objects in the COSMOS2015 catalog for each FMOS object (a column {\tt ID\_LAIGLE16}).}.   The fraction with respect to the Broad-L16 sample is 7.3\%, and we detected \Ha\ at $\ge 3\sigma$ for 1014 of these, thus the success rate is 35\% (1014/2878).  

Next we imposed additional limits onto the Broad-L16 sample while simultaneously trying to keep the sampling rate as high as possible and not to lose \Ha-detected objects.  For this purpose, we use the stellar mass and SFR estimates from SED-fitting given in the COSMOS2015 catalog.  We computed predicted \Ha\ fluxes as follows:
\begin{equation}
F^\mathrm{pred}_\mathrm{H\alpha} = \frac{1}{4\pi d_\mathrm{L}(z_\mathrm{phot})^2} \frac{\mathrm{SFR}/(M_\odot~\mathrm{yr^{-1}})}{4.6\times10^{-42}} \times 10^{-0.4A_\mathrm{H\alpha}}
\label{eq:predfHa}
\end{equation}
This is the modified version of Equation (2) of \citet{1998ARA&A..36..189K} for the use of a \citet{2003PASP..115..763C} IMF.  Dust extinction is taken into account with $A_\mathrm{H\alpha} = k_\mathrm{H\alpha}E(B-V)/f_\mathrm{neb}$, where $k_\mathrm{H\alpha}=2.54$ is the wavelength dependence of extinction \citet{1989ApJ...345..245C}.  The extinction $E(B-V)$ is taken from the COSMOS2015 catalog \footnote{Although we here assume a single attenuation curve for nebular emission from \citet{1989ApJ...345..245C}, different attenuation curves for stellar emission have been applied for different objects in the COSMOS2015 catalog, which induces systematic uncertainties in the estimates of $E(B-V)$.}, and is multiplied by a factor of $1/f_\mathrm{neb}=1/0.5$ to account for enhancement of extinction towards nebular lines (see Section \ref{sec:dust})\footnote{This factor makes the value of $k_\mathrm{H\alpha}/f_\mathrm{neb}$ nearly the same as one with the \citet{2000ApJ...533..682C} law ($k_\mathrm{H\alpha}=3.325$) and $f_\mathrm{neb}=0.66$, as used in our target selection and past papers.}.  Finally, we define the {\it Selected-L16} sample by imposing the criteria $K_\mathrm{S} \le 23.0$, $1.43 \le z_\mathrm{phot} \le 1.74$, $\log M_\ast/M_\odot \ge 9.6$, and the predicted \Ha\ flux $F^\mathrm{pred}_\mathrm{H\alpha} \ge 1 \times10^{-16}~\mathrm{erg~s^{-1}~cm^{-2}}$, in addition to the criteria on the Broad-L16 sample.  As a consequence, the Selected-L16 sample includes 3714 objects.

Cross-matching the Selected-L16 sample with the FMOS $H$-long sample, we find 1209 objects, and 628 of these have a successful detection of \Ha\ ($\ge3\sigma$) in the $H$-long spectra.  Thus, the sampling rate is 33\% (1209/3714), and the success rate is 52\% (628/1209).  It is worth noting that the rate of failing detection can be reasonably explained: if the redshift distribution is uniform, approximately $20\%$ of those within $1.43 \le z_\mathrm{phot} \le 1.74$ may fall outside the redshift range covered by the $H$-long grating for their uncertainty on the photometric redshift ($\delta z_\mathrm{phot}\approx 0.06$; see below).  Moreover, $\approx35\%$ of potential objects may be lost due to severe contamination by OH skylines (see Figure \ref{fig:histz}).  In Table \ref{tb:re_L16}, we summarize the selection and the sizes of the samples defined in this section.  We note that it is not guaranteed that all objects in the Selected-L16 sample were included in the input sample for the fiber allocation software.  

%-----------------------------------------------------------------------------------------
\subsection{Sampling and detection biases \label{sec:re_L16_bias}}

\begin{figure*}[htbp] 
   \centering
   \includegraphics[width=7.in]{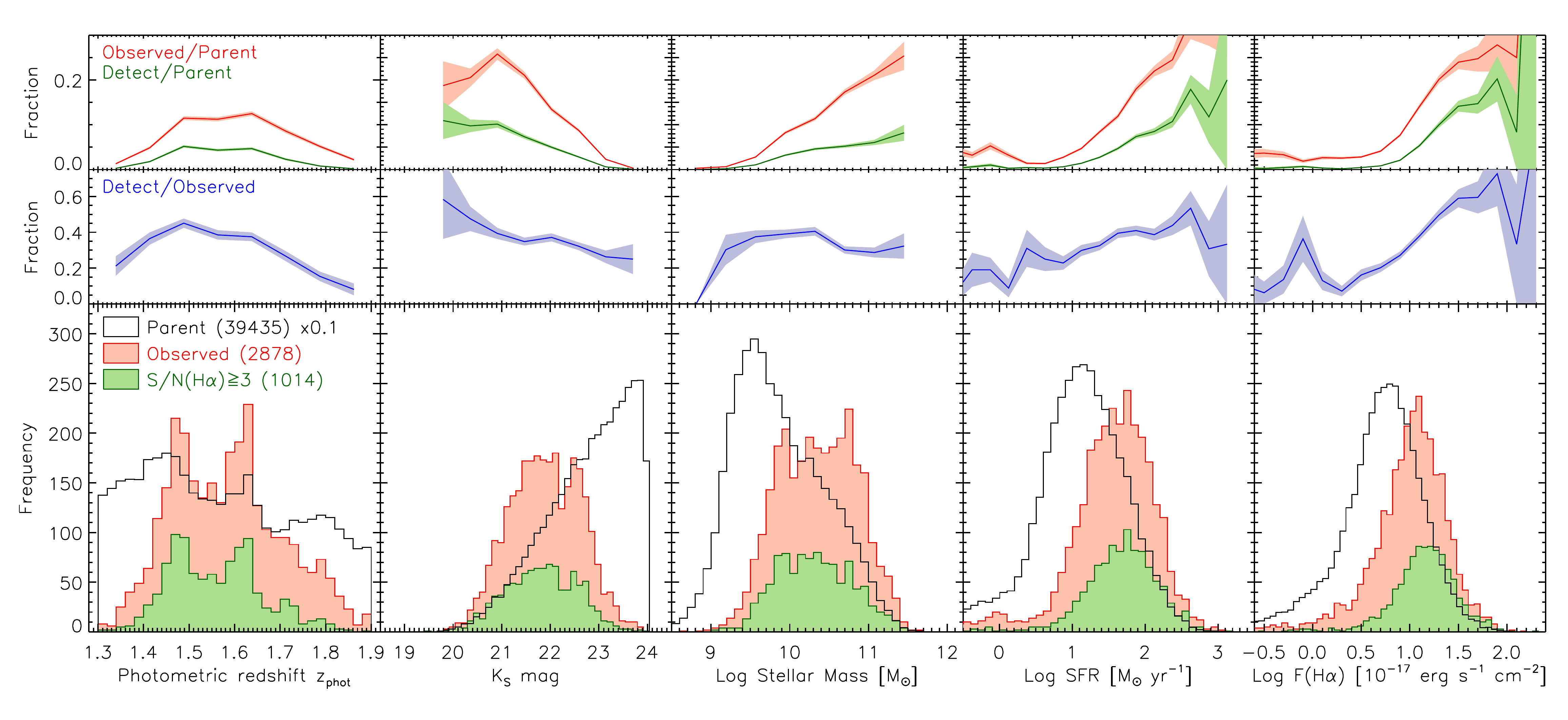}
   \caption{Distribution of the SED-based photometric redshifts, $K_\mathrm{S}$ magnitudes, stellar masses, SFRs and predicted \Ha\ fluxes from left to right.  The empty histograms show the distributions of the parent Broad-L16 sample ($N=39435$), scaled by a factor of 0.1, while the filled red and green histograms indicate the observed ($\cap$ FMOS HL, $N=2878$) and \Ha-detected ($\cap$ FMOS HL $+$ \Ha, $N=1014$) samples (see Table \ref{tb:re_L16}).  In the top panels, the fraction of the observed and \Ha-detected samples relative to the parent sample is shown for each quantity.  In the middle panel, the fraction of \Ha-detected sample to the observed sample is shown.  In the top and middle panels, the shaded region indicate the Poisson errors in each bin.}
   \label{fig:hist_L16_par}
\end{figure*}

\begin{figure*}[htbp] 
   \centering
   \includegraphics[width=7.in]{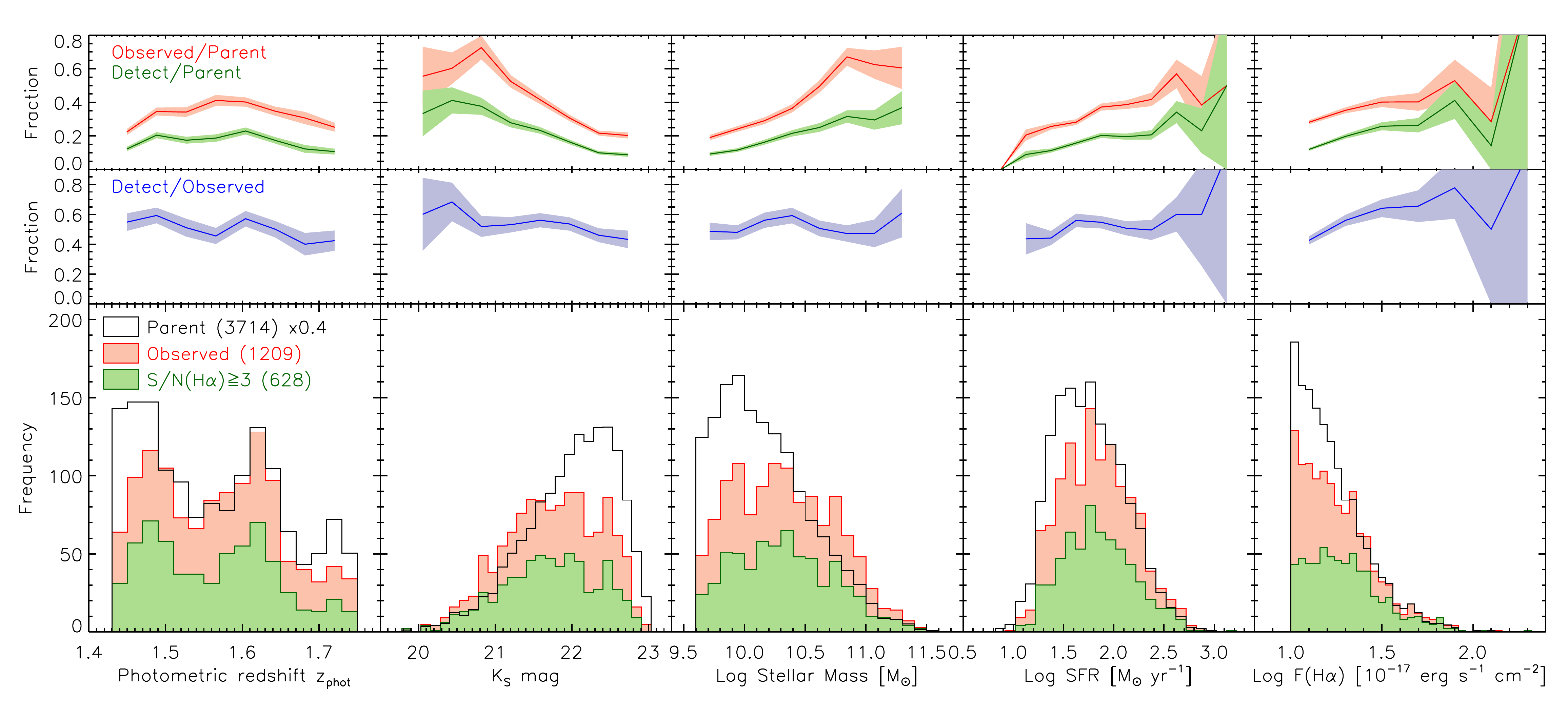}
   \caption{Same as Figure \ref{fig:hist_L16_par} but for the Selected-L16 sample ($N=3714$), and the corresponding ``$\cap$ FMOS HL'' ($N=1209$) and \Ha-detected ($\cap$ FMOS HL $+$ \Ha, $N=628$) samples (see Table \ref{tb:re_L16}).  The empty histograms are scaled by a factor of 0.4.}
   \label{fig:hist_L16_sel}
\end{figure*}

We investigate possible biases in the FMOS sample as functions of the properties of galaxies.  In Figure \ref{fig:hist_L16_par}, we show the distribution of the SED-based quantities ($z_\mathrm{phot}$, $K_\mathrm{S}$, $M_\ast$, SFR, and $F^\mathrm{pred}_\mathrm{H\alpha}$ from left to right) for galaxies matched in the Broad-L16 sample.  As shown in the top panels of Figure \ref{fig:hist_L16_par}, the sampling rates of both observed- and \Ha-detected sample depend on these quantities.  We note that the non-uniform sampling in terms of $z_\mathrm{phot}$ is trivial because we preferentially selected galaxies within a narrower range of $z_\mathrm{phot}$ ($ 1.46 \le z_\mathrm{phot} \le 1.72$), within which the sampling is nearly uniform.  In the middle panel in each column, we show the success rate, which is the fraction of the \Ha-detected objects relative to the observed objects at given $x$-axis value.  It is clear that, not only the sampling rates, but also the success rates depend on these galaxy properties, e.g., as shown by the trends with $K_\mathrm{S}$ and $F^\mathrm{pred}_\mathrm{H\alpha}$.

Next we show the Selected-L16 sample in the same manner in Figure \ref{fig:hist_L16_sel}.  At first glance, the distribution of the observed-/\Ha-detected FMOS objects is more similar to that of the parent sample.  Correspondingly, it is also clear that the sampling rate is now more uniform against any quantity of these than those of the Broad-L16 sample shown in Figure \ref{fig:hist_L16_par}.  However, the sampling rate still varies substantially as a function of some of these galactic properties.  In particular, the sampling rate increases rapidly around $\log M_\ast/M_\odot \approx 10^{10.5}$.  Given a tight correlation between $M_\ast$ and $K_\mathrm{S}$ magnitude, this trend with $M_\ast$ corresponds to the decrease in the sampling rate with increasing $K_\mathrm{S}$.  This is partially because the latter part of our observations were especially dedicated to increase the sampling rate of most massive galaxies.  However, the success rate shows no significant trend as a function of $M_\ast$ and $K_\mathrm{S}$-band magnitude.  As opposed to $M_\ast$ (or $K_\mathrm{S}$), not only the sampling rate, but also the success rate appear to increase as $F^\mathrm{pred}_\mathrm{H\alpha}$ increases (rightmost panels).  This is naturally expected because stronger lines are more easily detected.  From these results, we conclude that the spectroscopic sample (even after applying the criteria defined in this section) is biased towards massive galaxies and having higher $F^\mathrm{pred}_\mathrm{H\alpha}$.

\begin{figure}[htbp] 
   \centering
   \includegraphics[width=3.5in]{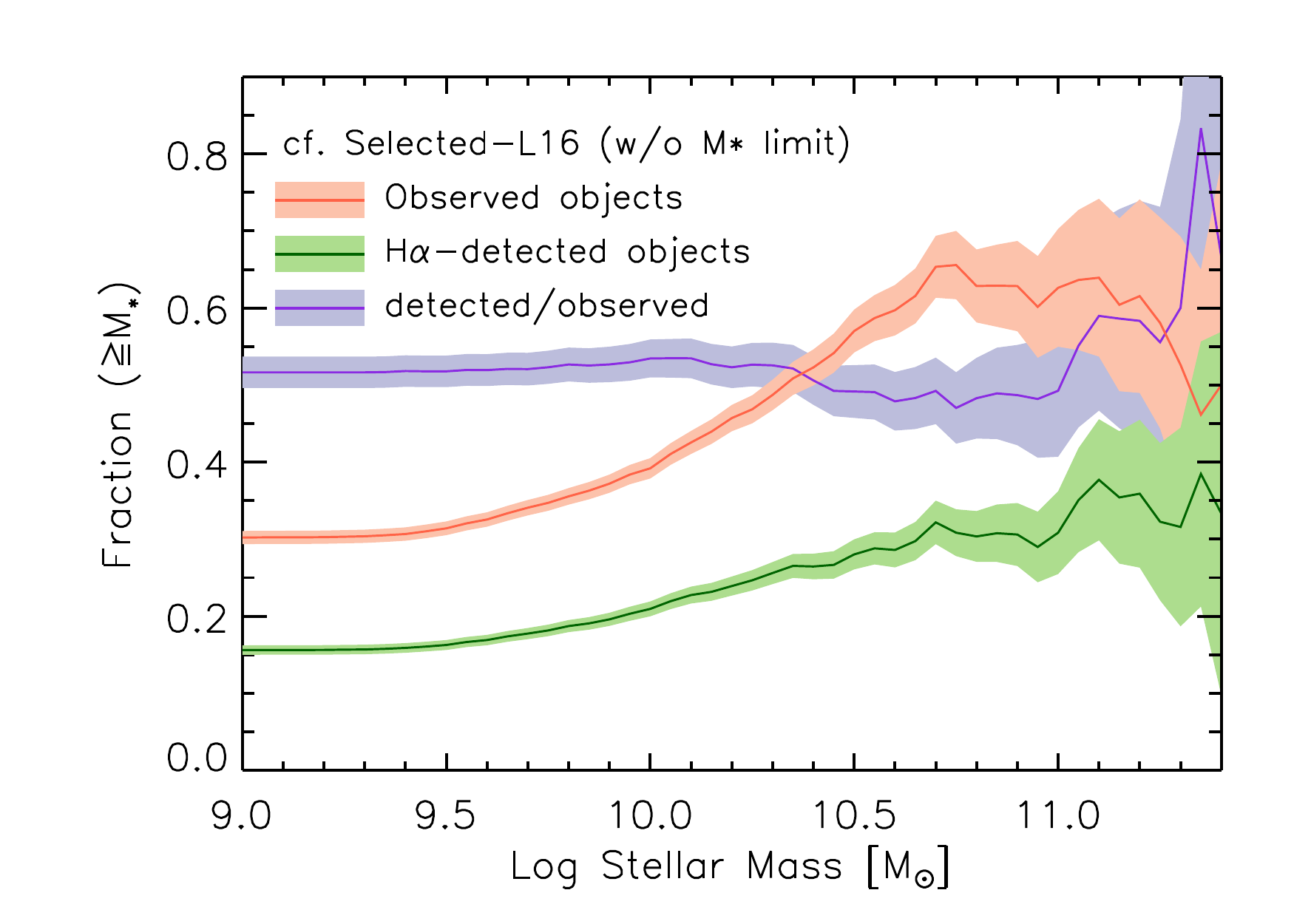}
   \includegraphics[width=3.5in]{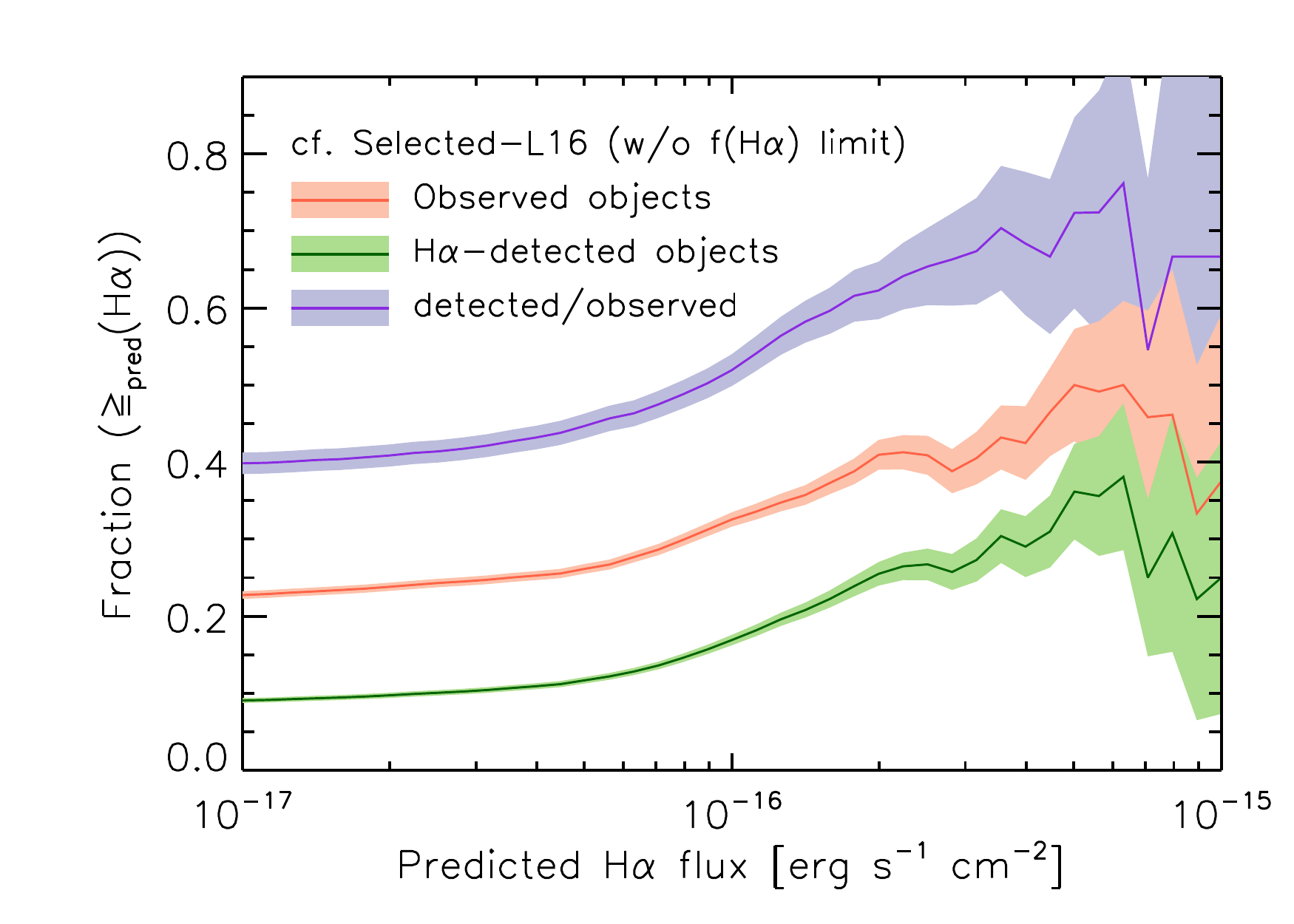}
   \caption{Cumulative sampling rate and success rate above a given $M_\ast$ (upper panel) and predicted \Ha\ flux (lower panel).   Red line indicates the the sampling fraction of galaxies within the Selected-L16 sample, but without the limit on the $x$-axis value in each panel.  Green line indicates the fractions of the \Ha-detected objects with respect to the Selected-L16 sample.  Purple line indicates the cumulative success rate, which is the fraction of \Ha-detected objects with respect to the observed galaxies above a given value.}
   \label{fig:rates}
\end{figure}

To quantify these trends, we show in Figure \ref{fig:rates} the cumulative sampling and success rates as a function of $M_\ast$ (upper panel) and predicted \Ha\ flux (lower panel).  The cumulative sampling rate is defined as the fraction of observed and/or \Ha-detected galaxies above a given $M_\ast$ or $F^\mathrm{pred}_\mathrm{H\alpha}$ with respect to the Selected-L16 sample, but without the limit on the quantity corresponding to the $x$-axis.  The cumulative success rate is defined in the same manner between the \Ha-detected and observed galaxy samples.  In the upper panel, it is shown that the cumulative sampling rate of the observed galaxies ($\cap$ FMOS-HL; red line) increases at $M_\ast \ge 10^{10}~M_\odot$, and reaches a level of $\approx~60$\% ($\approx~35$\%) at $M_\ast = 10^{10.7}~M_\odot$.  The cumulative sampling rate of the \Ha-detected subsample (green line) shows a similar trend, increasing monotonically with $M_\ast$ from 17\% at the lower $M_\ast$ limit to 35\% at $10^{11}~M_\odot$.  In contrast, the cumulative success rate (purple line) is nearly uniform across the entire $M_\ast$ range.  In the lower panel of Figure \ref{fig:rates}, the cumulative sampling rate for the \Ha-detected subsample (green line) increases from $\sim10\%$ at $F^\mathrm{pred}_\mathrm{H\alpha} < 10^{-16}~\mathrm{erg~s^{-1}~cm^{-2}}$ to 35\%.  The cumulative success rate (purple line) also increases slowly from 40\% to 70\% as the threshold $F^\mathrm{pred}_\mathrm{H\alpha}$ increases.  

% ----------------------------------------------------------------------------------------------------
\subsection{Comparison with the spectroscopic measurements \label{sec:re_L16_comp}}

\begin{figure}[htbp] 
   \centering
   \includegraphics[width=3.5in]{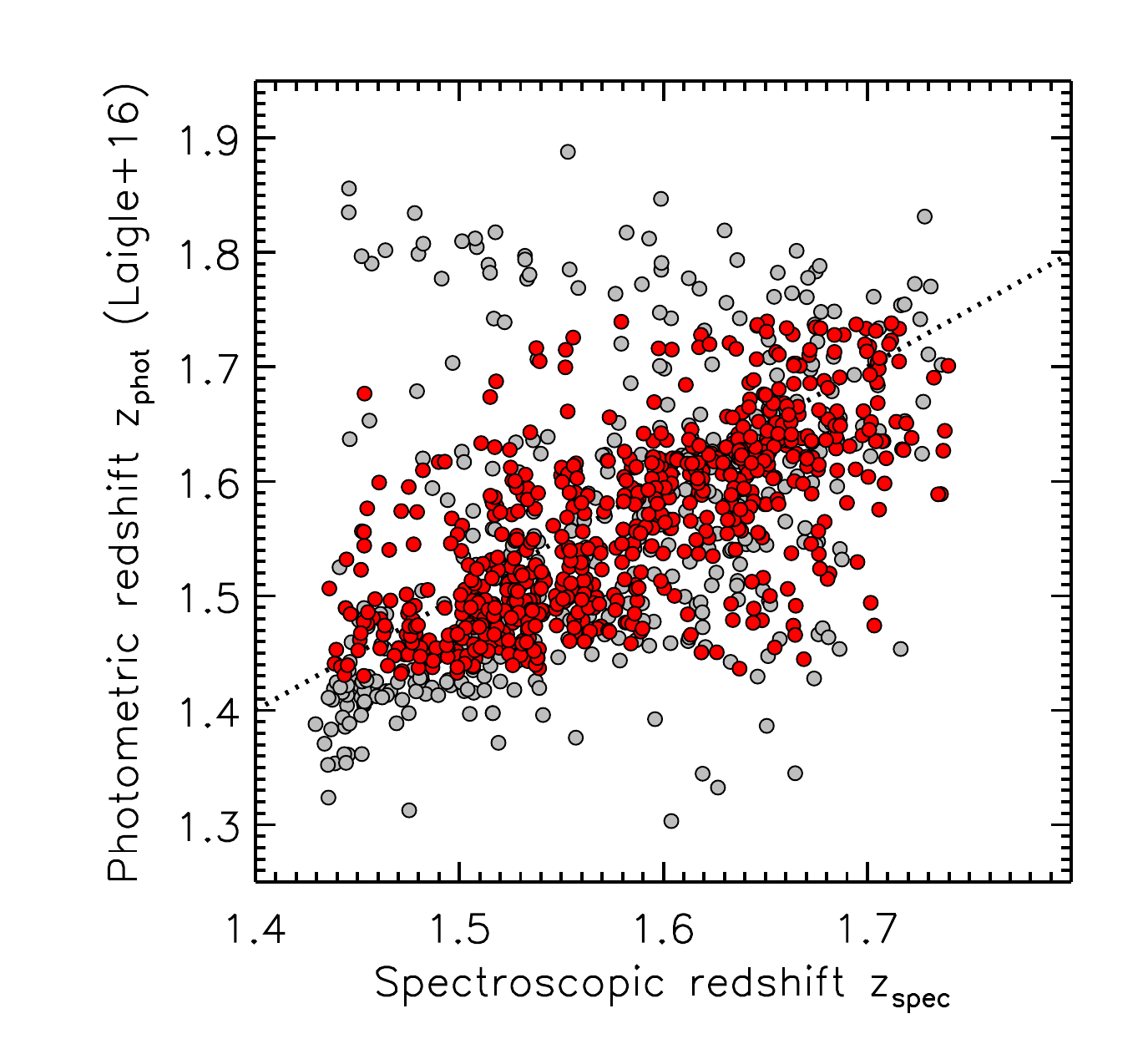} 
   \includegraphics[width=3.5in]{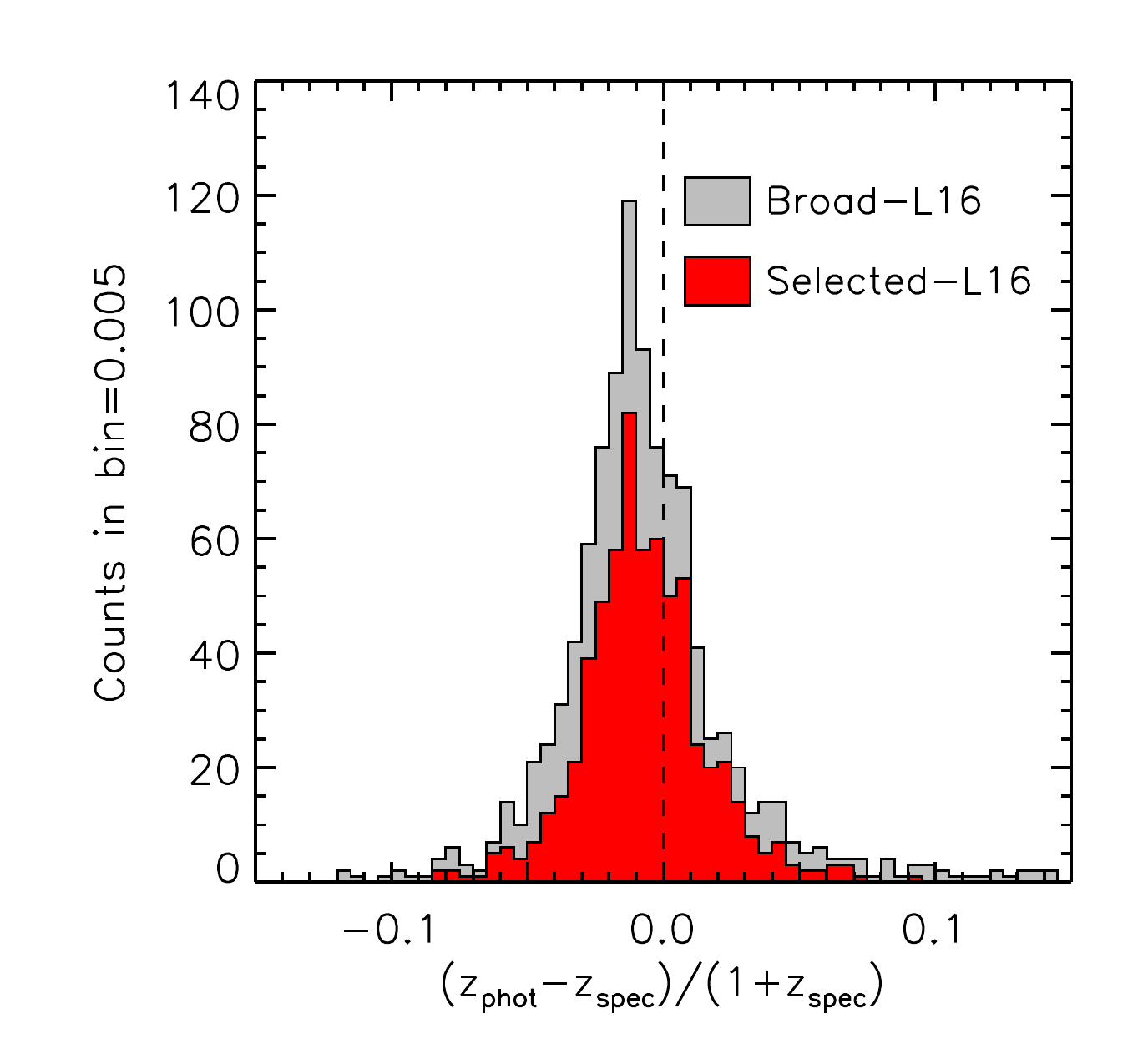}
   \caption{Upper panel: Comparison of spectroscopic and photometric redshifts.  Gray circles indicate the objects in the Broad-L16 sample while the red circles indicate the objects in the Selected-L16 sample.  Horizontal dotted and dashed lines indicate the imposed limits on the photometric redshifts: $1.3\le z_\mathrm{phot}\le 1.9$ for Broad-L16 and $1.43 \le z_\mathrm{phot} \le 1.74$ for Selected-L16.   Lower panel: the distribution of $(z_\mathrm{phot}-z_\mathrm{spec})/(1+z_\mathrm{spec})$ from the upper panel.  Gray and red histograms indicate the objects from the Broad-L16 and Selected-L16 samples, respectively.  }
   \label{fig:zs_vs_zp_L16}
\end{figure}

We compare our spectroscopic measurements with those based on the SED fits from COSMOS2015.  In Figure \ref{fig:zs_vs_zp_L16}, we compare the spectroscopic redshifts with the photometric redshifts.  Limiting those in the Selected-L16 sample (red circles and red histogram), we find that the median and the standard deviation of $(z_\mathrm{phot}-z_\mathrm{spec})/(1+z_\mathrm{spec})$ are $-0.0112$ ($-0.0086$) and 0.0264 (0.0237), respectively, after (before) taking into account the effect of limiting the range of photometric redshifts ($1.43 \le z_\mathrm{phot} \le 1.74$).  The level of the uncertainties in the photometric redshifts is very similar to that in the older version \citep{2013A&A...556A..55I}, as described in Section \ref{sec:redshifts_Primary-HL}.  Because there is only a little systematic offset in our $z_\mathrm{spec}$ estimates in comparison with the MOSDEF and 3D-HST surveys (Section \ref{sec:assessment}), the median offset between $z_\mathrm{spec}$ and $z_\mathrm{phot}$, which is significant compared to the scatter, should be regarded as the systematic uncertainty in the photometric redshifts.

\begin{figure}[htbp] 
   \centering
   \includegraphics[width=3.5in]{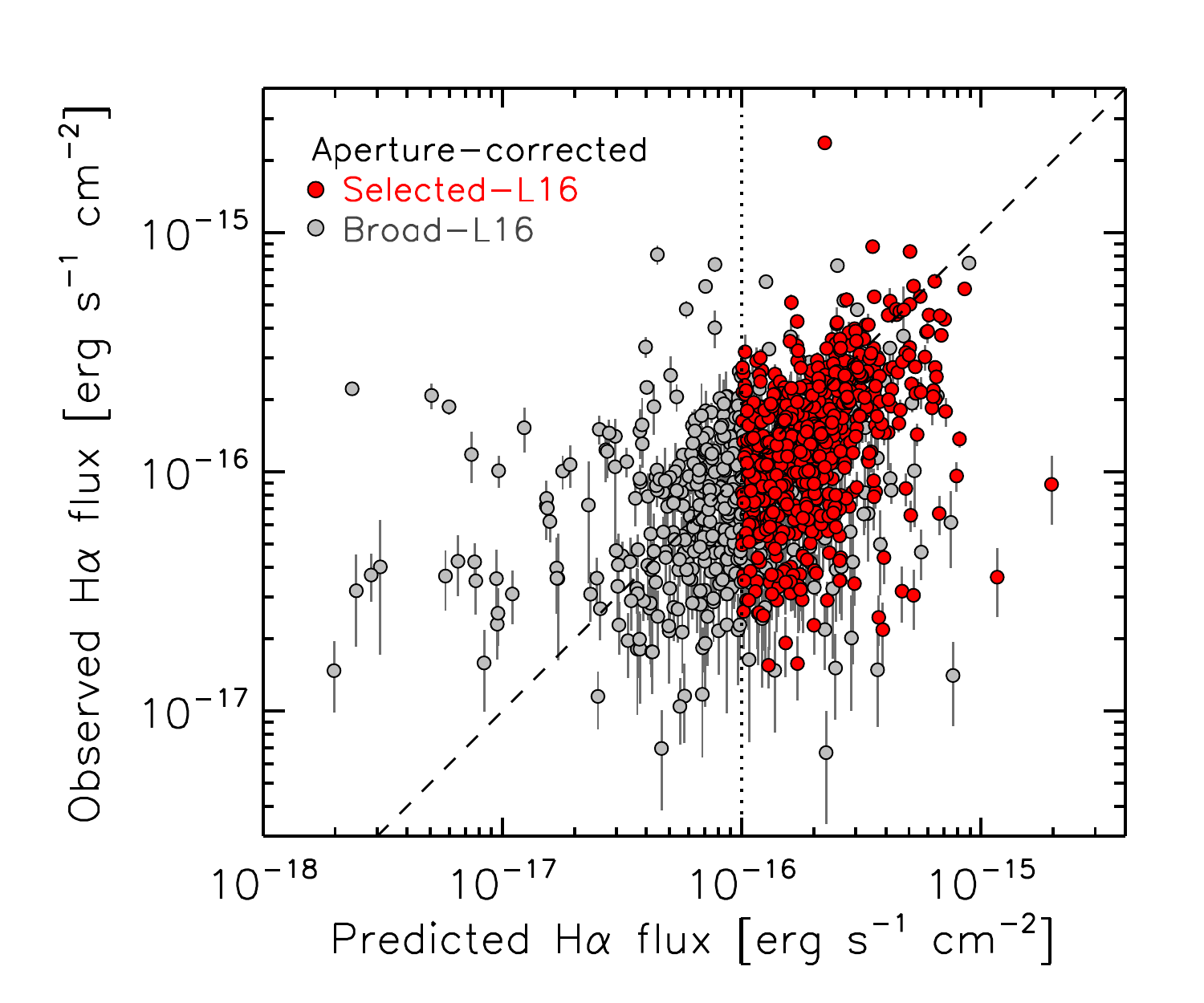} 
   \caption{Observed (aperture-corrected) vs. predicted \Ha\ flux.  Gray circles indicate objects in the Broad-L16 sample while red circles indicate objects in the Selected-L16 sample.  No error estimates are given for the predicted \Ha\ flux.  The vertical dotted line indicates the threshold $F^\mathrm{pred}_\mathrm{H\alpha} = 10^{-16}~\mathrm{erg~s^{-1}~cm^{-2}}$ for the Selected-L16 sample, and the dashed line indicates the one-to-one relation.   }
   \label{fig:fobsHa_vs_fpreHa_L16}
\end{figure}

We next compare in Figure \ref{fig:fobsHa_vs_fpreHa_L16} the observed \Ha\ fluxes to the predicted \Ha\ fluxes.  The observed fluxes are converted to the total fluxes by applying the aperture correction (see Section \ref{sec:apercorr}).  Note that the observed fluxes are not corrected for extinction, while the predicted fluxes include the reduction due to extinction.  It is shown that the observed fluxes are in broad agreement with the predicted values.  Limiting those in the Selected-L16 sample (red circles), we found a small systematic offset of $\log F^\mathrm{obs}_\mathrm{H\alpha} / F^\mathrm{pred}_\mathrm{H\alpha} = -0.16~\mathrm{dex}$ (median).  This offset may be attributed to the application of inaccurate dust extinction.  We revisit the dust extinction by using the new estimates of galaxy properties with spectroscopic redshifts in Section \ref{sec:dust}.

%-------------------------------------------------------------------------------------------------------
\section{Stellar mass and SFR estimation \label{sec:lephare}}
\subsection{SED-fitting with LePhare}

For FMOS galaxies with a spectroscopic redshift based on an emission-line detection, we re-derived stellar masses based on SED fitting using LePhare \citep{2002MNRAS.329..355A,2006A&A...457..841I}.  The stellar mass is defined as the total mass contained in stars at the considered age without the mass returned to the interstellar medium.  Our procedure follows the same method as in \citet{2015A&A...579A...2I} and \citet{2016ApJS..224...24L}, i.e., our estimation is consistent with the COSMOS2015 catalog.  The SED library contains synthetic spectra generated using the population synthesis model of \citet{2003MNRAS.344.1000B}, assuming a \citet{2003PASP..115..763C} IMF.  We considered 12 models combining the exponentially declining star formation history (SFH; $e^{-t/\tau}$ with $\tau/\mathrm{Gyr}=\left\{0.1:30\right\}$) and delayed SFH ($te^{-t/\tau}$ with $\tau=1$ and 3 Gyrs) with two metallicities (solar and half solar) applied.  We considered two attenuation laws, including the \citet{2000ApJ...533..682C} law and a curve $k(\lambda) = 3.1(\lambda/5500~\mathrm{\AA})^{-0.9}$ with $E(B-V)$ being allowed to take values as high as 0.7.  

For SED fitting, we used photometry from the COSMOS2015 catalog measured with 30 broad-, intermediate-, and narrow-band filters from {\it GALEX NUV} to {\it Spitzer}/IRAC ch2 (4.5~$\mu$m), as listed in Table 3 of \citet{2016ApJS..224...24L}.  Note that IRAC ch3 and ch4 were excluded since the photometry in these bands may affected by the PAH emissions, which are not modeled in our templates.  For CFHT, Subaru, and UltraVISTA photometry, we used measurements in $3^{\prime\prime}$-aperture fluxes and applied the offsets provided in the catalog to convert them to the total fluxes.

\begin{figure}[t] 
   \centering
   \includegraphics[width=3.5in]{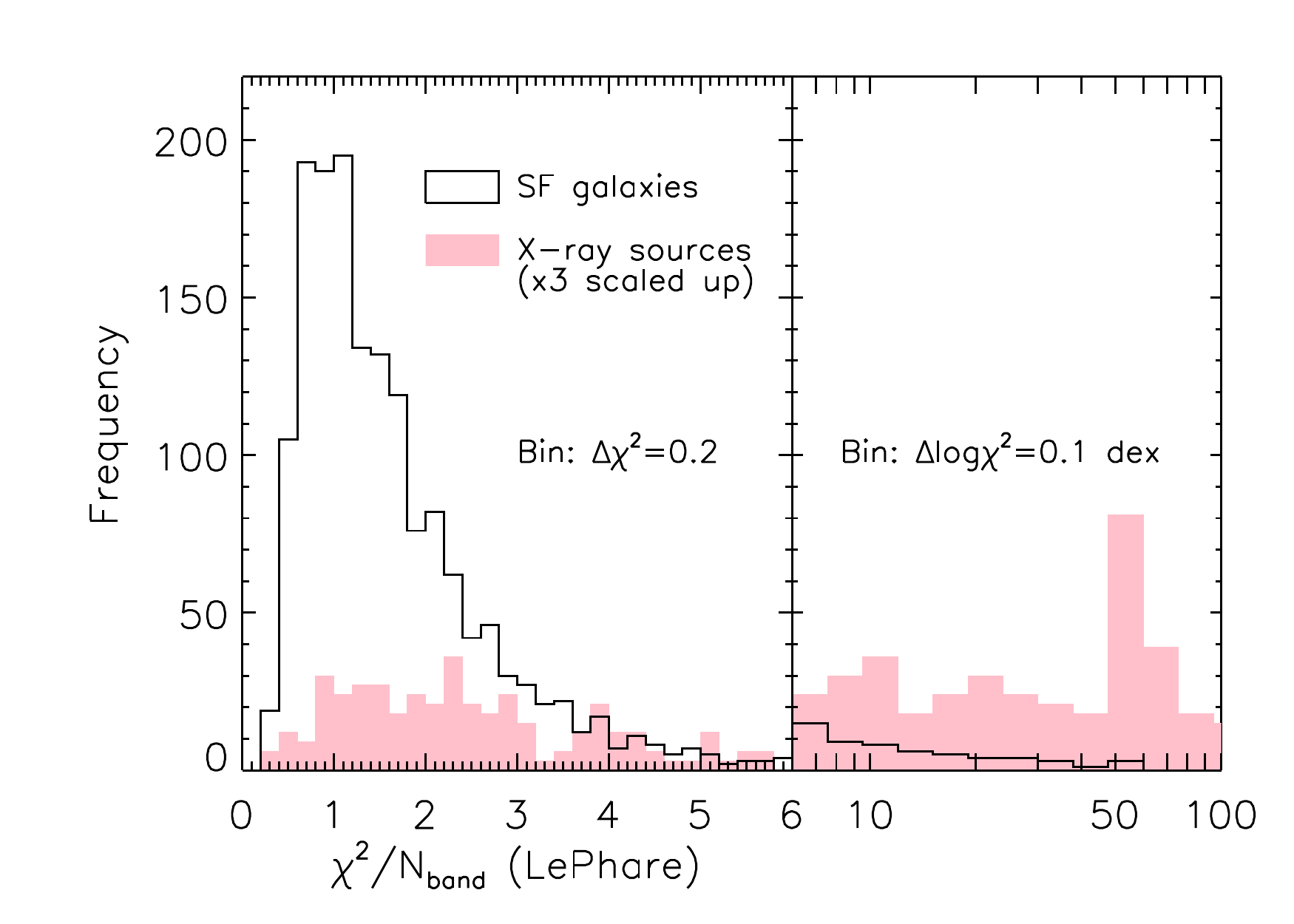} 
   \caption{Histograms of $\chi^2/N_\mathrm{band}$ values for the best-fit SEDs.  For non-Xray sources, the values are concentrated around $\chi^2/N_\mathrm{band}=1$ (empty histogram), indicating that the fitting is successful for the majority of the sample.  In contrast, for X-ray sources, the values are widely spread out to $\sim100$ (filled histogram, $\times3$ scaled up for display purpose), indicating a large fraction of those for which the fitting has failed (see text).}
   \label{fig:hist_chi2}
\end{figure}

In Figure \ref{fig:hist_chi2}, we show the histograms of the resulting $\chi^2/N_\mathrm{band}$ values (where $N_\mathrm{band}$ is the number of bandpasses that were used for fitting) for the best-fit SEDs, separately for non-X-ray and X-ray-detected sources (see Section \ref{sec:Chandra}).  It is shown that the $\chi^2/N_\mathrm{band}$ values are concentrated around $\chi^2/N_\mathrm{band}=1$ for non-X-ray star-forming galaxies, which indicates that the fitting has reasonably succeeded for the majority of the sample.  In contrast, the fitting may be unreasonable for many of X-ray sources, as indicated by their $\chi^2/N_\mathrm{band}$ distribution, which is widely spread out to $\chi^2/N_\mathrm{band} \sim100$.  The main reason for such lower goodness-of-fit is the additional emission from an AGN at the rest-frame UV and near-to-mid IR wavelengths.  The derivation of SED properties for X-ray sources, accounting for the AGN emission component, is postponed to a future companion paper (Kashino et al, in prep.).  Throughout the paper, we disregard the LePhare estimates for all X-ray-detected sources and those with $\chi^2/N_\mathrm{band}\ge6$.  

\subsection{SFRs from the UV luminosity\label{sec:SFRUV}}

We estimated the total SFR of our sample galaxies directly from the UV continuum luminosity in order to compare with those estimated from \Ha\ luminosity.  Dust extinction is accounted for based on the slope $\beta_\mathrm{UV}$ of the rest-frame UV continuum spectrum \citep[e.g.,][]{1999ApJ...521...64M}.  The UV slope $\beta_\mathrm{UV}$ is defined as $f_\mathrm{\lambda} \propto \lambda^{\beta_\mathrm{UV}}$.  We measured the rest-frame FUV (1600~\AA) flux density and $\beta_\mathrm{UV}$ by fitting a power-law function to the broad- and intermediate-band fluxes within $1200~\mathrm{\AA} \le \lambda_\mathrm{eff}/(1+z) \le 2600~\mathrm{\AA}$ where $\lambda_\mathrm{eff}$ is the effective wavelength of the corresponding filters.  The slope $\beta_\mathrm{UV}$ is converted to the FUV extinction, $A_\mathrm{1600}$, as well as to the reddening value, $E_\mathrm{star}(B-V)$, with the following relations from \citet{2000ApJ...533..682C}:
\begin{eqnarray}
A_{1600} = 4.85+2.31~\beta_\mathrm{UV}, \\
E_\mathrm{star} (B-V) = A_{1600} / k_{1600},
\end{eqnarray}
where $k_{1600}=10.0$.  We set the lower and upper limits to be $E_\mathrm{star} (B-V)=0$ and 0.8, respectively.  The extinction-corrected UV luminosity, $L_{1600}$, is then converted to SFR using a relation from \citet{2004ApJ...617..746D}:
\begin{equation}
\mathrm{SFR}_\mathrm{UV} (M_\odot~\mathrm{yr^{-1}}) = \frac{L_{1600} (\mathrm{erg~s^{-1}~Hz^{-1}})}{1.7 \times 8.85 \times 10^{27}}
\end{equation}
where a factor of $1/1.7$ is applied to convert from a \citet{1955ApJ...121..161S} IMF to a \citet{2003PASP..115..763C} IMF.  We disregard the measurements with poor constraints of either the UV luminosity ($<5\sigma$) or the UV slope $\sigma(\beta_\mathrm{UV})>0.5$ (only 6\% of the sample of \Ha-detected ($\ge3\sigma$) galaxies).  

\begin{figure}[htbp] 
   \centering
   \includegraphics[width=3.5in]{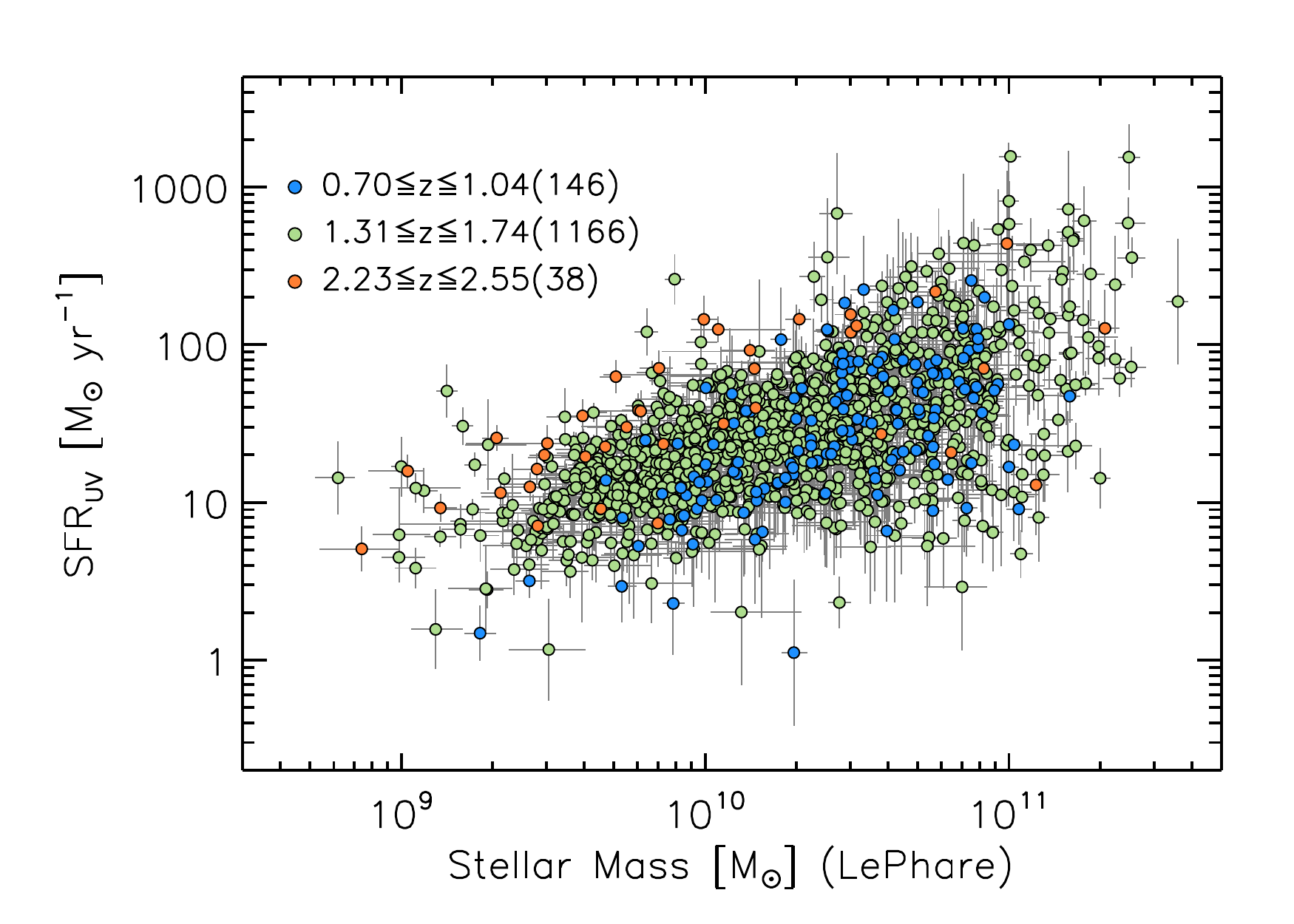} 
   \includegraphics[width=3.5in]{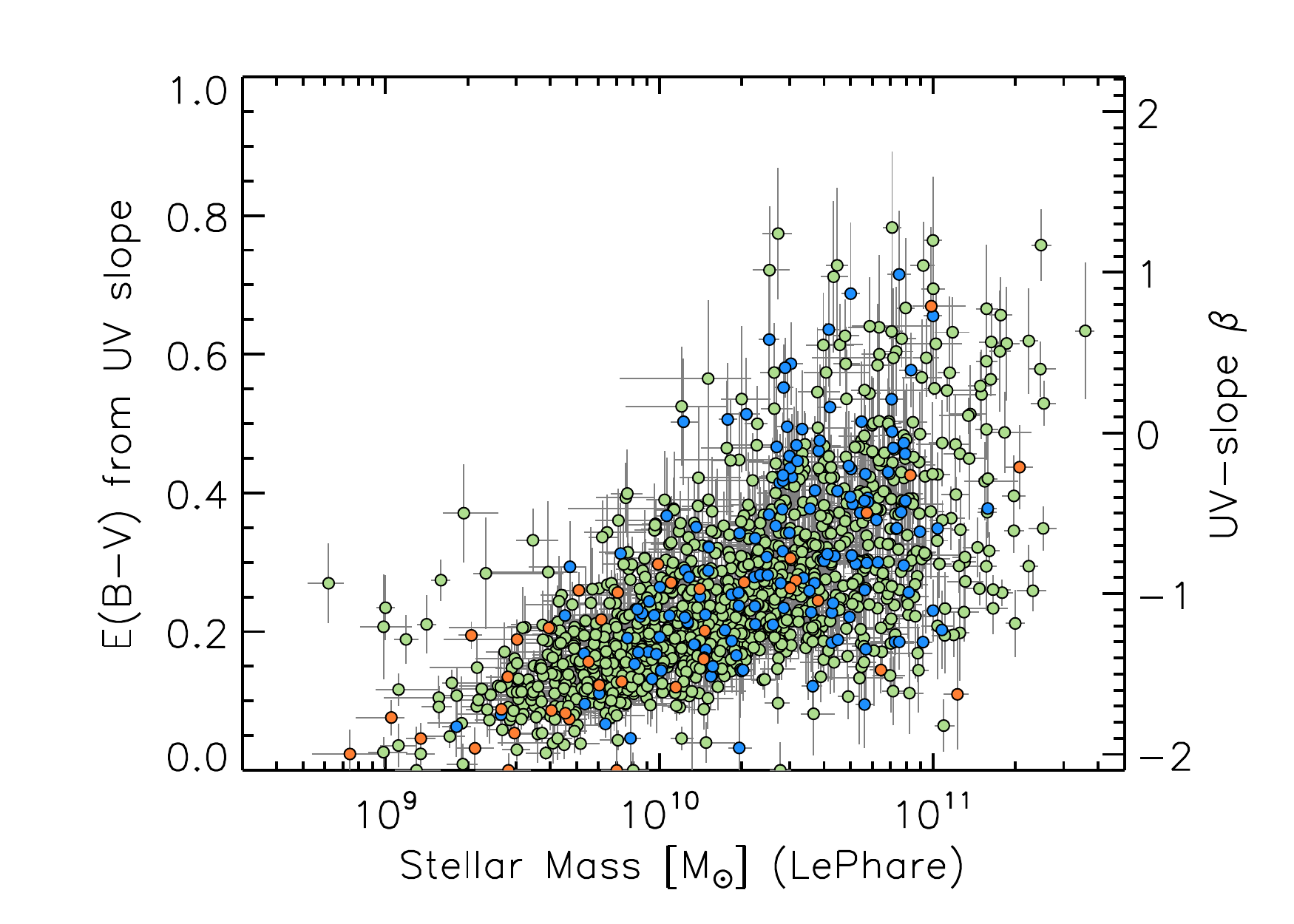}
      \caption{$M_\ast$ vs. SFR$_\mathrm{UV}$ (upper panel) and $E_\mathrm{star}(B-V)$ (lower panel) estimated from the rest-frame UV continuum.  FMOS galaxies are divided into three redshift intervals as labeled.  The number of objects shown here are indicated in parentheses.}
   \label{fig:LePhare_M_vs_UVSFR}
\end{figure}

In the top panel of Figure \ref{fig:LePhare_M_vs_UVSFR}, we show the distributions of the estimates of $M_\ast$ and UV-based SFR for the entire FMOS sample, removing those with a resultant $\chi/N_\mathrm{band}^2\ge6$ and X-ray objects.  Objects are shown in the figure, separated into three redshift ranges as labeled.  In the lower panel of Figure \ref{fig:LePhare_M_vs_UVSFR}, we show the reddening $E_\mathrm{star}(B-V)$, estimated from $\beta_\mathrm{UV}$, as a function of $M_\ast$ for the same objects shown in the upper panel.  It is clear that the average and the scatter in $E_\mathrm{star}(B-V)$ increase with increasing $M_\ast$.

\subsection{\Ha-based SFR and extinction correction\label{sec:dust}}

Next we compute the intrinsic SFR from the  observed \Ha\ flux, applying correction for aperture loss and dust extinction, through Equation (\ref{eq:predfHa}).  It is known that the extinction of the nebular emission is enhanced on average relative to the extinction toward the stellar component, which is expressed with a factor $f_\mathrm{neb}$ as $E_\mathrm{neb}(B-V)=E_\mathrm{star}(B-V)/f_\mathrm{neb}$.  In the local Universe, a factor $f_\mathrm{neb}=0.44$, derived by \citet{2000ApJ...533..682C}, has been widely applied, whereas observations at higher redshifts have measured larger values ($\sim0.5\textrm{--}1$; e.g., \citealt{2013ApJ...777L...8K,2014ApJ...788...86P}).  There remains large uncertainties in the constraints on the $f_\mathrm{neb}$ factor because these results may depend on the method used to determine the level of extinction and the extinction laws applied.  In the remainder of the paper, we adopt the \citet{1989ApJ...345..245C} ($R_V=3.1$) and the \citet{2000ApJ...533..682C} ($R_V=4.05$) extinction laws, respectively, for the nebular and stellar extinction, following the analysis in the original work of \citet{2000ApJ...533..682C}, where they used a similar law by \citet{1999PASP..111...63F} for the nebular emission.

We here determine the $f_\mathrm{neb}$ by comparing the SFRs estimated from the observed \Ha\ and UV luminosities, both not corrected for dust extinction as done in \citet{2013ApJ...777L...8K}.  For this investigation, we limit our sample to 702 galaxies having \Ha\ detection ($\ge 5\sigma$) and the estimates of $M_\ast$, SFR$_\mathrm{UV}$, and $E_\mathrm{star}(B-V)$ (see Section \ref{sec:lephare}).  We excluded all X-ray detected objects and possible AGNs flagged by the emission-line width of $\ge 1000~\mathrm{km~s^{-1}}$ and/or their emission line ratios of \OIII$\lambda$5007/\Hb\ and \NII$\lambda$6584/\Ha\ \citep[see][for details]{2017ApJ...835...88K}. 

Assuming that the appropriate dust correction equalizes the UV-based and \Ha-based SFRs, the dust-uncorrected ratio $\mathrm{SFR}_\mathrm{H\alpha}^\mathrm{uncorr}/\mathrm{SFR}_\mathrm{UV}^\mathrm{uncorr}$ is expressed as a function of $E_\mathrm{star}(B-V)$ with a parameter $f_\mathrm{neb}$ as follows:
\begin{equation}
\log \left(\frac{\mathrm{SFR}_\mathrm{H\alpha}^\mathrm{uncorr}}{\mathrm{SFR}_\mathrm{UV}^\mathrm{uncorr}}\right) = -0.4 E_\mathrm{star}(B-V) \left( \frac{k_\mathrm{H\alpha}}{f_\mathrm{neb}} - k_\mathrm{1600}\right), \label{eq:SFRHa/SFRUV}
\end{equation}
where $k_{1600}=10.0$ \citep{2000ApJ...533..682C} and $k_\mathrm{H\alpha}=2.54$ \citep{1989ApJ...345..245C}.  The observed \Ha\ flux is converted to SFR$_\mathrm{H\alpha}^\mathrm{uncorr}$ following Equation (\ref{eq:predfHa}) without the extinction term, and aperture correction is applied. The values of $E_\mathrm{star}(B-V)$ were estimated from the UV slope $\beta_\mathrm{UV}$ using the \citet{2000ApJ...533..682C} law (Section \ref{sec:SFRUV}).

\begin{figure}[t] 
   \centering
   \includegraphics[width=3.5in]{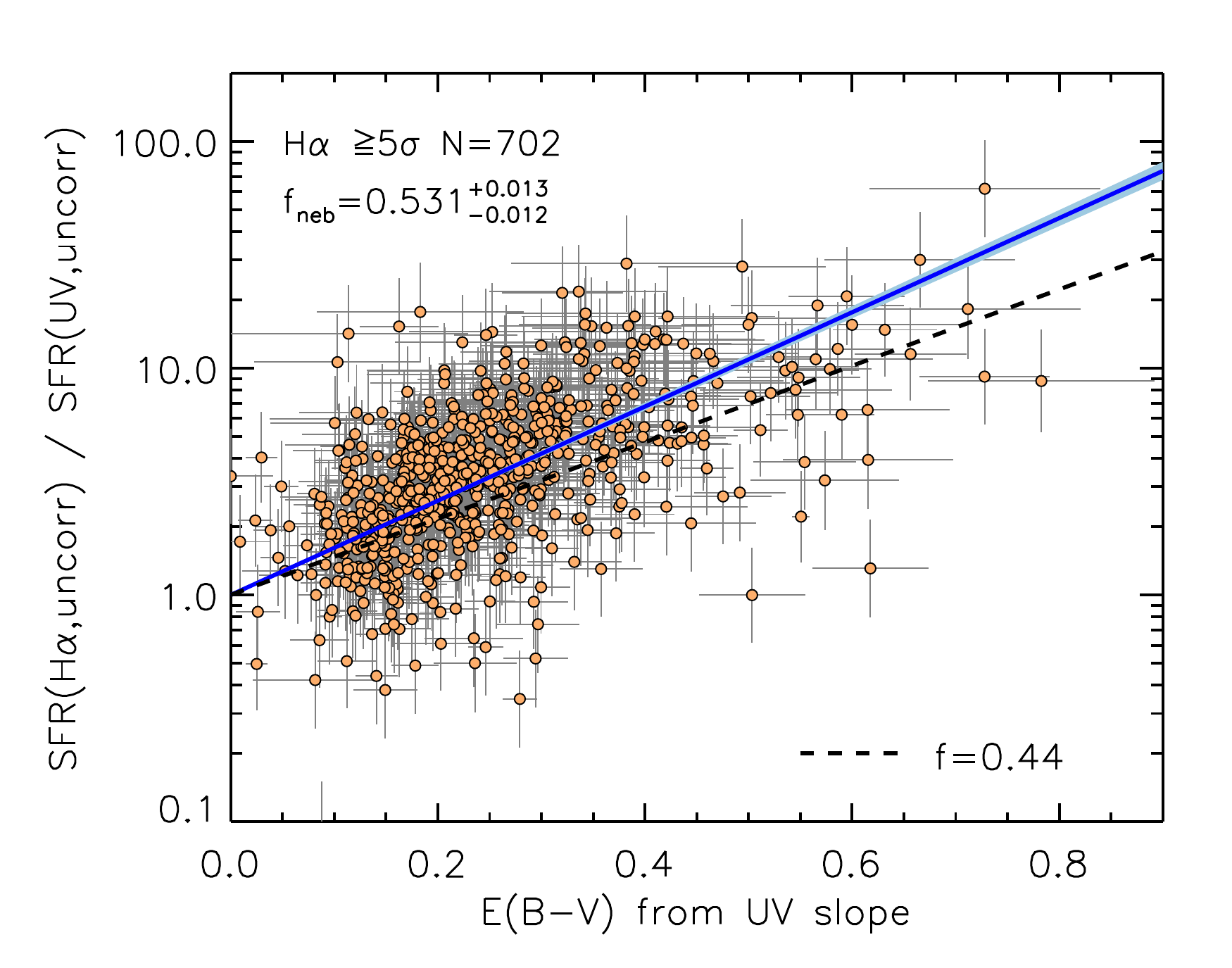}    
   \caption{The ratio of \Ha- and UV-based dust-uncorrected SFRs as a function of $E_\mathrm{star}(B-V)$.  The \Ha-based SFRs are corrected for aperture loss.  Solid line indicates the best-fit regression expressed by Equation (\ref{eq:SFRHa/SFRUV}) with $f^\mathrm{best}_\mathrm{neb}=0.53\pm0.01$.  The dashed line indicate the relation with $f_\mathrm{neb}=0.44$.}
   \label{fig:SFRHa/SFRUV}
\end{figure}

In Figure \ref{fig:SFRHa/SFRUV}, we plot the ratio $\mathrm{SFR}_\mathrm{H\alpha}^\mathrm{uncorr}/\mathrm{SFR}_\mathrm{UV}^\mathrm{uncorr}$ as a function of $E_\mathrm{star}(B-V)$.  There is a clear correlation between the SFR ratio and reddening.  It is apparent that more than half the data points falls above the line with the conventional value $f_\mathrm{neb}=0.44$ for local galaxies.  We found that the best-fit of Equation (\ref{eq:SFRHa/SFRUV}) yields a value of $f_\mathrm{neb}=0.53\pm0.01$, with a scatter of 0.15~dex after accounting for the individual errors.  Using the \citet{2000ApJ...533..682C} curve for both stellar and nebular reddening (i.e., replacing $k_\mathrm{H\alpha}$ with 3.33) results in a higher value ($f_\mathrm{neb}=0.69$), which is in agreement with \citet{2013ApJ...777L...8K}, where we did so.

\begin{figure}[t] 
   \centering
   \includegraphics[width=3.5in]{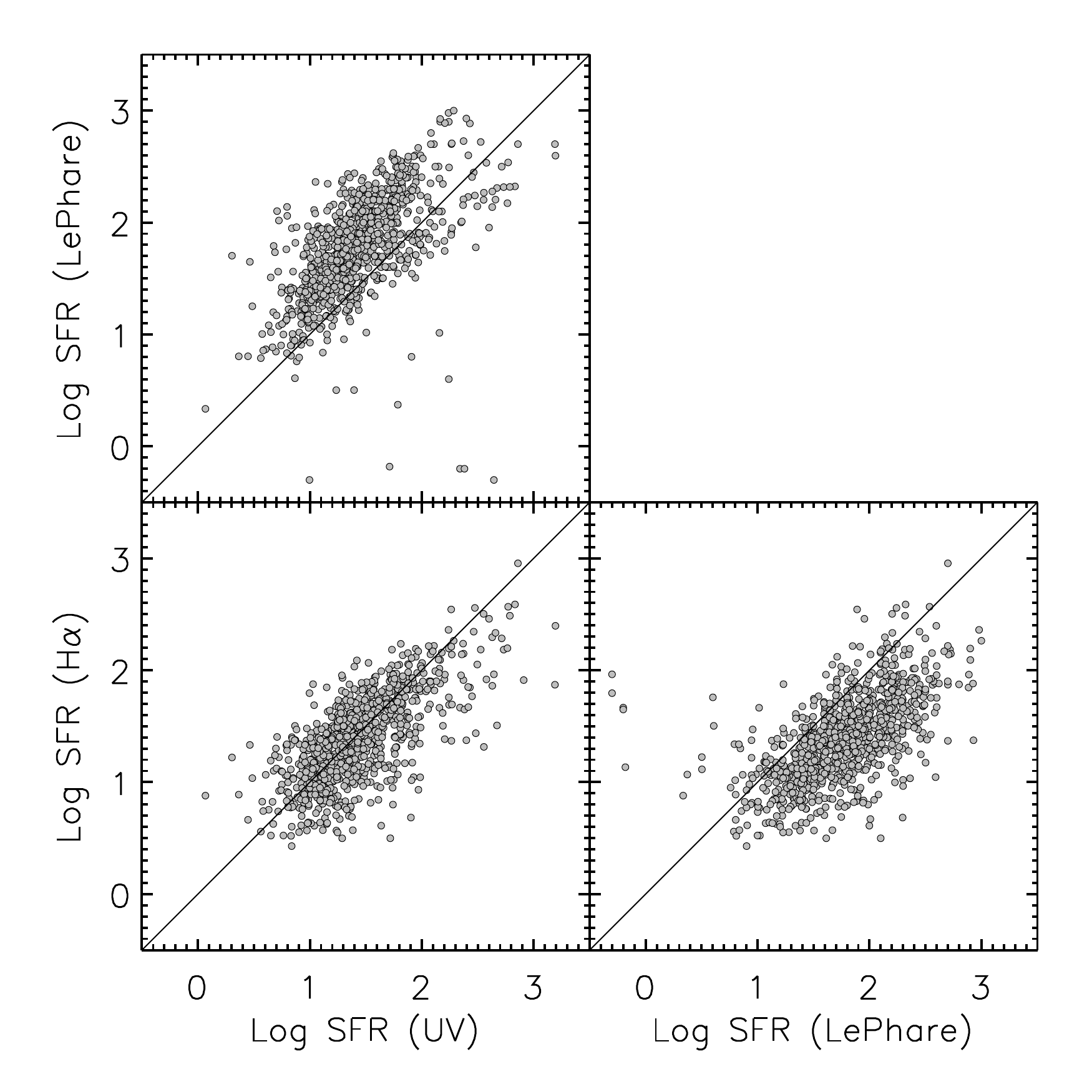}    
   \caption{Comparison of SFR estimates obtained from the rest-frame UV luminosity, \Ha\ luminosity and through the SED fitting with LePhare.  Diagonal lines indicate a one-to-one relation.}
   \label{fig:SFR_comp}
\end{figure}

Lastly, we compare in Figure \ref{fig:SFR_comp} different SFR indicators, the rest-frame UV, \Ha, and those obtained from the SED-fitting (i.e., the same procedure with LePhare as for the stellar mass).  Because the $f_\mathrm{neb}$ factor is adjusted so that the \Ha-based SFR matches the UV-based SFR on average, these two SFRs shows good agreement.  In contrast, we find systematic offset (median 0.4~dex) in comparison with the SFR through the SED fitting.  Similarly, some bias ($\sim 0.25$~dex) was found by \citet{2015A&A...579A...2I} from a comparison with SFR from IR$+$UV flux.

%=======================================

\section{The stellar mass--SFR relation, Revisited \label{sec:MS}}

Star-forming galaxies are known to form a tight sequence in the $M_\ast$--SFR plane, which is referred to as the main sequence of star-forming galaxies \citep{2007ApJ...660L..43N}.  It has been established that the normalization of the sequence increases with increasing redshift up to $z \sim 4$ or more \citep[e.g.,][]{2014ApJS..214...15S}.  However, the normalization and the slope vary from one study to another, depending on the sample selection and the methodology of the $M_\ast$ and SFR estimation \citep[e.g.,][]{2014MNRAS.443...19R}.  Moreover, there have also been studies on a possible bending feature seen at $\log (M_\ast/M_\odot) \approx 10\textrm{--}10.5$ \citep[e.g.,][]{2011ApJ...730...61K,2014ApJ...795..104W,2015ApJ...801...80L,2015A&A...575A..74S}.  

In \citet{2013ApJ...777L...8K}, we established the main sequence at $z \sim 1.6$ based on the \Ha-based SFRs using 271 s$BzK$-selected galaxies, a subset of the FMOS-COSMOS galaxy sample.  In this section, we re-define the main sequence based on the \Ha-based SFRs by using the complete FMOS-COSMOS sample, and discuss the bending feature and intrinsic scatter of the main sequence.

\subsection{Stellar mass vs. \Ha-based SFR}
\label{sec:M_vs_SFR_Ha}

\begin{figure*}[htbp] 
   \centering
   \includegraphics[width=5in]{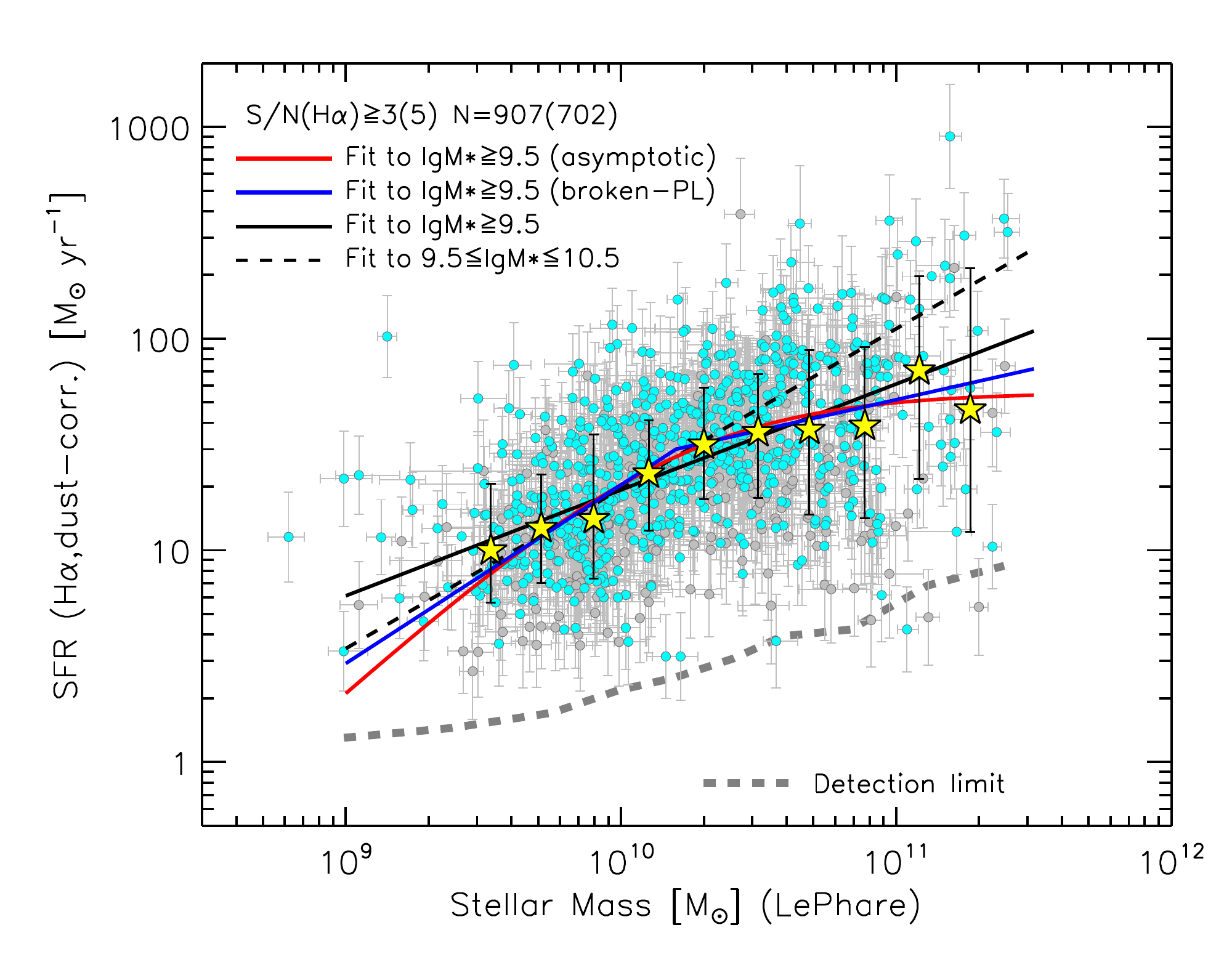} 
   \caption{Stellar mass vs. SFR from the observed \Ha\ fluxes, corrected for dust extinction and aperture effects.  907 objects with a detection of \Ha\ ($\ge3\sigma$) in the FMOS-$H$-long are shown, separately by their S/N (gray circles -- $3.0\le S/N < 5.0$; cyan circles -- $S/N\ge5.0$ ($N=702$); X-ray objects are excluded).  Yellow star symbols indicate the median \Ha-based SFRs in $M_\ast$ bins from $10^{9.4}~M_\odot$ to $10^{11.4}~M_\odot$ with a constant interval of 0.2~dex, with the central 68 percentiles indicated by the vertical error bars.  Black solid and dashed lines indicate a linear regression of our data limited to $M_\ast\ge10^{9.5}~M_\odot$ and t$10^{9.5}\le M_\ast/M_\odot \le 10^{10.5}$, respectively.  Blue and red solid lines indicates the best fits with a broken power-law (Equation \ref{eq:MxSFR_BPL}) and an asymptotic function (Equation \ref{eq:MxSFR_asymptotic}).  Thick gray dashed line indicates the typical detection limit, accounting for the dust extinction and aperture correction.}
   \label{fig:M_vs_SFR_Ha}
\end{figure*}

\begin{deluxetable*}{lccccc}
\tabletypesize{\scriptsize}
\tablecaption{Best-fit parameters for the $M_\ast$--SFR relation \label{tb:MxSFR}}
\tablehead{\colhead{Parameters}&
		 \colhead{$N$}&
		 \multicolumn{3}{c}{Parameters}&
		 \colhead{$\chi^2$ ($\chi^2/\mathrm{dof}$) \tablenotemark{a}}}
\startdata
\multicolumn{2}{l}{Power-law (Eq.~\ref{eq:MxSFR_powerlaw})} & $\alpha$ & $\beta$ & & \\
$M_\ast / M_\odot \ge 10^{9.5}$ & $876$ & $1.285\pm0.008$ & $0.500\pm0.017$ & & 2261 (2.59) \\
$10^{9.5}\le M_\ast / M_\odot \le 10^{10.5}$ & $609$ & $1.290\pm0.009$ & $0.755\pm0.035$ && 1187 (1.95) \\
\hline
\multicolumn{2}{l}{Broken power-law (Eq.~\ref{eq:MxSFR_BPL})} & $a_\mathrm{low}$ & $a_\mathrm{high}$ & $b$ & \\
$M_\ast / M_\odot \ge 10^{9.5}$ & $876$ & $0.844 \pm 0.055$ & $0.292 \pm 0.037$ & $1.476\pm 0.016$ & 2200 (2.52)  \\
\hline
\multicolumn{2}{l}{Asymptotic function (Eq.~\ref{eq:MxSFR_asymptotic})} & $\mathcal{S}_0$ & $\log \mathcal{M}_0$ & $\gamma$ & \\
$M_\ast / M_\odot \ge 10^{9.5}$ & $876$ & $1.74 \pm 0.033$ & $10.205 \pm 0.068$ & $ 1.17\pm0.10$ & 2206 (2.53)  \\
\enddata
\tablenotetext{a}{Chi-square statistics are computed including both errors on $\log \mathrm{SFR}$ and $\log M_\ast$.  The degree of freedom (dof) is $N - n_p$, where $n_p$ is the number of parameters.}
\end{deluxetable*}

In Figure \ref{fig:M_vs_SFR_Ha}, we show 907 galaxies with an \Ha\ detection ($\ge3\sigma$) in the $H$-long spectral window.  Of them, 702 galaxies have an \Ha\ detection at $\ge 5\sigma$ (cyan circles).  Observed \Ha\ fluxes\ are corrected for dust extinction by using $A_\mathrm{H\alpha} = 2.54 E_\mathrm{star} (B-V) / 0.53$ (see Section \ref{sec:dust}).  Vertical error bars include in quadrature the individual formal errors on the flux measurements (i.e., errors from line fitting) and a common uncertainty of 0.17~dex for aperture correction (see Section \ref{sec:apercorr}), as well as individual measurement errors on $E_\mathrm{star}(B-V)$, while not including any systematic uncertainty in the extinction law.  The average detection limit is estimated by assuming the $3\sigma$ detection limit of \Ha\ flux of $6 \times 10^{-18}~\mathrm{erg~s^{-1}~cm^{-2}}$ (see Figure \ref{fig:flux_SN_Ha}) and taking into account the $M_\ast$-dependent aperture correction and dust extinction.  The detection limit increases with $M_\ast$, mainly because the level of extinction increases on average with $M_\ast$ (see the lower panel of Figure \ref{fig:LePhare_M_vs_UVSFR}).  The observed data points are well above this line across the whole $M_\ast$ range, indicating that the observed distribution of SFR at fixed $M_\ast$ is less biased by the detection limit.

The correlation between $M_\ast$ and SFR is evident.  The Spearman's rank correlation coefficient is $\rho=0.52$ for all objects with $S/N(\textrm{\Ha})\ge3.0$ shown here.  To illustrate the behavior of the observed sequence more clearly, we separated the data points into bins of $M_\ast$ from $10^{9.4}~M_\odot$ to $10^{11.4}~M_\odot$ with a constant interval of 0.2~dex.  We indicate in Figure \ref{fig:M_vs_SFR_Ha} the median values and the central 68 percentiles in each bin.  These median points indicate possible bending of the main sequence, as reported by several authors \citep[e.g.,][]{2014ApJ...795..104W,2015ApJ...801...80L,2015A&A...575A..74S}.  

We parametrize the observed $M_\ast$--SFR relation.  In doing so, we excluded the objects below $10^{9.5}~M_\odot$, and used the individual points taking into account errors on both $\log M_\ast$ and $\log \mathrm{SFR}$.  We first employ a power-law function to fit the data as follows:
\begin{equation}
\log \mathrm{SFR}/(M_\odot~\mathrm{yr^{-1}}) = \alpha + \beta \log \left[ \frac{M_\ast}{10^{10}M_\odot} \right].
\label{eq:MxSFR_powerlaw}
\end{equation}
We fit to two subsamples: one contains all 876 galaxies above $M_\ast \ge 10^{9.5}~M_\ast$, and the other is limited to 609 objects between $10^{9.5} \le M_\ast/M_\odot  \le 10^{10.5}$ to avoid the effect of the possible bending.  We summarize the results in Table \ref{tb:MxSFR}, and indicate the best-fit relations in Figure \ref{fig:M_vs_SFR_Ha}.  The best-fit relation for the former ($M_\ast \ge 10^{9.5}~M_\ast$) has the slope of $\beta=0.500\pm0.017$, which is shallower that the slope of $\beta=0.755\pm0.035$ for the latter subsample limited to $10^{9.6} \le M_\ast/M_\odot  \le 10^{10.5}$.  Hence it is obvious that the shallower slope is caused by the massive population.  

We next account for the bending feature of the $M_\ast$--SFR relation at $M_\ast \sim 10^{10.5}~M_\odot$.  We employ two functional forms, a broken power-law and an asymptotic function, proposed by \citet{2014ApJ...795..104W} and \citet{2015ApJ...801...80L}, respectively.  The broken power-law is parametrized as
\begin{equation}
\log \mathrm{SFR}/(M_\odot~\mathrm{yr^{-1}}) = a \left( \log M_\ast/M_\odot - 10.2\right) + b
\label{eq:MxSFR_BPL}
\end{equation}
where the value of $a$ is different above ($a_\mathrm{high}$) and below ($a_\mathrm{low}$) the characteristic mass of $\log M_\ast/M_\odot =10.2$.   The characteristic mass is fixed following the original paper \citep{2014ApJ...795..104W}, though the best-fit value is $\log M_\ast/M_\odot = 10.31\pm 0.04$ if we allow it to vary.  The asymptotic function is defined as 
\begin{equation}
\log \mathrm{SFR}/(M_\odot~\mathrm{yr^{-1}}) = \mathcal{S}_0 + \log \left[ 1+ \left( \frac{M_\ast}{\mathcal{M}_0} \right)^{-\gamma} \right],
\label{eq:MxSFR_asymptotic}
\end{equation}
where $\mathcal{S}_0$ is the asymptotic value of the $\log \mathrm{SFR}$ at high $M_\ast$, $\mathcal{M}_0$ is the characteristic mass for turnover, and $\gamma$ is a low-mass slope.

Table \ref{tb:MxSFR} gives the results of fits to the individual objects with $M_\ast \ge 10^{9.5}~M_\ast$.  With the asymptotic relation, the characteristic mass for turnover is constrained to be $\log \mathcal{M}_0/M_\odot = 10.205\pm0.068$, which is fully consistent to the fixed characteristic mass ($\log M_\ast/M_\odot =10.2$) of the broken power-law fit, as well as the result of \citet{2015ApJ...801...80L}, as discussed below.  We show in Figure \ref{fig:M_vs_SFR_Ha} the best fits with the broken power-law and asymptotic function.  The two functional forms yield almost identical $M_\ast$--SFR relations across the $M_\ast$ range probed, which both well fit the median SFRs.  

The tight constraint on the turnover characteristic mass $\mathcal{M}_0$, as well as the significant difference detected between $a_\mathrm{low}$ and $a_\mathrm{high}$, indicate the presence of bending of the sequence at $M_\ast\approx10^{10.2}$.  This is also supported by the fact that, compared with the simple power-low, the resultant $\chi^2$ is reduced by $\Delta \chi^2 \approx 60$ by invoking the bending feature (Table \ref{tb:MxSFR}).  Meanwhile, there is no significant difference between the fits with the broken power-law and the asymptotic function.

\begin{figure}[tbp] 
   \centering
   \includegraphics[width=3.5in]{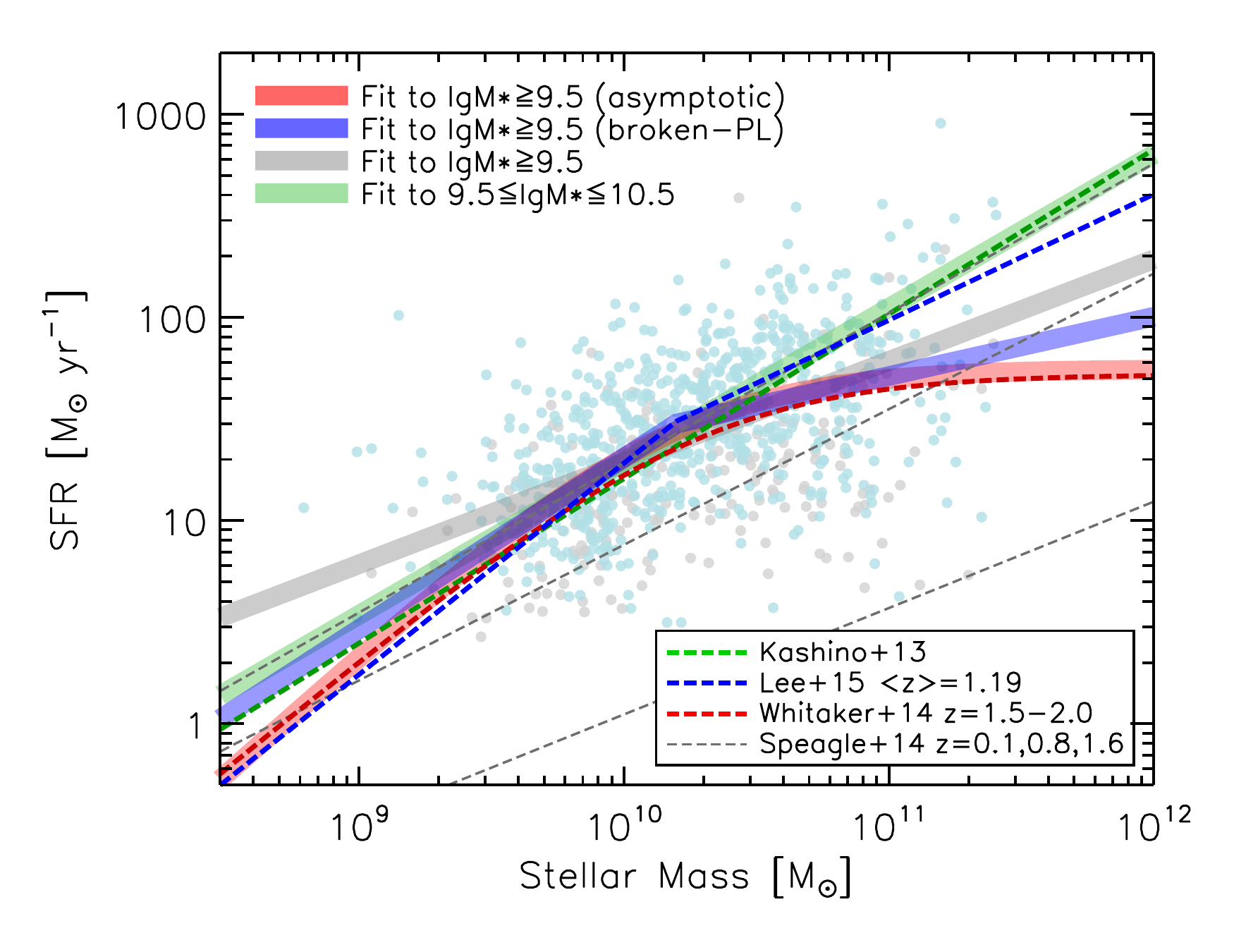} 
   \caption{The best-fit $M_\ast$ vs. SFR relations for the FMOS objects, in comparison to literature measurements.  Pale color circles indicate FMOS objects (same as in Figure \ref{fig:M_vs_SFR_Ha}).  Thick solid curves indicate the best-fit relations to the FMOS data: power-law fits to objects with $\log M_\ast/M_\odot \ge 9.6$ (gray) and $9.6 \le \log M_\ast/M_\odot \le 10.6$ (green), asymptotic function fit (red) and broken power-law fit (blue).  Thin colour dashed lines indicate literature measurements: fit to a subset of FMOS sample \citep[][green]{2013ApJ...777L...8K}, broken power-law fit at $1.5\le z\le2.0$ from \citet[][blue]{2014ApJ...795..104W}, asymptotic function fit at median $\left<z\right>=1.19$ from \citet[][blue]{2015ApJ...801...80L}, and three dotted lines indicate an empirically parametrized relation at $z=0.1,0.8,1.6$ (from top to bottom) derived by \citet{2014ApJS..214...15S}.}
   \label{fig:M_vs_SFR_Ha_relations}
\end{figure}

In Figure \ref{fig:M_vs_SFR_Ha_relations}, we compare our results with the $M_\ast$--SFR relations from the literature.  The power-law fit to the limited $M_\ast$ range ($9.5 \le \log M_\ast/M_\odot \le 10.5$) is fully consistent with our previous result \citep{2013ApJ...777L...8K}, and with the parametrization from a compilation across a wide redshift range derived by \citet{2014ApJS..214...15S}.  On the other hand, fit to the entire $M_\ast$ range ($\log M_\ast/M_\odot \ge 9.5$) yields a shallower slope.  The difference with \citet{2013ApJ...777L...8K} may be attributed to the increased weight of massive population and their different sample selection.  The broken power-law fit to the FMOS data yields a shallower high-mass slope ($a_\mathrm{high}=0.29$) than the result from \citet{2014ApJ...795..104W} ($a_\mathrm{high}=0.62$ for a sample at $1.5<z<2.0$).  In contrast, the best-fit with the asymptotic function (Equation \ref{eq:MxSFR_asymptotic}) is in good agreement with the result with the same parametrization by \citet{2015ApJ...801...80L} at $\left<z\right>=1.19$, with a similar characteristic mass $\log \mathcal{M}_0(=10.31)$ for turnover.  \citet{2015ApJ...801...80L} also found a high-mass power-law slope of 0.27 for $M_\ast>10^{10}~M_\odot$, which is rather similar to our $a_\mathrm{high} (=0.29)$.  The  steeper high-mass slope found by \citet{2014ApJ...795..104W} may be attributed, at least partially, to the fact that the authors derived total SFR using IR luminosity estimated from {\it Spitzer}/MIPS 24~$\mu m$ flux with a luminosity-independent conversion.  \citet{2015ApJ...801...80L} showed that total SFRs estimated in this way are overestimated at $\log \mathrm{SFR}/(M_\odot~\mathrm{yr^{-1}})\gtrsim 2$.  This effect is thus more important at high masses, and would artificially make the high-mass slope steeper.

\subsection{Scatter of the $M_\ast$--SFR relation\label{sec:scatter_MS}}

\begin{figure*}[tbp] 
   \centering
   \includegraphics[width=6.5in]{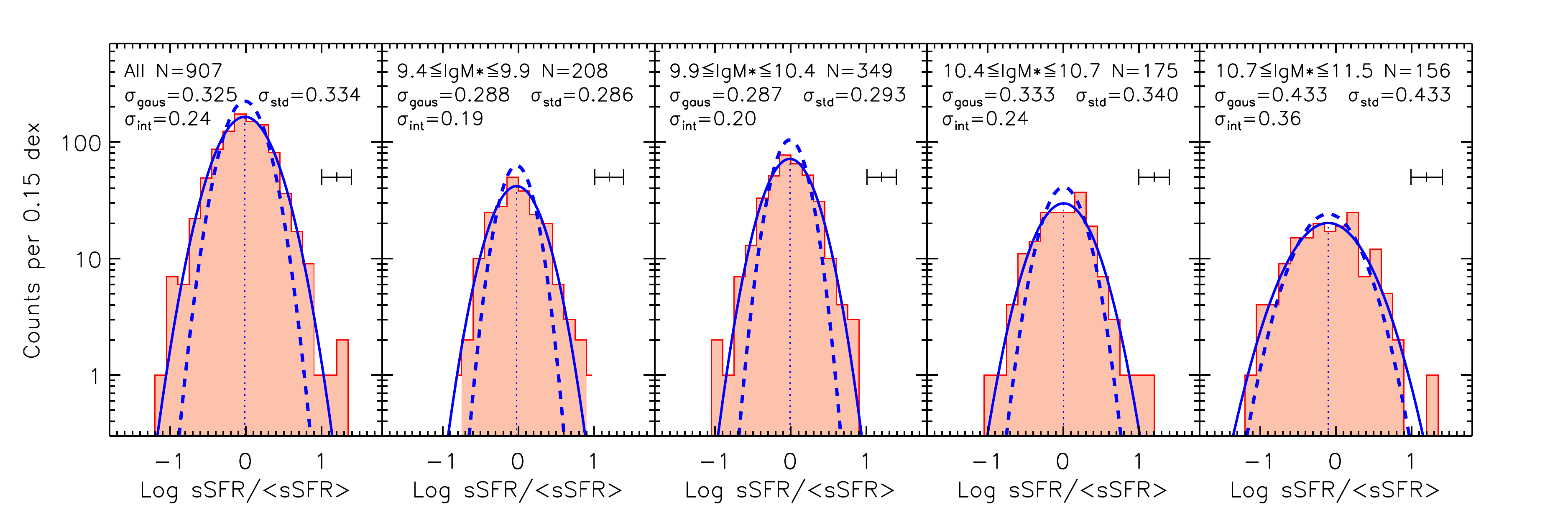} 
   \caption{Distribution of $\log \mathrm{sSFR}/\left<\mathrm{sSFR}\right>$ in bins of stellar mass: the entire $M_\ast$ range, $\log M_\ast/M_\odot = \left[ 9.4: 9.9\right]$, $\left[9.9:10.4\right]$, $\left[10.4:10.7\right]$, and $\left[10.7:11.5\right]$ from left to right.  The average $\left< \mathrm{sSFR} \right>$ is taken from the best fit with the asymptotic function (Equation \ref{eq:MxSFR_asymptotic}) at each $M_\ast$.  Blue solid lines indicate the best-fit functions assuming a log-normal profile with mean values indicated by vertical dotted lines.  Blue dashed lines indicate log-normal functions with the estimated intrinsic scatter ($\sigma_\mathrm{int}$).  Median error of SFR$_\mathrm{H\alpha}$ in each bin is indicated by a horizontal error bar in each panel.}
   \label{fig:dist_delta_sSFR}
\end{figure*}

In Figure \ref{fig:M_vs_SFR_Ha}, the observed scatter in SFR$_\mathrm{H\alpha}$ appears to increase with $M_\ast$.  We estimate the intrinsic scatters of the $M_\ast$--SFR relation as a function of $M_\ast$.  For this purpose, we define the offset from the best-fit $M_\ast$--SFR relation at fixed $M_\ast$ as $\log \mathrm{sSFR}/\left< \mathrm{sSFR}\right>$, and divide the sample into four bins of $M_\ast$: $\log M_\ast/M_\odot = $ [9.4:9.9], [9.9:10.4], [10.4:10.7], and [10.7:11.5].  We here use the best fit with the asymptotic function (Equation \ref{eq:MxSFR_asymptotic}), but the use of another fit (i.e., simple or broken power-law fit for $M_\ast\ge10^{9.5}~M_\odot$) does not change the conclusions.

Figure \ref{fig:dist_delta_sSFR} shows the distributions of $\log \mathrm{sSFR}/\left< \mathrm{sSFR}\right>$ for the entire sample, and in the four $M_\ast$ bins.  These distributions are well fit with a log-normal profile (solid blue lines).  The standard deviation of these log-normal profiles ($\sigma_\mathrm{gaus}$) and the values directly computed from the sample after 3-$\sigma$ clipping ($\sigma_\mathrm{std}$) are indicated in each panel, which agree each other.

The intrinsic scatter is then estimated from $\sigma_\mathrm{std}$ by accounting for the individual uncertainties on $\log M_\ast$ and $\log \mathrm{SFR}$.  The individual errors were obtained by summing in quadrature the statistical uncertainties on the individual \Ha-based SFR (the formal errors), $\log M_\ast$, and a common 0.17~dex for aperture correction.  The uncertainty on $\log M_\ast$ was included by multiplying it by the slope of the relation at given $M_\ast$.  Systematic uncertainties and the error on $f_\mathrm{neb}=0.53$ were not included.  We also deconvolved the effect of the time evolution of average sSFR across $1.43\le z\le1.74$ ($\approx 0.04$~dex) by using the actual redshift distribution of the sample.  We adopted the scaling relation $\mathrm{sSFR} \propto (1+z)^{3.14}$ from \citet{2015A&A...579A...2I}.  

We found the intrinsic scatter to be 0.24~dex for the entire sample, and found it to increase with $M_\ast$ from 0.19 to 0.37~dex in the four bins shown in Figure \ref{fig:dist_delta_sSFR}.  The intrinsic scatter in $\mathrm{sSFR}/\left< \mathrm{sSFR}\right>$ found at low to intermediate masses is in good agreement with previous constraints of the width of the main sequence ($ \approx 0.2~\mathrm{dex}$) \citep[e.g.,][]{2012ApJ...754L..14S,2014ApJS..214...15S}.  \citet{2007ApJ...660L..43N} obtained an observed value of 0.35~dex, and put an upper limit of $<0.30~\mathrm{dex}$ on the intrinsic scatter.  It is also argued that the scatter around the main sequence is nearly redshift-independent across a wide redshift range ($0\le z\lesssim4$) \citep[e.g.,][]{2014ApJS..214...15S,2015A&A...575A..74S,2015A&A...579A...2I}.  \citet{2015A&A...579A...2I} parametrized the sSFR function in $M_\ast$ bins with a log-normal function convolved with the measurement uncertainties, and found the intrinsic scatter to be $\approx0.28$ to 0.46~dex, increasing with $M_\ast$.  The $M_\ast$ dependence they found is qualitatively consistent with our result, while the scatter they found is larger than our findings at all masses.

\citet{2015A&A...579A...2I} argue that, as a caveat, the dynamical range of sSFR covered by the data may be not enough large in many cases to correctly estimate the width of the main sequence.  Indeed, in our case, the criterion on the predicted \Ha\ flux in the pre-selection of the spectroscopic targets reduces the sampling rate, especially, of a population with low $M_\ast$ and low SFR (see Section \ref{sec:Primary-HL}).  Figure \ref{fig:M_vs_SFR_subsamples} shows this selection bias exists across the whole $M_\ast$ range, while being mitigated more or less at $M_\ast \gtrsim 10^{10.7}~M_\odot$.   Establishing the main-sequence based on \Ha\ at high redshifts may require further unbiased deep spectroscopic surveys with high multiplicity, which would be achieved finally by upcoming projects and instruments, such as Multi Object Optical and Near-infrared Spectrograph (MOONS).

% =====================================
\section{Emission-line ratio diagnostics, Revisited \label{sec:lineratio}}

\begin{deluxetable}{lcc}
\tablecaption{Summary of emission-line detections for the samples used in Section \ref{sec:lineratio}\tablenotemark{a} \label{tb:samples for lineratios}}
\tablehead{\colhead{Samples}&\colhead{Sample-H\tablenotemark{b}}&\colhead{Sample-HJ\tablenotemark{d}}}
\startdata
$z_\mathrm{spec}$ range & $1.43 \le z \le 1.74$ & $1.43\le z \le 1.67 $ \\
median $z_\mathrm{spec}$ & 1.579 & 1.557  \\ 
H$\alpha$ & 907 (702) & 648 (506) \\
$\left[ \textrm{N\,{\sc ii}} \right]$ & 551 (347) &  419 (272) \\
$\left[ \textrm{S\,{\sc ii}} \right]$ & 72 (19) & 62 (17) \\
H$\beta$ & - & 203 (136)  \\
$\left[ \textrm{O\,{\sc iii}} \right]$ & - & 242 (220)  \\
$\mathrm{H\alpha} + \left[ \textrm{N\,{\sc ii}} \right]$ \tablenotemark{e} & 551 (325) & 419 (254)  \\
$\mathrm{H\alpha} + \left[ \textrm{S\,{\sc ii}} \right]$  & 72 (18)    & 62 (16)  \\
$\left[ \textrm{O\,{\sc iii}} \right]+\mathrm{H\beta}$     & - & 170 (114)  \\
$\mathrm{H\alpha} + \left[ \textrm{N\,{\sc ii}} \right] + \left[ \textrm{O\,{\sc iii}} \right]+\mathrm{H\beta}$        & - & 118 (59)  \\
$\mathrm{H\alpha} + \left[ \textrm{S\,{\sc ii}} \right] + \left[ \textrm{O\,{\sc iii}} \right]+\mathrm{H\beta}$         & - & 19 (6) \\
\enddata
\tablenotetext{a}{The threshold S/N is 3 for H$\alpha$ and 1.5 for other lines.  In parentheses, the numbers of detections with higher S/N ($\ge5$ for H$\alpha$ and $\ge3$ for other lines) are listed.}
\tablenotetext{b}{Sample-H consists of 907 galaxies with H$\alpha$ ($\ge 3\sigma$).}
\tablenotetext{d}{Sample-HJ consists of 648 galaxies with both H$\alpha$ ($\ge 3\sigma$) and the additional $J$-long coverage.}
\tablenotetext{e}{The numbers of galaxies with multiple emission-line detections are listed in the 8th-12th rows.}
\end{deluxetable}

In \citet{2017ApJ...835...88K}, we have extensively investigated the physical conditions of the ionized gas in star-forming galaxies at $z\sim1.6$ from the FMOS-COSMOS survey.   We utilized various, commonly-used emission-line ratio diagnostics such as the Baldwin-Phillips-Terlevich (BPT; \citealt{1981PASP...93....5B}; see also \citealt{1987ApJS...63..295V}) diagram.  Hereafter we specially refer to the \NII/\Ha\ vs. \OIII/\Hb\ plot as the N2-BPT diagram, and to the \SII/\NII\ vs. \OIII/\Hb\ plot as the S2-BPT diagram.  We confirmed that star-forming galaxies at these redshifts have systematically larger \OIII/\Hb\ ratios, both at fixed $M_\ast$ and fixed metallicity, than their present-day counterparts, as having been indicated by several authors \citep[e.g.,][]{2014ApJ...785..153M,2014ApJ...795..165S,2015ApJ...801...88S}.  In this section, we revisit these diagnostic diagrams to confirm the average emission-line properties based on the final FMOS catalog.

\subsection{Sample definition}
For the following exercises, we selected galaxies in a similar way to \cite{2017ApJ...835...88K} as follows.  The sample is limited to have a detection of \Ha\ at $\ge 3\sigma$ in the $H$-long band and a stellar mass estimate (see Section \ref{sec:lephare}).  Any objects, either detected in X-ray or with the FWHM of \Ha\ greater than $1000~\mathrm{km~s^{-1}}$, are removed from the sample.  Furthermore, we removed a fraction of the individual galaxies as possible AGNs by using their observed emission-line ratios, \Ha/\NII, and \OIII/\Hb.  We excluded objects that are located above the theoretical ``maximum starburst'' line derived by \citet{2001ApJ...556..121K} or have a line ratio of $\log \textrm{\NII}\lambda6584/\textrm{\Ha} \ge -0.1$ or $\log \textrm{\OIII}\lambda5007/\textrm{\Hb} \ge 0.9$.  

Finally, we have 907 galaxies, which are referred to as {\it Sample-H}.  Of these, there are 648 galaxies that have both a redshift between $1.43\le z_\mathrm{spec}\le 1.67$ and $J$-long coverage, which are referred to as {\it Sample-HJ}.  For Sample-HJ, the upper limit of the redshift range is slightly decreased to ensure that all the key emission lines including \SII$\lambda\lambda$ 6717, 6731 fall within the wavelength ranges of the FMOS $H$-long and $J$-long gratings.  In Table \ref{tb:samples for lineratios}, we summarize the numbers of galaxies and line detections in each subsample.  We group galaxies by the S/N of their emission-line detections: high quality (HQ) if S/N$\ge5$ for \Ha\ and S/N$\ge3$ for other lines, and low quality (LQ) if $3\le S/N < 5$ for \Ha\ and $1.5 \le S/N<3$ for other.  The typical range of the stellar mass is $\log M_\ast/M_\odot \approx 9.46 \textrm{--} 11.17$ (the central 95 percentiles).  Note that the sizes of the Sample-H and Sample-HJ increase by 265 and 365, respectively, relative to the corresponding samples in \citet{2017ApJ...835...88K}.

Both individual and stacked measurements were corrected for the Balmer absorption for the \Hb\ line as a function of $M_\ast$ and SFR by using a relation as follows (Kashino et al. in prep):
\begin{eqnarray}
\frac{F^\mathrm{int}_\mathrm{H\beta} - F^\mathrm{obs}_\mathrm{H\beta}}{F^\mathrm{int}_\mathrm{H\beta}} = \frac{1}{2}\left[ \mathrm{erf} \left(-0.626 \left(x + 0.248\right) \right) + 1\right]
\label{eq:balmerabs}
\end{eqnarray}
where 
\begin{equation}
x = \log \mathrm{SFR}/(M_\odot~\mathrm{yr^{-1}}) - 1.32 \left( \log M_\ast / M_\odot - 10\right).
\end{equation}
For this equation, we used $M_\ast$ from SED-fitting and SFR$_\mathrm{UV}$ (Section \ref{sec:lephare}) and substitute the median values in each bin for stacked measurements.  Although \Ha\ fluxes were not corrected for the Balmer absorption following our previous study \citep{2017ApJ...835...88K}, the effects ($\lesssim 3\%$) do not alter the conclusions.

For comparison, we extracted a sample of local galaxies from the SDSS.  The stellar mass and SFRs, from the MPA-JHU catalog \citep{2003MNRAS.341...33K,2004MNRAS.351.1151B,2007ApJS..173..267S}, are converted to a Chabrier IMF to match our sample.  We divided the galaxies into two categories -- star-forming and AGN -- by using the \citet{2003MNRAS.346.1055K} classification line in the BPT diagram, and excluded AGNs from the sample for the following analysis.  The SDSS comparison sample consists of 80,003 star-forming galaxies between $0.04 \le z\le 0.10$.  To illustrate the average relation between line ratios and stellar mass of local star-forming galaxies, we split the sample into bins of $M_\ast$ between $10^{8.6}\le M_\ast \le 10^{11.2}$ with a binsize of $0.2~\mathrm{dex}$, and computed pseudo stacked line ratios from the mean line fluxes in each bin (see \citealt{2017ApJ...835...88K}).  We refer the reader to \citet{2017ApJ...835...88K} for full description of the sample construction of local galaxies.

% ------------------------------------------------------
\subsection{N2-BPT and S2-BPT diagrams}

\begin{figure*}[t] 
   \centering
   \includegraphics[width=5in]{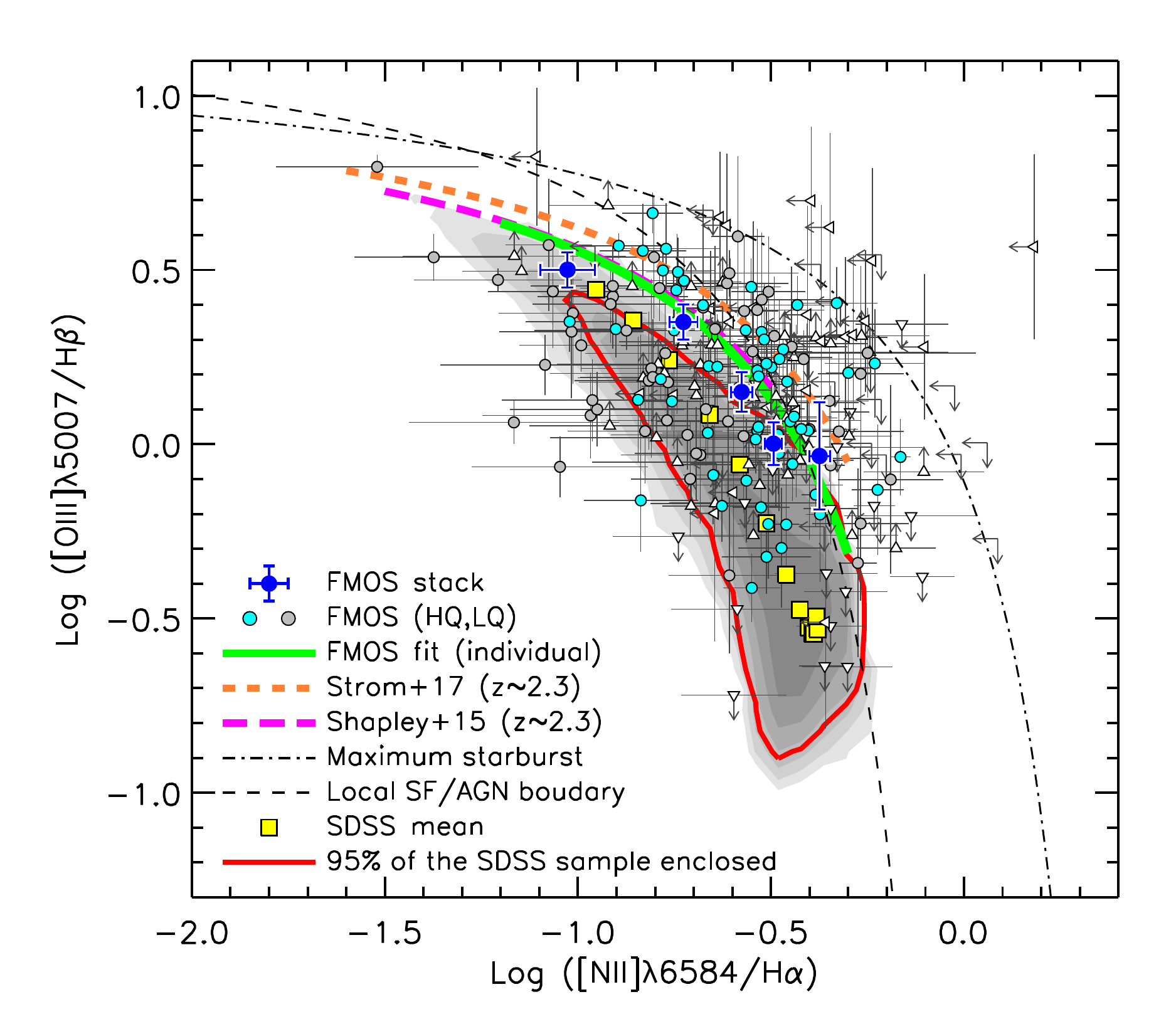} 
   \caption{N2-BPT diagram: $\log \textrm{\NII}\lambda6584/\textrm{\Ha}$ vs. $\log \textrm{\OIII}\lambda5007/\textrm{\Hb}$.  FMOS galaxies in the Sample-HJ are plotted.  Objects with detections of all four lines are shown with circles, being divided into two groups: high-quality objects (HQ, cyan) and low-quality objects (LQ, gray).  For other objects, the arrows and triangles indicate the 2-$\sigma$ upper and/or lower limits.   Large blue circles indicate the FMOS stacked measurements in five mass bins (with the median $M_\ast$ increasing from left to right).   Green line indicate the best-fit curve to the FMOS galaxies (Equation \ref{eq:BPT}).  The shaded contours indicate the distribution of the SDSS sample in log scale and the red contour encloses 90\% of the SDSS galaxies.  Yellow squares indicate the stacked line ratios of the SDSS galaxies in bins of $M_\ast$ between $10^{8.6}\le M_\ast \le 10^{11.2}$.  Thin dashed and dotted-dashed curves indicate the empirical separation between star-forming galaxies and AGNs for the SDSS sample \citep{2003MNRAS.346.1055K}, and the theoretical ``maximum starburst'' limit \citep{2001ApJ...556..121K}, respectively.   In addition, the best-fit relations at $z\sim1.4$ are shown (orange short-dashed line -- \citealt{2017ApJ...836..164S}; magenta long-dashed line -- \citealt{2015ApJ...801...88S}).}
   \label{fig:BPT}
\end{figure*}

Figure \ref{fig:BPT} shows FMOS galaxies in the N2-BPT diagram, in comparison with the SDSS galaxies and with average locations of samples at higher redshifts from the literature \citep{2015ApJ...801...88S,2017ApJ...836..164S}.  The distribution of the SDSS galaxies is represented by the red contour that encloses 95\% of the sample.  

As we originally reported in \citet{2017ApJ...835...88K},  star-forming galaxies at $z \sim 1.6$ in the FMOS sample are located, on average, offset from the sequence of the SDSS galaxies.  The N2-BPT locus of our FMOS galaxies can be described empirically using a simple functional form.  We fit the individual galaxies with detection of all four lines ($N=118$; S/N$\ge 3.0$ for \Ha\ and $\ge1.5$ for others).  The best-fit curve for the locus of the FMOS galaxies (green line in Figure \ref{fig:BPT}) takes the form as follows:
\begin{equation}
\log \left( \textrm{\OIII} /\mathrm{H\beta}\right) = \frac{0.61}{\log (\textrm{\NII}/\mathrm{H\alpha}) - (0.13\pm0.03)} + (1.09\pm0.04)
\label{eq:BPT}
\end{equation}
where the coefficient is fixed to 0.61 \citep{2001ApJ...556..121K}.  Here we accounted for the errors on both line ratios simultaneously.

The offset is further clearly seen by comparing the stacked line ratios between the FMOS (blue circles) and SDSS (yellow squares) samples.  The FMOS stacked points are located along the upper envelope of the red contour, and in agreement with the best-fit curve to the individual galaxies.

\begin{figure}[t] 
   \centering
   \includegraphics[width=3.5in]{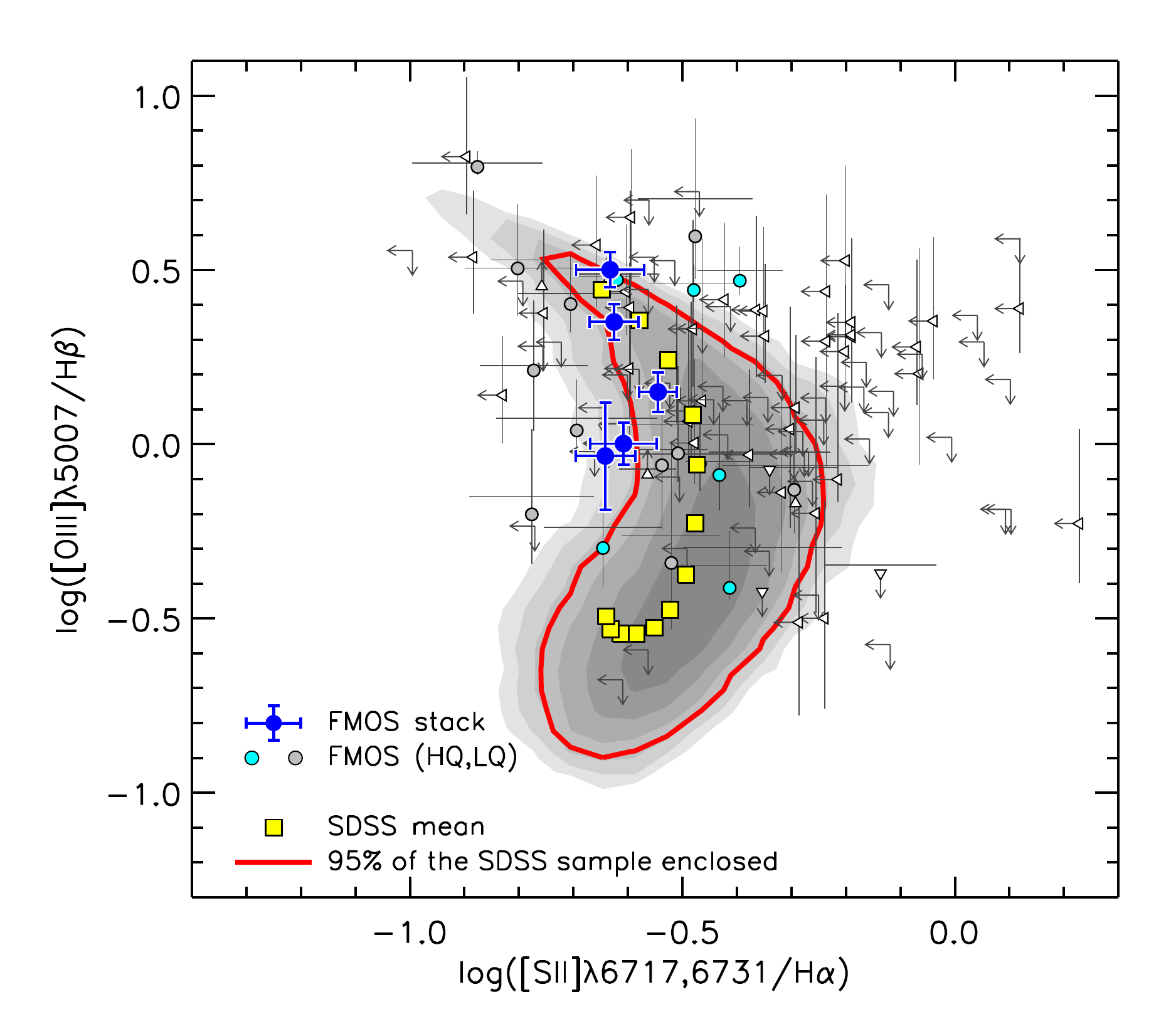} 
   \caption{S2-BPT diagram: $\log \textrm{\SII}\lambda\lambda6717,6731/\textrm{\Ha}$ vs. $\log \textrm{\OIII}\lambda5007/\textrm{\Hb}$.  FMOS galaxies in the Sample-HJ are plotted in comparison with the SDSS galaxies.  Symbols are the same as in Figure \ref{fig:BPT}.}
   \label{fig:BPT_S2}
\end{figure}

Figure \ref{fig:BPT_S2} shows the S2-BPT diagram that replaces the $x$-axis of the N2-BPT digram with the \SII/\Ha\ ratio.  While the \SII\ lines are not detected for the majority of the individual galaxies, it is evident that the stacked measurements certainly differ from the average locus of the local galaxies.  The data points of high $M_\ast$ bins are located near the left-hand envelope of the red contour that encloses 95\% of the SDSS sample.  

We previously reported the possible offset of the high-$z$ galaxies towards left-hand side of the diagram, i.e., lower \SII/\Ha\ at fixed \OIII/\Hb, and regarded this offset as a key observational feature to support our hypothesis that an increase in the ionization parameter is the primary origin of the evolution of the observed emission-line ratios. (see Figure 12 of \citealt{2017ApJ...835...88K}).  Our larger sample in this paper confirmed the offset towards a lower \SII/\Ha\ ratio at $\log \textrm{\OIII/\Hb} \sim 0$, in higher $M_\ast$ bins ($M_\ast \gtrsim 10^{10.3} M_\odot$).

% ---------------------------------------------------------
\subsection{Stellar mass--excitation diagram}

\begin{figure}[t] 
   \centering
   \includegraphics[width=3.5in]{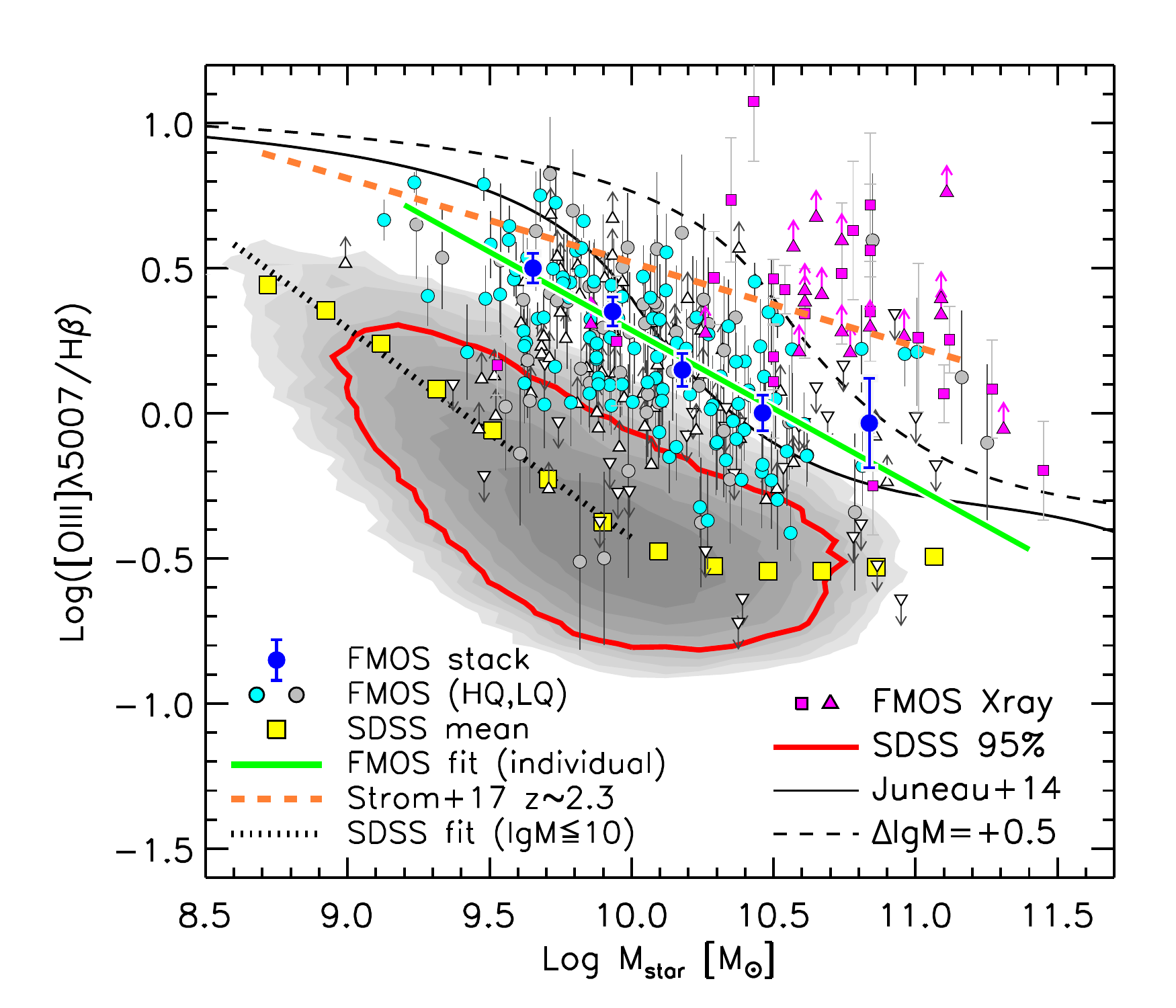} 
   \caption{Mass--excitation (MEx) diagram: $M_\ast$ vs \OIII$\lambda5007$/\Hb.  The FMOS galaxies in the Sample-HJ are compared with the SDSS sample.  Symbols are the same as in Figure \ref{fig:BPT}.  Solid green line indicates the best-fit linear relation for the FMOS galaxies, and the orange dashed line is the best-fit relation for the KBSS-MOSFIRE sample at $z\sim2.3$ \citep{2017ApJ...836..164S}.  The thin dashed curves indicate the divisions between star-forming/composite galaxies and AGN at $z\sim0$ \citep{2014ApJ...788...88J}.}
   \label{fig:MxO3}
\end{figure}

We show in Figure \ref{fig:MxO3} the \OIII$\lambda5007$/\Hb\ ratio as a function of $M_\ast$ for both FMOS Sample-HJ and the local SDSS sample.  This is known as the mass--excitation (MEx) diagram \citep{2011ApJ...736..104J}.  It is clear that the FMOS galaxies occupy a region distinct from the local galaxies, well above the upper envelope of the red contour enclosing 95\% of the local sample.  Across the entire $M_\ast$ range probed, the line ratio increases at fixed $M_\ast$ by $\approx 0.5~\mathrm{dex}$ from $z \sim 0.1$ to $z \sim 1.6$.  Similarly to the SDSS sample, the FMOS galaxies exhibit an inverse correlation between \OIII/\Hb\ and $M_\ast$.  We derived a linear fit to the locus of the FMOS galaxies, while limiting to 170 galaxies in the Sample-HJ having both \Hb\ and \OIII\ detections ($\ge 1.5\sigma$).  The best-fit relation (thick green line in Figure \ref{fig:MxO3}) is given as
\begin{equation}
\log(\textrm{\OIII/\Hb}) = 0.23 - 0.54 \times \left[ \log M_\ast/M_\odot -10 \right].
\end{equation}
The stacked measurements in five $M_\ast$ bins are in good agreement with the best-fit relation.  With respect to the best-fit linear relation, the intrinsic scatter is found to be $\sigma_\mathrm{int}=0.17$ (for only objects with both detections) after accounting for the individual measurements errors on $\log M_\ast$ and $\log$\OIII/\Hb.  

For comparison, we show the best-fit linear relation to the KBSS-MOSFIRE samples at $z\sim2.3$ \citep{2017ApJ...836..164S}, indicating further increase in the emission-line ratio at fixed $M_\ast$.  The best-fit relation to the higher redshift sample has a shallower slope ($-0.29$) than the FMOS sample at $z\sim1.6$ due to higher ratios at high $M_\ast$.  Turning to the SDSS sample, the stacked points show further steeper slope at lower masses.  Fitting to those at $M_\ast\le 10^{10}~M_\odot$, we find the slope to be $-0.72$.  The gradual change of the slope indicates that the \OIII/\Hb\ ratio evolves with redshift in a mass-dependent way: the more massive the systems are, the faster the \OIII/\Hb\ decrease with redshift.    

We overplot in Figure \ref{fig:MxO3} the empirically-calibrated division line between AGN and star-forming (or composite) galaxies at $z\sim0.1$ (thin solid line; \citealt{2014ApJ...788...88J}).  It is clear that the majority of the FMOS galaxies are located above this classification line.  For comparison to the star-forming population, Figure \ref{fig:MxO3} shows X-ray AGNs at $z\sim1.6$ from the full FMOS catalog.  For these X-ray detected objects, we estimated stellar masses using the SED3FIT package  \citep{2013A&A...551A.100B} based on the MAGPHYS software \citep{2008MNRAS.388.1595D}, including the emission from an AGN torus (full analysis of SEDs of X-ray sources are presented in Kashino et al., in prep).  These objects tend to have even higher \OIII/\Hb\ ratios than the star-forming population at fixed $M_\ast$, while roughly one-third of those are virtually mixed with the star-forming population.  We found that shifting the division line in the MEx diagram by $\Delta \log M_\ast = + 0.5~\mathrm{dex}$ yields a reasonable classification between the star-forming galaxies and X-ray sources (dashed line in Figure \ref{fig:MxO3}).   This is in agreement with the luminosity-dependent offset modeled by \citet[][see Appendix B]{2014ApJ...788...88J} for the threshold luminosity of the FMOS sample which is $L^\mathrm{thresh}_\mathrm{H\alpha} \approx 10^{41.5}~\mathrm{erg~s^{-1}}$ (see Figures \ref{fig:z_vs_LHa}).  \citet{2015ApJ...801...35C} found that a shift of $+0.75~\mathrm{dex}$ in $M_\ast$ is required to purely distinguish AGNs from star-forming galaxies at $z\sim2.3$ using the MOSDEF survey \citep{2015ApJS..218...15K}, while recently, \citet{2017ApJ...836..164S} argued that an even larger shift ($\sim1~\mathrm{dex}$) is needed for the KBSS-MOSFIRE sample.  

% ---------------------------------------------
\subsection{Stellar mass vs. \NII/\Ha \label{sec:MxN2}}

\begin{figure}[t] 
   \centering
   \includegraphics[width=3.5in]{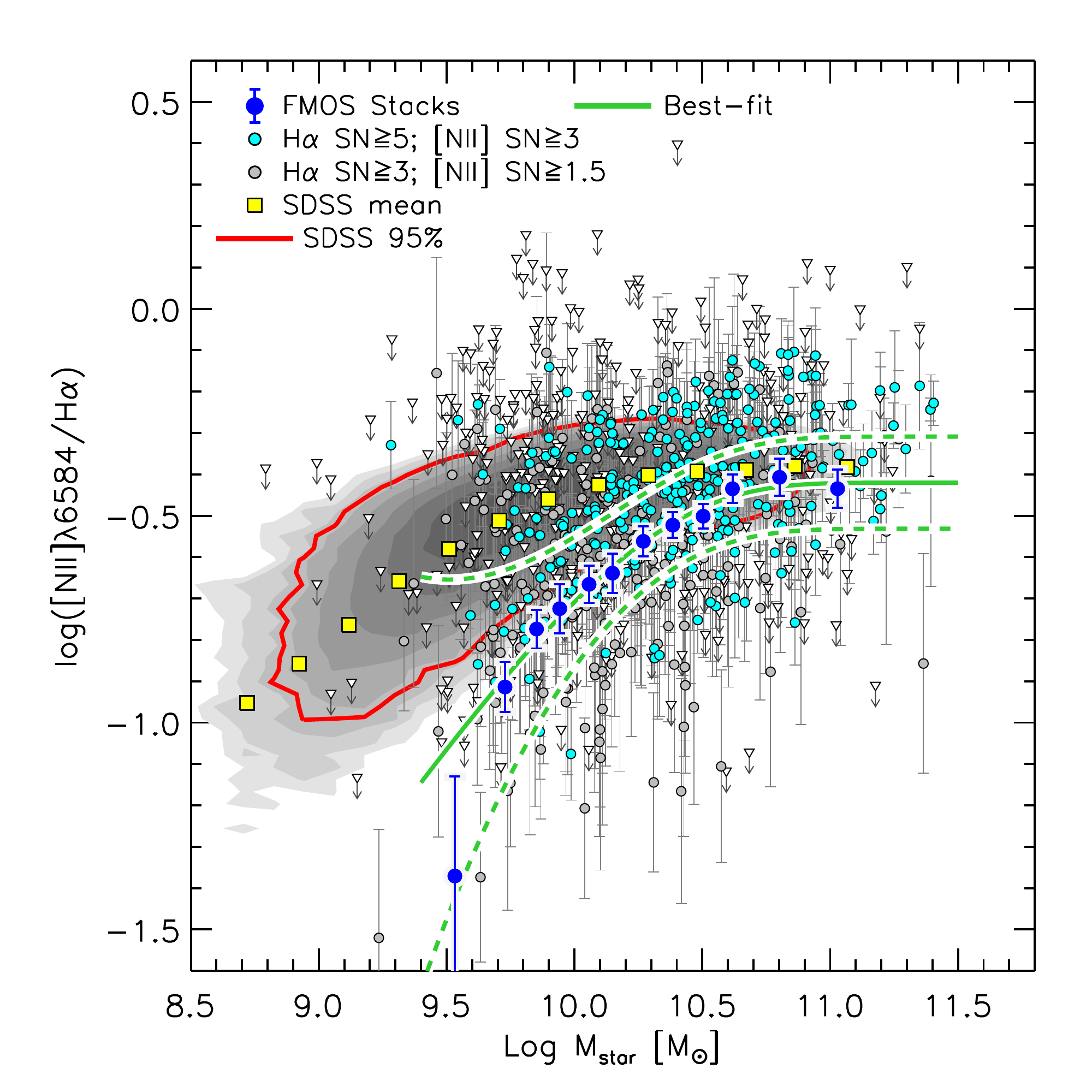} 
   \caption{$M_\ast$ vs. \NII$\lambda$6584/\Ha.  The FMOS galaxies in the Sample-H are compared with the local SDSS sample.  Cyan and gray circles show the HQ ($S/N(\mathrm{H\alpha})\ge5$ and $S/N(\textrm{\NII})\ge3$) and LQ ($S/N(\mathrm{H\alpha})\ge3$ and $S/N(\textrm{\NII})\ge1.5$) FMOS galaxies.  The solid and dashed green curves indicate the best-fit linear relation for the FMOS galaxies with the estimated intrinsic scatters (see Section \ref{sec:N2res}).}
   \label{fig:MxN2}
\end{figure}

In Figure \ref{fig:MxN2}, we show the observed \NII$\lambda$6584/\Ha\ ratios as a function of $M_\ast$ for the FMOS Sample-H and the local SDSS sample.  We plot 557 galaxies with both \Ha\ and \NII\ detections, divided into two groups: high-quality objects (HQ, $N=325$; cyan circles) if S/N(\Ha)$\ge5$ and S/N(\NII)$\ge3$, and low-quality objects (LQ, $N=226$; gray circles) if S/N(\Ha)$\ge3$ and S/N(\NII)$\ge1.5$.  For others, the upper limits are shown by downward arrows.  The region occupied by the FMOS galaxies are largely overlapped with the locus of the SDSS sample, while a number of objects have a lower \NII/\Ha\ ratio than the lower envelope of the red contour enclosing 95\% of the SDSS sample.  The stacked measurements (large blue circles), however, are off clearly from the average locus of the SDSS galaxies.  The amount of the offset is a strong function of $M_\ast$, from $\ge0.5~\mathrm{dex}$ at $M_\ast \sim 10^{9.7}~M_\odot$ to $<0.1~\mathrm{dex}$ at the massive end ($M_\ast \ge 10^{11}~M_\odot$).  

To analytically describe the the average $M_\ast$ vs. \NII/\Ha\ relation, we used a functional form proposed by \citet{2014ApJ...791..130Z} originally to parametrize the mass-metallicity relation:
\begin{equation}
N_2 (M_\ast) = \mathcal{R}_0 - \log \left[ 1 - \exp \left( \left[\frac{M_\ast}{\mathcal{M}_0}\right]^\gamma \right) \right]
\label{eq:MxN2}
\end{equation}
where $N_2$ denotes $\log \textrm{\NII}\lambda6584/\textrm{\Ha}$, $\mathcal{R}_0$ is the asymptotic value of the line ratio in log scale at the high-mass end, $\mathcal{M}_0$ is the characteristic mass at which the line ratio begins to saturate, and $\gamma$ is the low-mass end slope.  The best-fit to the stacked line ratios of the FMOS Sample-H takes $\mathcal{R}_0= -0.42 \pm 0.04$, $\log \mathcal{M}_0/M_\odot= 10.16 \pm 0.09$, and $\gamma=0.90 \pm 0.14$.  The best-fit relation well traces the FMOS stacked points.  For the local SDSS sample, we obtained the best-fit parameters of $\mathcal{R}_0= -0.39$, $\log \mathcal{M}_0/M_\odot= 9.50$, and $\gamma=0.66$ with the same procedure.  

Supposing that the \NII/\Ha\ ratio is sensitive to the gas-phase metallicity, the behavior of the stacked points, as well as the best-fit relation, support with higher confidence our past statement that the majority of massive ($M_\ast\gtrsim10^{10.6}~M_\odot$) galaxies are already chemically mature at $z\sim1.6$ as much as local galaxies with the same stellar masses \citep{2014ApJ...792...75Z,2017ApJ...835...88K}.  In Figure \ref{fig:MxN2},  we also indicate estimated intrinsic scatter of the FMOS sample around the average best-fit $M_\ast$ vs. \NII/\Ha\ relation (green dashed lines).  Though the derivation of the scatter is described in detail in the next subsection,  the sequence is tight almost entire $M_\ast$ range, except the lowest $M_\ast$, where the constraint is poor due to the small number of detections.  It is seen that the amount of the redshift evolution of the average \NII/\Ha\ is comparable to twice the intrinsic scatter at $M_\ast\sim10^{10}~M_\odot$.

\subsection{Intrinsic scatter of the $M_\ast$--\NII/\Ha\ relation \label{sec:N2res}}

Comparing to the local SDSS sample, it seems that the FMOS galaxies show a larger scatter in the \NII/\Ha\ ratio at given $M_\ast$.  To estimate the intrinsic scatter in the line ratio, we define $\Delta \log \textrm{\NII/\Ha}$ as the offset of the \NII/\Ha\ ratios with respect to $N_2(M_\ast)$, i.e., the best-fit $M_\ast$--\NII/\Ha\ relation at given $M_\ast$.  

\begin{figure}[t] 
   \centering
   \includegraphics[width=3.5in]{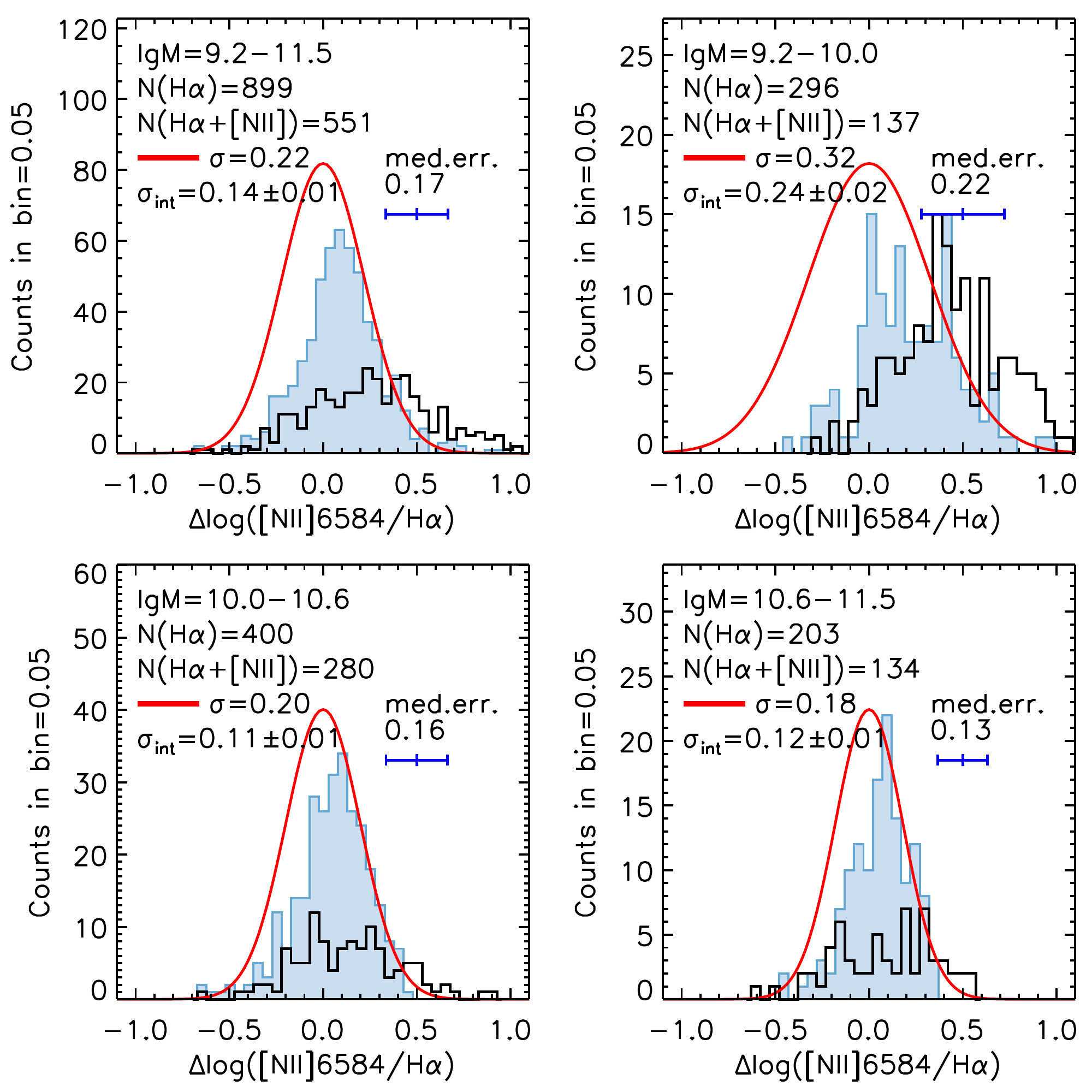}    
   \caption{Distribution of the \NII$\lambda$6584/\Ha\ ratios normalized to the best-fit relation at given $M_\ast$.  Objects in the FMOS Sample-H, having both \Ha\ and \NII\ detections, are shown by filled histograms in four ranges of stellar mass: $\log M_\ast/M_\odot = $ [9.2:11.5] (upper left), [9.2:10.0] (upper right), [10.0,10.6] (lower left), and [10.6:11.5] (lower right).  Black empty histograms indicate the distribution of upper limits.  The numbers of all \Ha-detected galaxies (regardless of \NII\ detection; $N(\textrm{\Ha})$) and those with both \Ha\ and \NII\ detections ($N(\textrm{\Ha}+\textrm{\NII})$) within the mass ranges are given in each panel.  Red curve indicates a zero-mean normal function with a broadened standard deviation of $\sigma=\sqrt{\sigma_\mathrm{int}^2 + \left<\delta\right>^2}$.  The values of $\sigma$, $\sigma_\mathrm{int}$, and the median errors $\left<\delta\right>$ are given in each panel.}
   \label{fig:hist_N2res}
\end{figure}

In Figure \ref{fig:hist_N2res}, we show the distribution of the $\Delta \log \textrm{\NII/\Ha}$ for objects with both \Ha\ and \NII\ detections in different ranges of $M_\ast$ as labeled in each panel.  The upper-left panel is for the almost entire stellar mass range between $10^{9.2}\le M_\ast/M_\odot \le 10^{11.5}$, while other three panels for three partial mass bins ($\log M_\ast/M_\odot = $ [9.2:10.0], [10.0:10.6], and [10.6:11.5]).  It is clear that the distributions of the $\Delta \log \textrm{\NII/\Ha}$ values of the \Ha+\NII-detected objects are skewed towards higher values because the objects with both detections are biased towards having a higher \NII/\Ha\ ratio with respect to the best-fit $M_\ast$--\NII/\Ha\ relation, as seen in Figure \ref{fig:MxN2}.

To estimate the true distribution and the scatter, we first assumed that the line ratios of all galaxies follow a zero-mean normal distribution with respect to the best-fit $M_\ast$--\NII/\Ha\ relation.  We then estimated the intrinsic scatter in each $M_\ast$ bin including upper limits.  We computed the likelihood $\mathcal{L}(\sigma_\mathrm{int})$ for grids of $\sigma_\mathrm{int}$ as follows:
\begin{equation}
\mathcal{L}(\sigma_\mathrm{int}) \propto \prod_\mathrm{dec} F(x_i, \tilde{\sigma}_i) \prod_\mathrm{sup} (1-S(c_i, \tilde{\sigma}_i))
\end{equation}
where $x_i$ and $c_i$ are the detection values and upper limits of $\Delta \log \textrm{\NII/\Ha}$, respectively.  The probability functions $F$ and $S$ are a zero-mean normal distribution function and a zero-mean normal survival function, respectively.  The standard deviation $\tilde{\sigma}_i$ is computed for each object by summing in quadrature the uncertainties on $\log \textrm{\NII/\Ha}$, $\log M_\ast$, and the intrinsic scatter $\sigma_\mathrm{int}$.  The uncertainties on $\log M_\ast$ were included by multiplying them by the slope of the best-fit $M_\ast$--\NII/\Ha\ relation at a given $M_\ast$.  In the figure, we also show the distribution of the upper limits for the \NII-undetected objects.

For the subsamples shown in Figure \ref{fig:hist_N2res}, we obtained a tight constraint on the intrinsic scatter $\sigma_\mathrm{int}$, as indicated in each panel.  We found $\sigma_\mathrm{int}=0.14\pm 0.01$ for the entire $M_\ast$ range, and the largest value ($\sigma_\mathrm{int}=0.24\pm0.02$) in the lowest $M_\ast$ bin.  In the figure, we overplot the normal distribution function (red curves) with a standard deviation convolved with the median error $\left<\delta\right>$ (including the $M_\ast$ uncertainties) in each bin (i.e., $\sigma=\sqrt{\sigma_\mathrm{int}^2 + \left< \delta\right>^2}$).  These model distribution functions well trace the high-value tail of the histograms of the detected $\log \textrm{\NII/\Ha}$ values, while there are no upper limits beyond the low-value tail of the model functions.  This indicates that our estimates of $\sigma_\mathrm{int}$ are robust.

\begin{figure}[t] 
   \centering
   \includegraphics[width=3.5in]{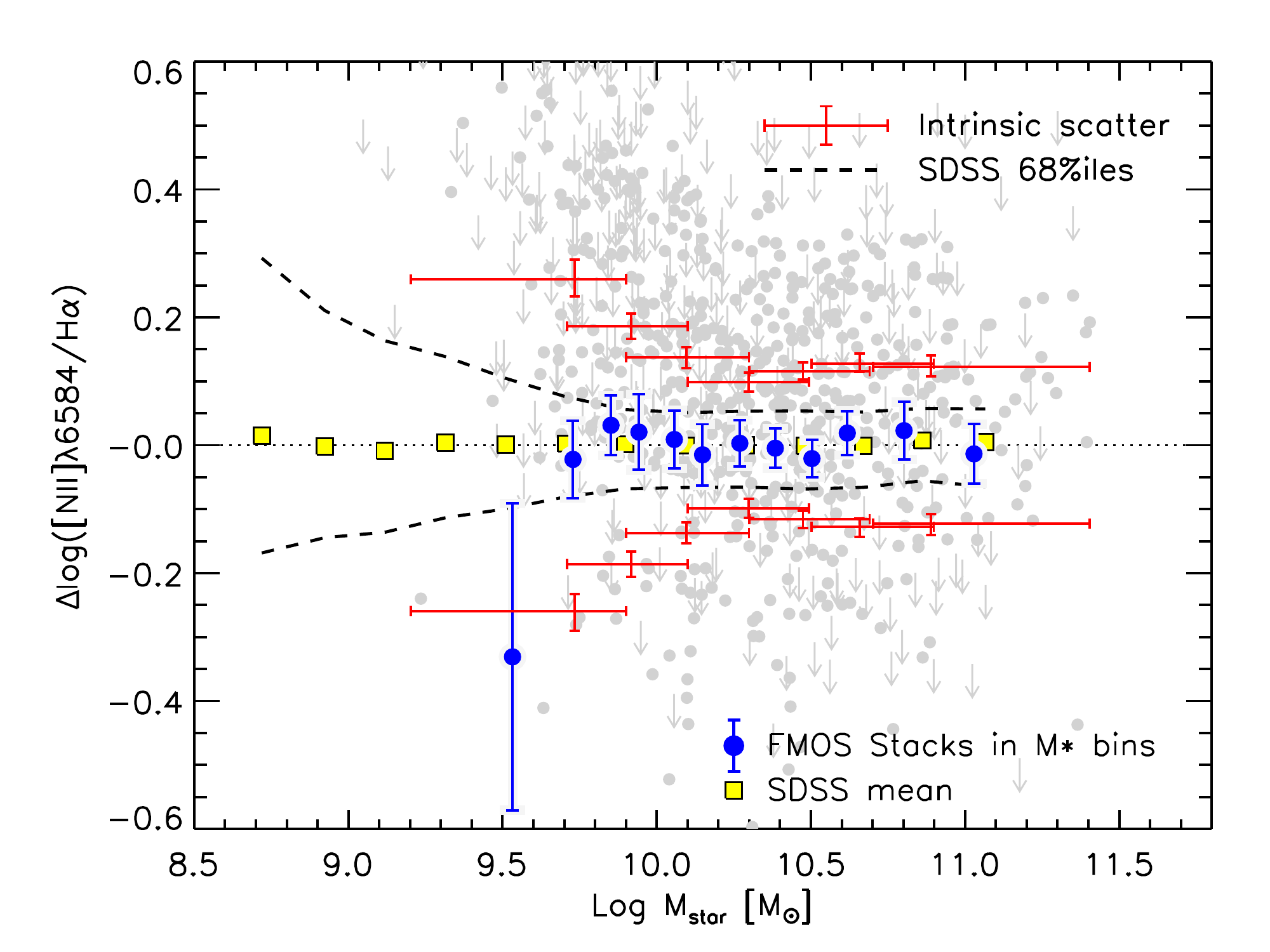}
   \includegraphics[width=3.5in]{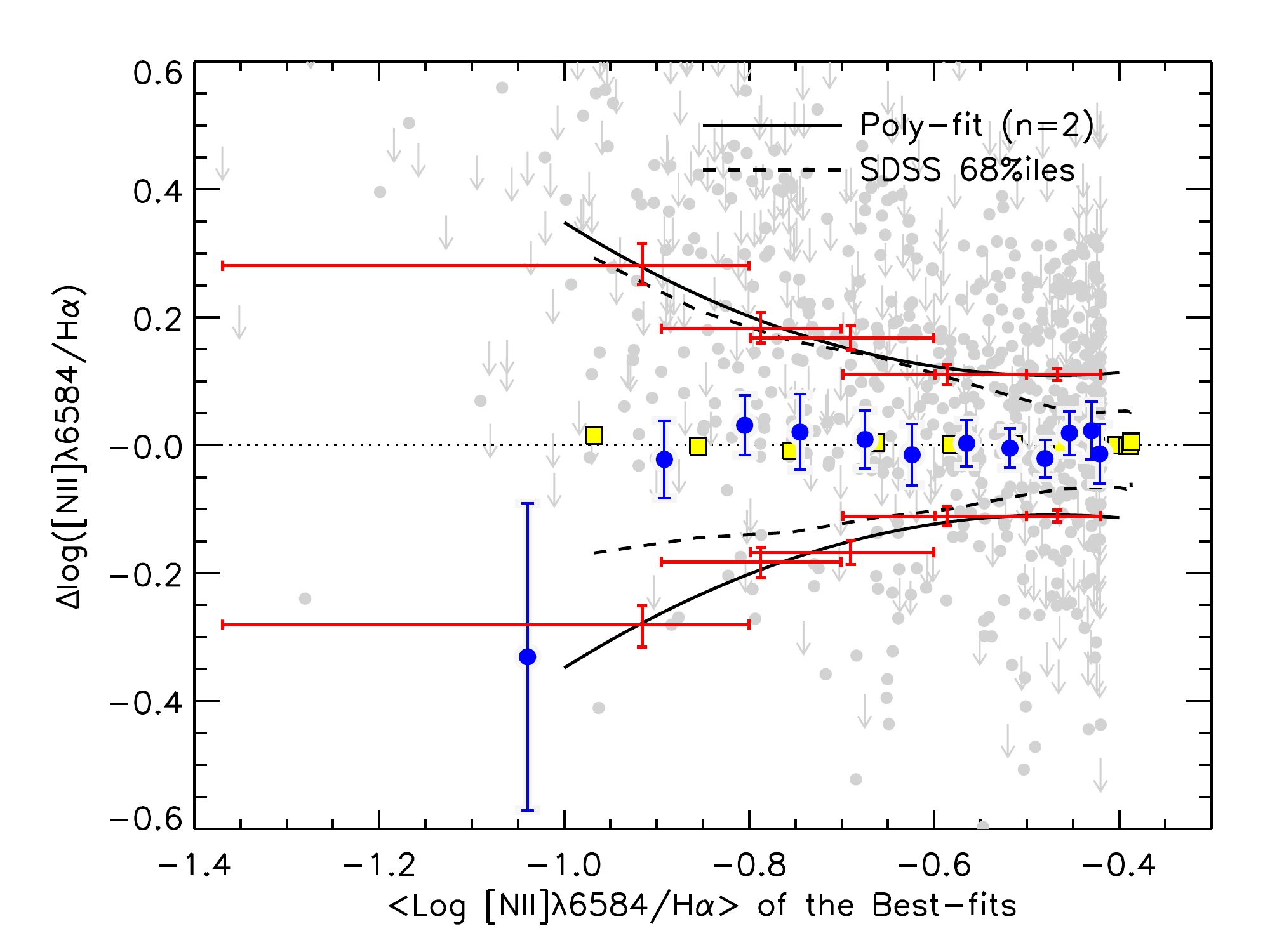} 
   \caption{Upper panel: the \NII/\Ha\ ratio normalized by the best-fit $M_\ast$--\NII/\Ha\ relation as a function of $M_\ast$.  FMOS objects are shown by gray circles and downarrows (upper limits).  Red error bars mark the estimated intrinsic scatter in overlapping $M_\ast$ bins: the vertical bars indicate the central 68\% confidence levels at the median $M_\ast$ values and the horizontal bars indicate the widths of the $M_\ast$ bins.  The black dashed lines indicate the central 68 percentiles of the individual SDSS galaxies in bins of $M_\ast$.  Lower-panel:  same as the upper panel, but $\Delta \log \textrm{\NII/\Ha}$ as a function of $N_2(M_\ast)$, i.e., the ``best-fit'' \NII/\Ha\ ratio at given $M_\ast$, which are taken from the best-fit relation at given stellar mass for the FMOS and SDSS samples each.  The solid curve indicates a second-order polynomial fit to the $\sigma_\mathrm{int}$ estimates of the FMOS sample.}
   \label{fig:MxN2res}
\end{figure}

For further investigation of the trend of the intrinsic scatter and comparison with the SDSS galaxies, we repeated the likelihood analysis with narrower, overlap binning.  In the upper panel of Figure \ref{fig:MxN2res}, we show the estimated $\sigma_\mathrm{int}$ as a function of $M_\ast$, with the individual $\Delta \log \textrm{\NII/\Ha}$ values, in comparison with the SDSS sample.  For the FMOS and SDSS samples, $\Delta \log \textrm{\NII/\Ha}$ are computed, separately, with their own best-fit relation.  For the local galaxies, we show the stacked ratios normalized to the best-fit relation and the central 68 percentiles in the $M_\ast$ bins (black dashed lines).  Note that the increase in the scatter due to the individual measurement errors on \NII/\Ha\ are negligible ($\lesssim 5\%$) for the SDSS sample.  It is clear that the intrinsic scatter of the FMOS galaxies increases with decreasing $M_\ast$, while being almost constant $\sigma_\mathrm{int}\approx0.1$ at $M_\ast\gtrsim10^{10.3}~M_\odot$.  The similar trend is seen in the SDSS sample, although the intrinsic scatter of the local sample is smaller than those of the FMOS sample at fixed $M_\ast$.  

Next we compare the scatters at fixed $N_2(M_\ast)$ values, which are taken from the best-fit $M_\ast$--\NII/\Ha\ relations for the FMOS and SDSS samples, respectively (lower panel in Figure \ref{fig:MxN2res}).  Now the trends of the intrinsic scatter are rather in good agreement between the local and FMOS samples, though the scatter of the FMOS sample is about twice the SDSS sample at the highest $N_2(M_\ast)$.  With respect to this, a caveat is that the local star-forming galaxies are limited to those below the \citet{2003MNRAS.346.1055K} division line in the BPT diagram, which is more strict than the maximum starburst limit adopted for the FMOS galaxies, and hence may effectively reduce the scatter, especially, at high masses where the line ratio is nearly saturated.  

Comparing the two panels, it seems that the scatter in \NII/\Ha\ is more directly related to $N_2(M_\ast)$, rather than $M_\ast$ itself, and thus $\sigma_\mathrm{int}$ varies more continuously with $N_2$ across its whole range.  We thus parametrized $\sigma_\mathrm{int}$ as a function of $ N_2 $ by a second-order polynomial:
\begin{equation}
\sigma_\mathrm{int} = 0.299 + 0.807 N_2(M_\ast) + 0.856 N_2(M_\ast)^2,
\end{equation}
which is shown in the lower panel of Figure \ref{fig:MxN2res}.  We used this fit to indicate the estimated intrinsic scatter around the average $M_\ast$ vs. \NII/\Ha\ relation in Figure \ref{fig:MxN2}.

The \NII/\Ha\ ratio is known to reflect the gas-phase metallicity of the galaxy \citep[e.g.,][]{2004MNRAS.348L..59P}, though it is also affected by other IGM conditions such as ionization parameter, the shape of ionizing spectra \citep[e.g.,][]{2013ApJ...774..100K}, and the intrinsic ratio of N/O \citep{2014ApJ...785..153M}.  Therefore, interpreting the scatter in \NII/\Ha\ as a result of only variation in metallicity may lead to inaccurates insights.  Though, our result likely indicate that there is no large difference in the amount of metallicity variation {\it at fixed average metallicity} between the local SDSS and $z\sim1.6$ FMOS samples.  We note that the physical time across the redshift range of the FMOS sample ($1.43 \le z\le 1.74$; 0.69~Gyr) is similar to that of the local sample ($0.04\le z\le 0.10$; 0.76~Gyr).  Therefore, the effects of the time evolution of metallicity within the redshift ranges of the two samples should be small.

% =======================================================
\section{Comparison between \Ha\ and \OIII\ emitters \label{sec:Ha_vs_OIII}}

The \OIII$\lambda$5007 emission line is one of the strongest lines in the rest-frame optical window, being comparable to \Ha.  Therefore, it has been used as a tracer of galaxies \citep{2015ApJ...806..208S,2015MNRAS.452.3948K}.  Meanwhile, it is well known that the intensity of the \OIII\ line is sensitive to metallicity at fixed SFR, as well as more affected by dust extinction than \Ha.  \citet{2016MNRAS.462..181S} compared the narrow-band selected \Ha- and \OIII-emitter samples at $z\sim2$, and argue that the \OIII-emitters trace almost the same galaxy populations as the \Ha-emitters.  However, the contamination of remaining AGNs would be not negligible since there is no way to see the BPT line ratios (i.e., \NII/\Ha\ and \OIII/\Hb) and line profiles with their narrow-band observations.  Moreover, the contamination of \Hb-emitters misidentified as an \OIII-emitter with no \Ha\ detection may lead to inaccurate results.  We thus use our FMOS sample to study the population of \OIII-emitting galaxies in comparison with the \Ha-emitting sources, and examine their claims. 

%Although the authors attempted to exclude AGNs using the X-ray and the {\it Spitzer}/IRAC data, the contamination of remaining AGNs would be not negligible since there is no way to see the BPT line ratios (i.e., \NII/\Ha\ and \OIII/\Hb) and line profiles with the narrow-band observations.  Moreover, the contamination of \Hb-emitters misidentified as an \OIII-emitter with no \Ha\ detection may lead to inaccurate results.  We thus use the FMOS sample to study the population of \OIII-emitting galaxies in comparison with the \Ha-emitting sources, and examine their claims. 

\begin{deluxetable}{lcl}
\tablecaption{Summary of the subsamples used in Section \ref{sec:Ha_vs_OIII} \label{tb:Ha_vs_OIII}}
\tablehead{\colhead{Subsamples}&
		\colhead{$N$}&
		\colhead{Note}}
\startdata
\Ha-emitters            & 682 & \Ha\ ($\ge3\sigma$) and \OIII\ coverage \\
\Ha-single-emitters & 439 & \Ha-emitters with no \OIII\ detection\tablenotemark{a}  \\
\OIII-emitters           & 270 & \OIII\ ($\ge3\sigma$) and \Ha\ coverage \\
\OIII-single-emitters & 27 & \OIII-emitters with no \Ha\ detection\tablenotemark{a}  \\
\Ha+\OIII-emitters   & 243 & Both \Ha\ and \OIII\  detections ($\ge3 \sigma$) \\
\enddata
\tablenotetext{a}{These two single-emitter samples have spectral coverage for the other emission line.  Detections at $1.5\le S/N <3$ are regarded as non-detection through this section.}
\end{deluxetable}

For our purposes, we define subsamples of $z\sim1.6$ galaxies as listed in Table \ref{tb:Ha_vs_OIII}.  These objects are limited to have secure estimates of $M_\ast$ and $E_\mathrm{star}(B-V)$ (Section \ref{sec:lephare}), and X-ray objects and possible AGNs were excluded in the same way as in Section \ref{sec:MS}.  The \Ha-emitter sample contains 682 objects with detection of \Ha\ at $\ge3\sigma$ and the $J$-long coverage of the \OIII\ emission line (i.e., we have either detection or estimate of the upper limit of \OIII).  Of these, we detected \OIII\ line at $\ge3\sigma$ for 243 objects, which are referred to as the \Ha+\OIII\ emitter sample.  The detection failed for the remaining 439 objects ($<3\sigma$ or upper limit on \OIII\ flux), which are categorized into the \Ha-single-emitters.  We also defined the the \OIII-emitter sample containing 270 objects with a detection of \OIII\ ($\ge3\sigma$) and the $H$-long coverage, and the \OIII\ single-emitter ($<3\sigma$ or upper limit on \Ha) sample of 27 objects.   

\subsection{\OIII\ flux vs. \Ha\ flux}

\begin{figure}[t] 
   \centering
   \includegraphics[width=3.5in]{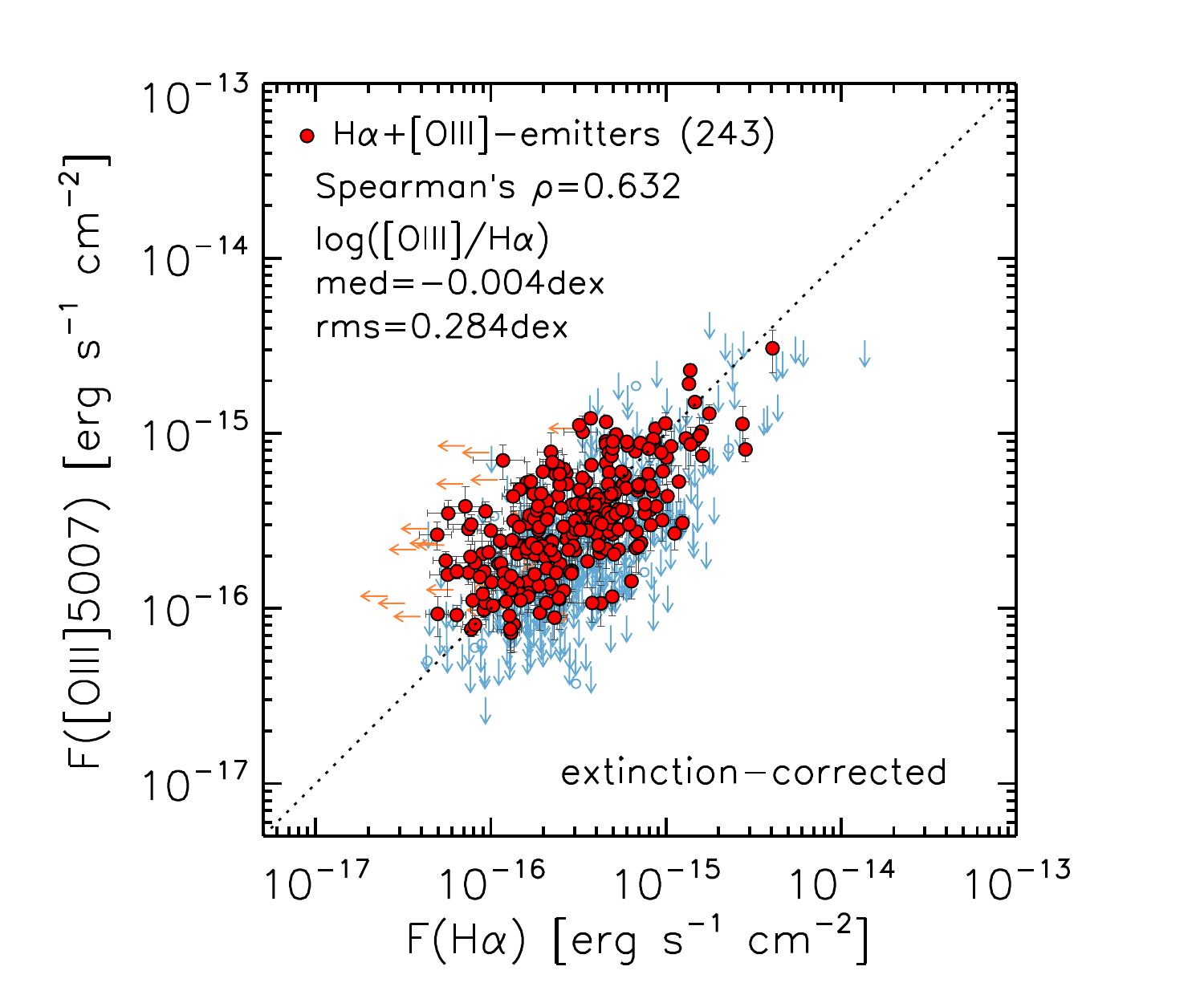} 
   \caption{Correlation between \Ha\ and \OIII$\lambda$5007 fluxes, after correcting for dust extinction (both corrected for aperture loss).  Red circles indicate 243 objects in the \Ha+\OIII-emitter sample (\Ha\ at $\ge3\sigma$ and \OIII\ at $\ge 3\sigma$).  \Ha-single emitters are shown by light blue circles ($1.5\le S/N(\textrm{\OIII})<3$) or downward arrows ($2\sigma$ upper limits).  \OIII-single emitters are shown by orange circles  ($1.5\le S/N$(\OIII)$<3$) or leftward arrows ($2\sigma$ upper limits).  Dotted line indicates a one-to-one relation.}
   \label{fig:flux_Ha_OIII}
\end{figure}

In Figure \ref{fig:flux_Ha_OIII}, we show the correlation between \Ha\ and \OIII$\lambda$5007 fluxes after correcting for dust extinction and aperture loss.   Extinction correction is applied by assuming the \citet{1989ApJ...345..245C} extinction law with $f_\mathrm{neb}=0.53$ (Section \ref{sec:dust}).  Limiting the \Ha+\OIII-emitter sample, a strong correlation and good agreement exist between these quantities.  We found the Spearman's rank correlation coefficient to be $\rho=0.63$, excluding the null hypothesis of no correlation.  The scatter of $\log \textrm{\OIII/\Ha}$ is found to be 0.28~dex with a small median offset $\log \textrm{\OIII/\Ha} = -0.004$.  %Our result is in good agreement with \citet{2016MNRAS.462..181S}, who selected galaxies based on narrow-band excesses corresponding to \Ha\ and \OIII.  

\begin{figure}[t] 
   \centering
   \includegraphics[width=3.5in]{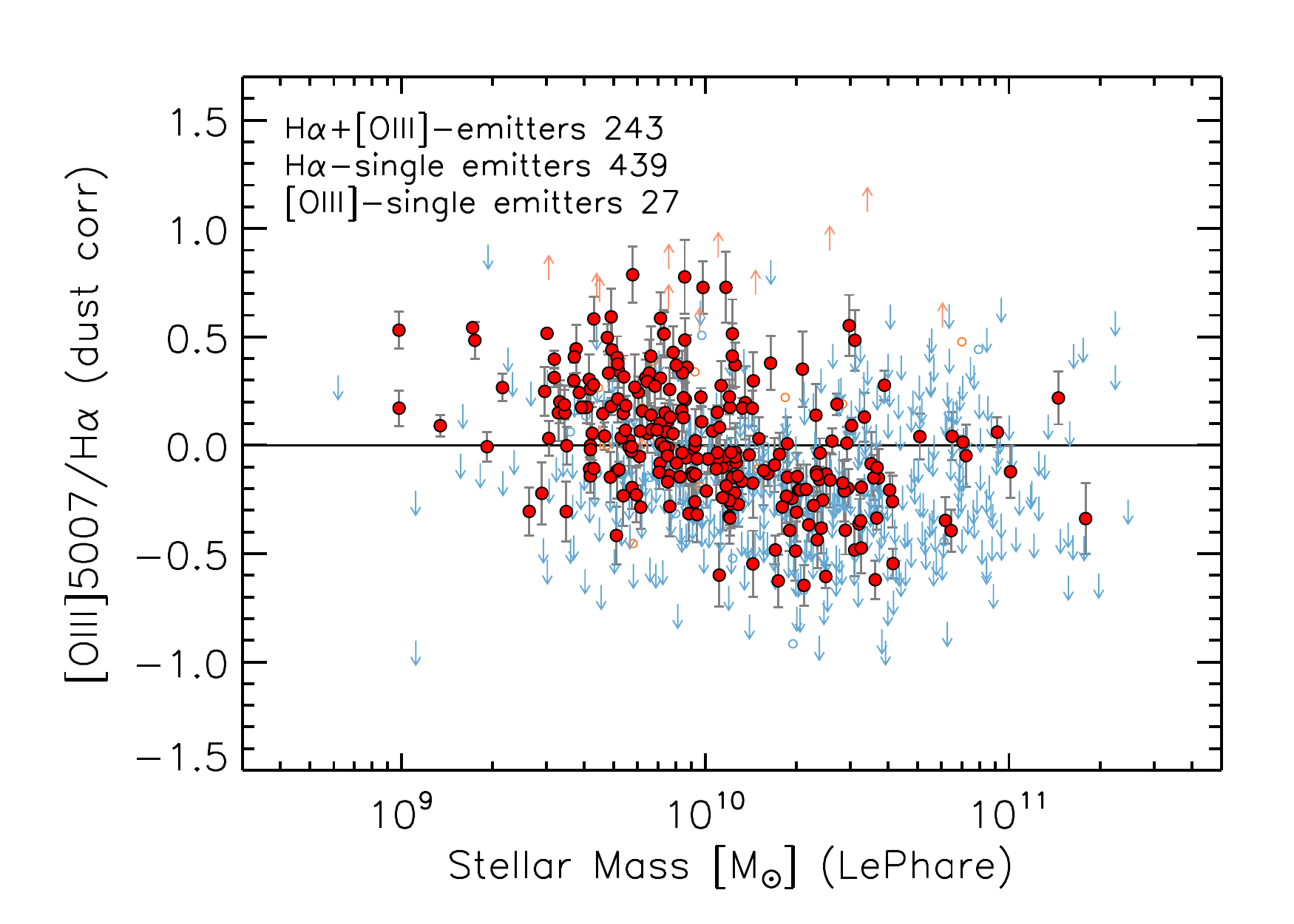} 
   \caption{Dust-corrected \OIII/\Ha\ ratio as a function of $M_\ast$.  Red circles indicate 261 objects in the \Ha+\OIII-emitter sample (\Ha\ at $\ge3\sigma$ and \OIII\ at $\ge 1.5\sigma$), and downward arrows indicate the upper limits for 416 \Ha-single-emitters ($\ge3\sigma$, but only upper limits on the \OIII\ flux).}
   \label{fig:M_vs_OIII_Ha}
\end{figure}

The \OIII/\Ha\ ratio is expected to depend on stellar mass because metallicity and dust extinction increase on average with $M_\ast$.  In Figure \ref{fig:M_vs_OIII_Ha}, we show the dust-corrected (aperture as well) \OIII/\Ha\ ratio as a function of $M_\ast$ for the subsamples.  Limiting to the \Ha+\OIII\ emitters, it is clear that the line ratio decreases with increasing $M_\ast$, and that at high $M_\ast$ ($\gtrsim 10^{10.5}~M_\odot$) the majority of the \Ha-detected objects have no significant detection of the \OIII\ line.  We note that the extinction-corrected \OIII/\Ha\ ratio has essentially the same information as the \OIII/\Hb\ ratio (i.e., Figure \ref{fig:MxO3}).  %Converting \OIII/\Hb\ ratios measured at intermediate redshifts ($z\sim2$), we can find the intrinsic \OIII/\Ha\ ratios to be within $\sim \pm0.4$ \citep[e.g.,][]{2015ApJ...801...88S,2017ApJ...836..164S}.

\subsection{Comparison of the subsamples}

\begin{figure}[t] 
   \centering
   \includegraphics[width=3.5in]{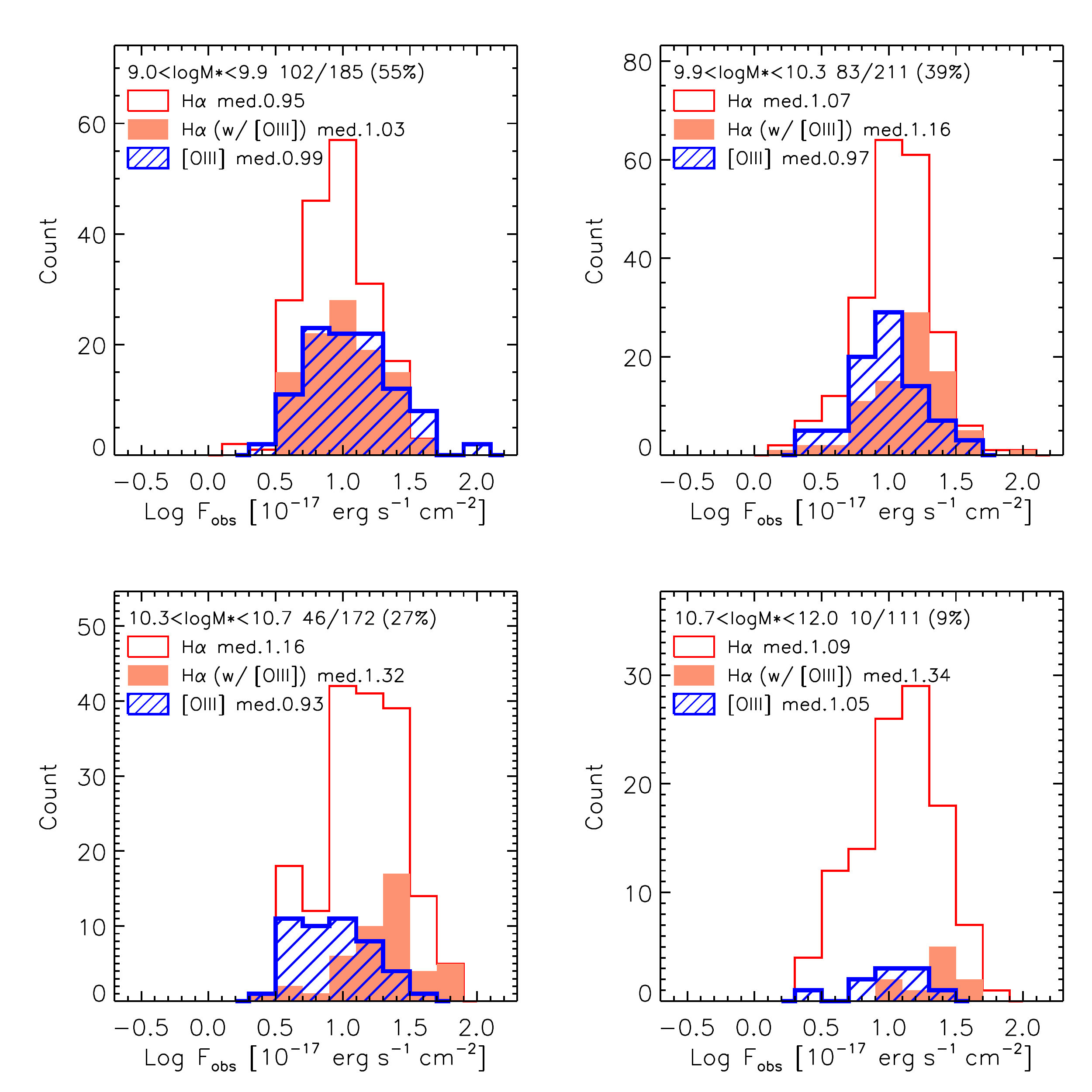} 
   \caption{Distribution of \Ha\ and \OIII$\lambda 5007$ fluxes (corrected for aperture loss, but not for dust extinction) in different bins of $M_\ast$: $\log M_\ast/M_\odot = [9.0:9.9]$ (upper-left); $[9.9:10.3]$ (upper-right); $[10.3:10.7]$ (lower-left); $\ge 10.7$ (lower-left).  Empty histograms indicate the $F_\mathrm{H\alpha}$ distribution for the \Ha-emitters.  Red filled and blue hatched histograms indicate distributions of \Ha\ and \OIII$\lambda5007$ fluxes, respectively, for the \Ha+\OIII-emitter sample.  In each panel, we give the numbers of \Ha-emitters and \Ha+\OIII-emitters (e.g., in the upper-left panel, the numbers of \Ha-emitters and \Ha+\OIII-emitters within the bin are 185 and 102 (55\%), respectively), as well as the median values of the observed \Ha\ and \OIII\ fluxes in log scale in units of $10^{-17}~\mathrm{erg~s^{-1}~cm^{-2}}$.}
   \label{fig:fobs_hist_Mbins}
\end{figure}

\begin{figure}[t] 
   \centering
   \includegraphics[width=3.5in]{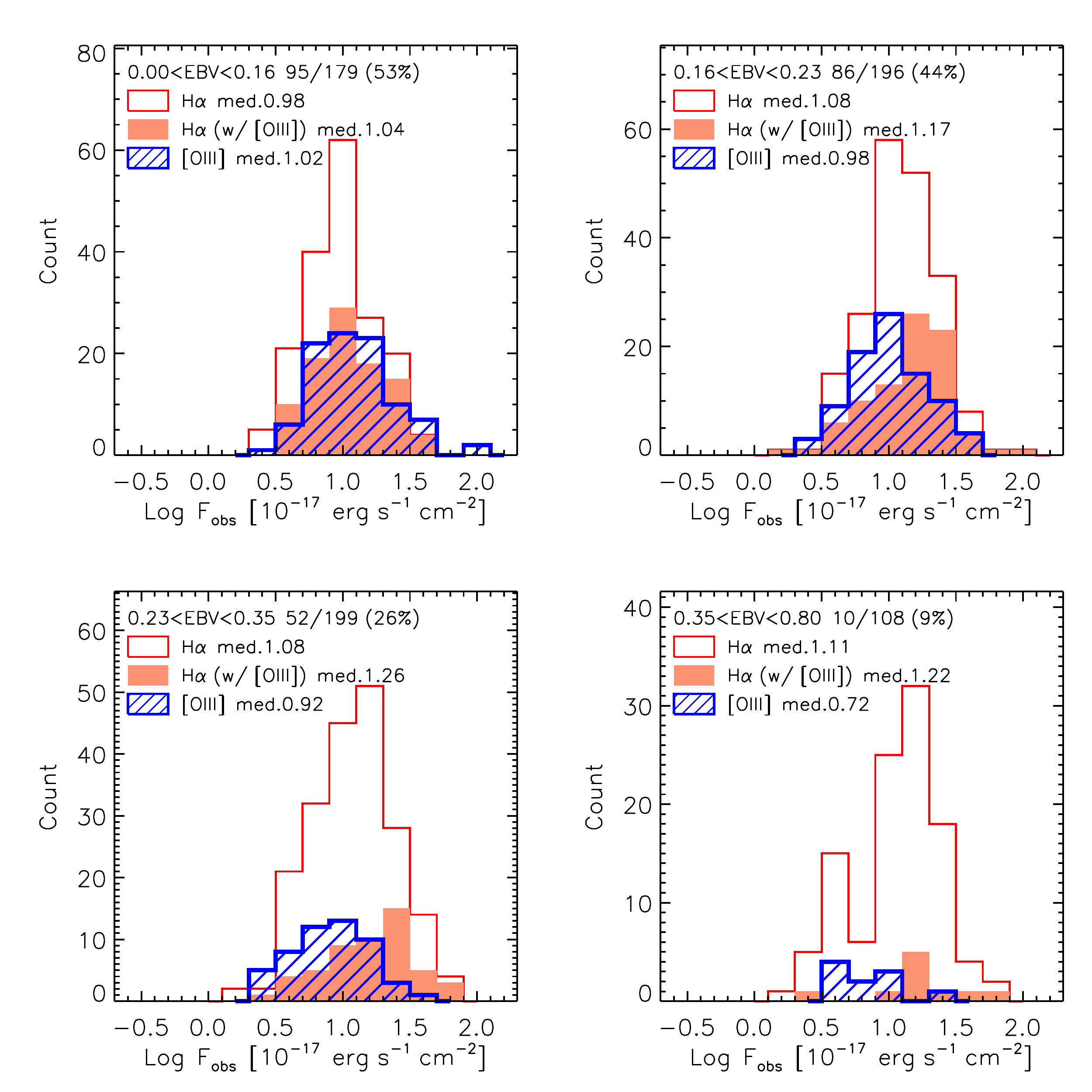} 
   \caption{Same as Figure \ref{fig:fobs_hist_Mbins}, but in bins of extinction: $ E_\mathrm{star}(B-V) = [0.0:0.16]$ (upper-left); $[0.16:0.23]$ (upper-right); $[0.23:0.35]$ (lower-left); $\ge0.35$ (lower-left).}
   \label{fig:fobs_hist_EBVbins}
\end{figure}

The lack of \OIII\ detection at high $M_\ast$ indicates possible biases that exist between the \Ha- and \OIII-selected populations.  The \OIII\ flux decreases more rapidly with increasing metallicity and the level of dust extinction, and thus the \OIII/\Ha\ ratio depends significantly on these properties.  To indicate biases between the \Ha- and \OIII-emitter subsamples, we separate the objects into different $M_\ast$ bins, and compute the fraction of objects with \OIII\ detection in each bin.  We then compare the distribution of the observed \Ha\ and \OIII\ fluxes between the subsamples in each bin.  In Figure \ref{fig:fobs_hist_Mbins}, we show the distribution of the aperture-corrected \Ha\ fluxes (not corrected for dust) for the \Ha-emitter and \Ha+\OIII-emitter samples in four bins of stellar mass ($\log M_\ast/M_\odot < 9.9$,  $9.9\le \log M_\ast/M_\odot < 10.3$, $10.3\le \log M_\ast/M_\odot < 10.7$, $\log M_\ast/M_\odot \ge 10.7$).  For the \Ha+\OIII-emitter sample, we also plot the distribution of observed \OIII\ fluxes.  In each panel, we give the numbers of objects in the \Ha-emitter and \Ha+\OIII-emitter samples, as well as the median values of the observed \Ha\ and \OIII\ fluxes in log scale in units of $10^{-17}~\mathrm{erg~s^{-1}~cm^{-2}}$.

In the lowest $M_\ast$ bin (upper-left panel in Figure \ref{fig:fobs_hist_Mbins}), we detected the \OIII\ line ($\ge 3\sigma$) for more than half (55\%) the \Ha-emitter sample.  The median \Ha\ fluxes of the \Ha- and \Ha+\OIII-emitter samples are similar to each other.  At higher $M_\ast$ bins, however, the fraction of the \Ha+\OIII-emitters is lower: 39, 27, and 9\%\ in the second, third, and the forth bins, respectively.  It is also clear that a difference becomes apparent between the median \Ha\ fluxes of the \Ha-emitter and \Ha+\OIII-emitter samples: the \Ha+\OIII-emitters are biased towards higher \Ha\ flux.  

For further insights, we separate the sample into bins of the level of extinction $E_\mathrm{star}(B-V)$ estimated from the UV slope (see Section \ref{sec:SFRUV}).  In Figure \ref{fig:fobs_hist_EBVbins}, we show the distribution of observed fluxes in four $E_\mathrm{star}(B-V)$ bins ($0 \le \log E_\mathrm{star}(B-V) < 0.16$,  $0.16\le E_\mathrm{star}(B-V) < 0.23$, $0.23 \le E_\mathrm{star}(B-V) < 0.35$, $E_\mathrm{star}(B-V)\ge 0.35$).  It is evident that, similarly to the trend with $M_\ast$, the fraction of the \Ha+\OIII-emitters decreases with increasing extinction value, from 53\%\ in the lowest bin to 9\%\ in the highest bin.

Our results clearly indicate that the \OIII\ emission line traces more preferentially lower $M_\ast$ galaxies and/or objects less obscured by dust.  We note that there is a strong correlation between $M_\ast$ and $E_\mathrm{star}(B-V)$ (Figure \ref{fig:LePhare_M_vs_UVSFR}).  In less massive galaxies, the higher \OIII\ fluxes are associated with lower metallicities.  We thus conclude that the use of \OIII\ for the FMOS-COSMOS sample comes with biases towards lower $M_\ast$, lower metallicity, and/or less-obscured populations than those traced by \Ha\ at the same flux limit.  Even so, the \OIII\ emission line is a powerful tool for galaxy surveys at high redshifts, since low-mass and low-metallicity galaxies may be a dominant population in the early Universe.

Note that the \OIII- and \OIII-single emitter samples are not purely selected by the \OIII\ line because we included the criterion on the predicted \Ha\ flux in the pre-selection of the spectroscopic targets to achieve a high success rate of detecting \Ha.  Indeed, the majority of the \OIII-emitter sample have a detection of \Ha, and thus there are a small number of \OIII-single emitters in our FMOS catalog.  We do not see any significant trends either in the \Ha-detection fraction or the average \OIII\ flux of the \Ha+\OIII\ emitters relative to the \OIII-emitters.

\section{Summary}
\label{sec:summary}

In this paper, we presented our analyses of near-IR spectra collected through the FMOS-COSMOS survey and the basic properties of spectroscopic measurements of star-forming galaxies based on the full catalog that contains 5427 galaxies.  The full FMOS-COSMOS catalog contains spectroscopic measurements of redshift and line fluxes for 1931 objects, including 1204 \Ha\ detections at $3\sigma$ at $1.43\le z\le 1.74$, down to the in-fiber flux limit of about $\sim1\times10^{-17}~\mathrm{erg~cm^{-2}~s^{-1}}$.  The full sample combines the main population of star-forming galaxies along the main sequence at $z\sim1.6$ with the stellar mass range of $9.5\lesssim \log M_\ast/M_\odot \lesssim 11.5$, and other specific subsamples of infrared-luminous galaxies at $z\sim0.9$ and $z\sim1.6$ and {\it Chandra} X-ray sources.  The success rate of the spectroscopic measurement achieves 43\% for the primary sample (Section \ref{sec:redshifts_Primary-HL}).  The full version of the catalog is publicly available online
\footnote{The full FMOS-COSMOS catalog is available here: \\
\url{http://member.ipmu.jp/fmos-cosmos/fmos-cosmos_catalog_2019.fits}\\ 
For more information, please refer to the README file:\\
\url{http://member.ipmu.jp/fmos-cosmos/fmos-cosmos_catalog_2019.README}}.

The precision of the redshift measurement is estimated to $\approx 70~\mathrm{km~s^{-1}}$ in standard deviation (Section \ref{sec:assessment}).  Compared to other spectroscopic campaigns, the probability of line misidentification is expected to be less than 10\% for $z\mathrm{Flag}\ge2$.
For all objects, we estimated the correction factor for the aperture loss of the observed fluxes.  The typical uncertainty in the absolute flux calibration, including aperture correction, is found to be $\approx 0.17~\mathrm{dex}$.  We found that our total aperture-corrected flux measurements are in excellent agreement with slitless measurements from the 3D-HST survey (Section \ref{sec:assessment}).

We used the latest sample to update our past analyses.  The enhancement of the extinction toward nebular emission was measured from comparisons between \Ha- and UV-based dust-uncorrected SFRs.  We found that $f_\mathrm{neb}=E_\mathrm{star}(B-V)/E_\mathrm{neb}(B-V)=0.53$, which is consistent with our previous result in \citet{2013ApJ...777L...8K} after taking into account the difference of the extinction laws applied (Section \ref{sec:dust}).

In Section \ref{sec:MS}, the $M_\ast$--SFR relation was remeasured using the recomputed stellar masses (based on $z_\mathrm{spec}$) and dust-corrected \Ha\ luminosities.  The result is in good agreement with an analytical form derived from a compilation of measurements across a wide redshift range.  We found the data is better fit with a parametrization invoking a bending feature of the sequence with a characteristic mass $M_\ast\approx10^{10.2}~M_\odot$.   The estimated scatter in the \Ha-based SFRs with respect to the best-fit $M_\ast$--SFR relation is found to be increases with increasing $M_\ast$, though the sample selection including a limit on the predicted \Ha\ fluxes may result in a reduction of the scatter of the spectroscopic sample.

In Section \ref{sec:lineratio}, we updated the emission-line diagnostic diagrams, and especially fond a significant offset in the S2-BPT (\SII/\Ha\ vs. \OIII/\Hb) diagram relative to low-$z$ galaxies, as originally reported in our previous study.  With this observational feature, we confirmed with higher confidence that the ionization parameter increases in high redshift star-forming galaxies relative to low-$z$ objects.  We redefined the $M_\ast$--\NII/\Ha\ relation and confirmed that the massive ($\ge10^{10.6}~M_\odot$) galaxies have a level of the line ratio, i.e., the gas-phase metallicity, similar to the local galaxies with the same masses.  Furthermore, we evaluated the intrinsic scatter of the $M_\ast$ vs. \NII/\Ha\ relation, and found that the scatter is small ($\approx 0.1$ dex) at high $M_\ast$ (or high \NII/\Ha), while increasing to $\approx0.3$ at low $M_\ast$ (or low \NII/\Ha).  The behavior of the intrinsic scatter is similar to that of the local galaxies when comparing them as a function of the average \NII/\Ha\ ratio at given $M_\ast$.

Comparing subsamples of \Ha- and \OIII-emitters, we found that there is little bias in the observed \Ha\ line flux between the \Ha-single and \Ha+\OIII-detected samples at low masses and/or low exticntion ($\lesssim 10^{10}~M_\odot$, $E_\mathrm{star}(B-V) \lesssim 0.2$).  In contrast, it has been shown that, at higher masses/extinction, the detection of \OIII\ becomes more biased towards a population having higher \Ha\ fluxes (Section \ref{sec:Ha_vs_OIII}).

To conclude, our large spectroscopic survey has established a large (the order of $10^3$) sample of star-forming galaxies at $1.4<z<1.7$, fully filling the redshift desert.  Combining with the rich panchromatic resources in the COSMOS field, the FMOS-COSMOS catalog offers the means to comprehensively learn how galaxies evolve across the cosmic noon era, as well as to elaborate survey strategies with a new generation of multi-fiber spectrographs such as MOONS, or Prime Focus Spectrograph.

\acknowledgements
We thank the Subaru telescope staff, especially K. Aoki, for their expertise in the observations.  This paper is based on the data collected at the Subaru telescope, which is operated by the National Astronomical Observatory of Japan.  We appreciate the MPA/JHU team for making their catalog public.  This work was supported in part by KAKENHI (DK 14J03216; JDS 26400221) through Japan Society for the Promotion of Science (JSPS) and the World Premier International Research Center Initiative (WPI), MEXT, Japan.  
C.M, G.R. and A.R. acknowledge support via the grant INAF PRIN-SKA 2017.
NA is supported by the Brain Pool Program, which is funded by the Ministry of Science and ICT through the National Research Foundation of Korea (2018H1D3A2000902).

{\it Facility}: Subaru (FMOS)

\bibliographystyle{apj}
\bibliography{apj-jour,ads}

\end{document}